\documentclass[14pt]{article}
\pdfoutput=1
\usepackage{amsmath,amssymb,amsfonts,amsthm,bm,bbm,cancel,wasysym}
\usepackage{epsfig,graphics,graphicx,epstopdf}
\usepackage{array,booktabs,colortbl,colordvi,multirow}
\usepackage{colordvi,color,xcolor}
\usepackage{hyperref}
\usepackage{rotating}
\usepackage{appendix}

\newcommand{\emailto}[1]{E-mail: \href{mailto:#1}{\protect \nolinkurl{#1}}} 

\usepackage{verbatim}
\usepackage{cite}
\usepackage{subfig}
\usepackage{setspace}
\usepackage{url}
\usepackage[percent]{overpic}
\usepackage{slashed}
\usepackage{authblk}
\usepackage{xspace}
\usepackage{fullpage}
\usepackage{hyperref}

\def\ie{{\it i.e.}}
\def\eg{{\it e.g.}}
\def\etc{{\it etc}}

\def\to{\rightarrow}

\newskip\zatskip \zatskip=0pt plus0pt minus0pt
\def\matth{\mathsurround=0pt}
\def\lsim{\mathrel{\mathpalette\atversim<}}
\def\gsim{\mathrel{\mathpalette\atversim>}}
\def\atversim#1#2{\lower0.7ex\vbox{\baselineskip\zatskip\lineskip\zatskip
  \lineskiplimit 0pt\ialign{$\matth#1\hfil##\hfil$\crcr#2\crcr\sim\crcr}}}

\begin{document}


\begin{flushright}
SLAC-PUB-17472\\
\today\\
\end{flushright}
\vspace*{5mm}

\renewcommand{\thefootnote}{\fnsymbol{footnote}}
\setcounter{footnote}{1}

\begin{center}

{\Large {\bf Towards a UV-Model of Kinetic Mixing and Portal Matter}}\\

\vspace*{0.75cm}
{\bf Thomas D. Rueter${}^{1,2}$}~\footnote{\emailto{tdr38@stanford.edu}}
{\bf Thomas G. Rizzo${}^2$}~\footnote{\emailto{rizzo@slac.stanford.edu}} and

\vspace{0.5cm}

${}^1${Stanford University, Stanford, CA, USA}

${}^2${SLAC National Accelerator Laboratory, Stanford University, Menlo Park, CA, USA}

\end{center}
\vspace{.5cm}


%
%


{


\begin{abstract}
 
\noindent
The nature of dark matter (DM) and how it might interact with the particles of the Standard Model (SM) is an ever-growing mystery. It is possible that the existence of new `dark sector' forces, yet undiscovered, are the key to solving this fundamental problem, and one might hope that in the future such forces might even be `unified' with the ones we already know in some UV-complete framework. In this paper, following a bottom-up approach, we attempt to take the first steps in the construction of such a framework. The much-discussed possibility of the kinetic mixing (KM) of the `dark photon' with the hypercharge gauge boson of the SM via loops of portal matter (PM) fields, charged in both sectors, offers an attractive starting point for these efforts. Given the anticipated finite strength of the KM in a UV-complete theory, the absence of anomalies, and the lifetime constraints on the PM fields arising from CMB and nucleosynthesis constraints, PM must behave as vector-like copies of the known SM fermion fields, such as those which appear naturally in, \eg, $E_6$-type models. Within such a setup, the SM and their corresponding partner PM fields would be related by a new $SU(2)_I$ gauge symmetry. With this observation as a springboard, we construct a generalization of these ideas where $SU(2)_I$ is augmented by an additional $U(1)_{I_Y}$ factor so that the light dark photon is the result of a symmetry breaking analogous to the SM, \ie, $SU(2)_I\times U(1)_{I_Y}\to U(1)_D$, but with $U(1)_D$ now also broken at the $\lsim$ GeV scale. While SM fields are $U(1)_D$ singlets, as in the conventional dark photon approach, they transform nontrivially under the full $SU(2)_I\times U(1)_{I_Y}$ gauge group. This approach leads to numerous interesting signatures, both at low energies and at colliders, and can be viewed as an initial step in the construction of a more UV-complete framework.
\end{abstract}

\renewcommand{\thefootnote}{\arabic{footnote}}
\setcounter{footnote}{0}
\thispagestyle{empty}
\vfill
\newpage
\setcounter{page}{1}



\section{Introduction}

The nature of dark matter (DM) is one of our biggest mysteries and points to there being new physics beyond the Standard Model (SM). Historically, the favorite candidates for DM 
originated in top-down theories which addressed other outstanding problems within the SM such as axions~\cite{Kawasaki:2013ae,Graham:2015ouw} and WIMPs~\cite{Arcadi:2017kky}. 
While so far obtaining only null results, 
important experimental searches for such states are on-going. However, these as-yet null results, in addition to the realization of how large the parameter space for potential DM candidates might be, have inspired both increased model-building activity and the ever-widening scope of experimental DM 
searches\cite{Alexander:2016aln,Battaglieri:2017aum}.  In almost all 
DM scenarios, new forces carried by non-SM mediator particles are needed to pass through the SM-dark sector barricade thrown down by nature and 
convey the interaction between the DM and the SM that is responsible for the DM obtaining its observed relic density\cite{Aghanim:2018eyx}. Among these,
one of the most experimentally accessible and also well-studied scenarios is that of the dark photon/vector portal model~\cite{vectorportal}. In its simplest form, one imagines a new, 
broken dark $U(1)_D$ gauge group associated with a massive vector dark photon, $V$; here we will assume this symmetry is broken by a the vev of a SM singlet dark Higgs field, $S$, at 
the $\sim$ GeV scale. While DM carries a non-trivial dark charge, $Q_D$, it is assumed that the SM fields do not and so are not directly coupled to $V$. However, via the kinetic 
mixing (KM)~\cite{KM} of $V$ with the familiar SM $U(1)_Y$ weak hypercharge gauge boson, $B$, after electroweak symmetry breaking the SM fields pick up a small coupling to 
$V$ proportional to their electric charge, $\epsilon eQ_{em}$. Here $\epsilon$ describes the strength of the KM~\cite{KM} and thus the suppression of our interactions with the dark 
sector is due to the small, loop-induced value of this parameter in such scenario. While this is an extremely interesting setup, it is likely to only be an IR shadow 
of a much larger and more complex UV-complete theory which may have implications beyond just the, now potentially more intricate, dark sector. 

In our earlier work~\cite{Rizzo:2018vlb}, hereafter designated as {\bf I}, we took the first step on the way to the construction of a more UV-complete KM scenario; it is the purpose 
of this paper to 
take a further step in this direction. The point of view taken in {\bf I} is that the KM generating mechanism itself may provide the first hint to the direction we should follow. As is well known, 
KM is generated by the existence of sets of fields, here called portal matter (PM), that carry both dark charges as well as, at the very least, SM weak hypercharge thus linking the two 
$U(1)$ factors via a 1-loop vacuum polarization-like diagram.  In {\bf I} we argued that in a complete UV-theory the strength of KM, $\epsilon$, is a finite, in principle calculable, quantity as it 
is in GUTs\cite{GUT} with split multiplets. This hypothesis, when combined with the necessity that such new states be unstable since they carry 
SM quantum numbers, led to a number of restrictions on the 
nature of the possible portal matter fields. If such fields are fermionic, as posited in {\bf I}, they must be also be vector-like\cite{vlf,searches} with respect to the SM to avoid both 
gauge anomalies and any conflict with precision electroweak measurements\cite{Tanabashi:2018oca}. This led us to consider the rather unique possibility that portal matter can only 
transform as vector-like copies\cite{vlf} of one or more of the usual SM fermion fields{\footnote {We note that such PM states have gotten little attention in the literature~\cite{however}.}. 
In {\bf I}, in order to obtain a finite, radiatively generated $\epsilon$, we considered toy models wherein the portal matter consisted of two vector-like 
copies of a single SM field with opposite dark charges thus rendering $\epsilon$ finite in a rather trivial manner. However, there we also briefly discussed the more complex, and perhaps 
much more interesting, possibility that portal matter may consist of a set of different vector-like fields with various SM quantum numbers whose infinite contributions to $\epsilon$ 
`naturally' cancel amongst themselves, as might be the case if they formed, \eg, a $5+\bar 5$ of $SU(5)$,  as a potential way forward in constructing more realistic UV-complete theories. 
This is the direction that we will follow in this paper.

An extension of the SM by an additional $5+\bar 5$ of $SU(5)$ cannot help but remind (some of) us of grand unification based on the group $E_6$, where 
such an additional set of fields naturally occurs \cite{Hewett:1988xc}. Such models proved of great phenomenological interest some time ago in connection with the early developments in string theory\cite{Hewett:1988xc} 
in the late-80's. In 
such a framework, not only was the matter sector of the SM extended but also generally present were SM gauge sector augmentations by various species of $U(1)$ and/or $SU(2)$ 
factors increasing the complexity of the resulting phenomenology. Of course, in the case under discussion here, the obvious gauge sector extension of 
interest is $U(1)_D$ for which all of the SM fields are neutral, unlike what 
we find in the case of new the $U(1)$'s which appear in $E_6$. However, also with an eye towards future work, some of the basic 
ideas in $E_6$-type setups, as we will see below, can provide a good jumping-off point for moving forward. The fact that the SM fields all have $Q_D=0$ tells us that we will actually need to go beyond the simple $E_6$ picture to obtain a viable model, as has been recently noted in the literature \cite{gherghetta2019price}, and we will take a bottom up approach in extending the $E_6$ setup in what follows. The general setup which 
we present below is, to say the 
least, somewhat fine-tuned; how much that should disturb us at this point is up to one's tastes. However, we should be reminded that the work here presents only a preliminary step on 
the way to a more UV-complete theory involving the SM, DM and the portal matter responsible for their mutual interaction, which may result in a reduction of this fine-tuning or at least make such fine-tuning more palatable.

The outline of this paper is as follows: Section 2 contains a general background discussion and an overview of our model setup using arguments from both top-down and bottom-up 
approaches. Section 3 provides a discussion on the various components of our setup in the gauge and fermion sectors and the relevant couplings and mixings among the physical 
states relevant for the discussions that follow. Section 4 contains a mainly collider-oriented view of some of the physics associated with the new exotic particles we have introduced 
while Section 5 contains a discussion of some DM physics within the present setup. Finally, the last Section summarizes our results and conclusions and points us in the direction 
of our next steps.


\section{Background and Model Setup}

In order to attempt to construct a GUT-inspired, UV-complete theory of both the dark matter (DM) and portal matter (PM) sectors as well as their interactions with the SM it is necessary to 
have some preliminary target properties in mind that we would expect such a theory to possess. Similarly, we should have a parallel set of, potentially overlapping, expectations for the 
low(er)-energy theory operating at the $\sim$ few TeV scale and below as clearly one might also anticipate that the lower-energy theory will inherit some of the properties found in the 
high-scale theory. As was discussed briefly in {\bf I} and mentioned above in the Introduction, 
to allow for their instability, new vector-like (with respect to the SM gauge group) candidate fermionic portal matter fields in their simplest 
manifestation must transform in a manner similar to that of some subset of SM fermionic representations, \ie, $SU(3)_c$ color singlets and triplets which also fall into $SU(2)_L$ weak-isospin singlets or doublets.{\footnote{Note that this is already suggestive that a symmetry may exist relating these portal matter fields to their SM `partners'.}} As such, these portal 
matter states automatically carry a set of weak hypercharges, $Y_i$, matching those of the corresponding conventional SM fermions. If these portal matter fields are assigned dark charges, 
$Q_{D_i}\neq 0$, then a {\it finite}, 1-loop induced value for the KM mixing between a light, $\lsim GeV$, dark $U(1)_D$ gauge field and the corresponding $U(1)_Y$ hypercharge field, 
can be induced.  As described above, and in the notation of {\bf I}, the parameter describing this KM is essentially given by 
\begin{equation}
\epsilon \sim \frac{g_Dg_Y}{12\pi^2} \sum_i ~Y_iQ_{D_i}~ \Big({\rm pole}+ln \frac{m^2_i}{\mu^2}\Big)\,
\end{equation}
where $g_{D,Y}$ are the $U(1)_{D,Y}$ gauge couplings and `pole' is the usual singular plus mass-independent constant piece obtained in dimensional regularization. 
$\epsilon$ will be generated via 
the various contributing particle mass differences and will also prove to be {\it finite} and $\mu$-independent provided that the condition $\sum_i ~Y_iQ_{D_i}=0$ is satisfied. 
From a purely low-energy perspective, this requirement would appear to be {\it ad hoc} but should be expected to be a natural property of the UV-theory as was speculated in {\bf I}. 
As is well-known, in a GUT-like framework, for any given representation, $Tr~T_iT_j =0 ~(i\neq j)$ is found to hold automatically for any two distinct generators, $T_i$, which in the case 
at hand renders the KM parameter $\epsilon$ finite and calculable in such models\cite{GUT} if both $Y$ and $Q_D$ correspond to (linear combinations of distinct) generators of the 
UV gauge group. This calculability of the analogs of $\epsilon$ in GUT theories with split representations was already made use of long ago\cite{GUT} for model-building/phenomenological purposes. 

To gain some further insight into these issues and to provide a point of departure for the analysis that follows, let us briefly consider the anomaly-free GUT group $E_6$\cite{Hewett:1988xc} whose 
fundamental representation, 27, contains the usual fifteen SM chiral fermions together with twelve additional  `exotic', (almost) vector-like fermions (VLF). Of particular interest to us 
here is the decomposition of $E_6$ into one of its maximal subgroups $SU(2)_I \times SU(6)$\cite{London:1986dk} under which the fundamental 27 representation decomposes as 
$27 \to (2,\bar 6)+(1,15)$, where the first[second] number is the dimension of the corresponding $SU(2)_I[SU(6)]$ representation. This $SU(2)_I$ gauge group will play an important 
role in our discussion below. When the $SU(6)$ group then breaks to the familiar 
$SU(5)\times U(1)_6$, where $SU(5)$ will be used here in the following discussion as just a proxy for the SM, one then has $27\to (2,\bar 5)_{-1} +(2,1)_5 +(1,5)_{-4}+(1,10)_2$ where 
the second number now labels the $SU(5)$ 
representation and the numbers written as subscripts are the corresponding $U(1)_6$ charges.  Here we see the well-known result that the new exotic fermions lie in a pair of $SU(5)$ 
$5+\bar 5$ representations as well as a pair of $SU(5)$ singlets. Following\cite{Hewett:1988xc} we will denote the vector (left-handed) fields in the $\bar 5$ as $(N,E)_L^T, h_L^c$, with the corresponding 
conjugate fields in the $5$. While $h_L^c$ is a color (anti-)triplet, weak isosinglet, transforming similarly to $\sim d_L^c$,  $(N,E)_L^T$ is a color singlet, weak isodoublet, transforming 
similarly to $\sim (\nu,e)_L^T$, and together with the content of the 5 these fields form a set of fermions which are vector-like with respect to the SM. The additional SM singlets in 
the $(2,1)_5$ are usually suggestively denoted as $\nu^c$ and $S^c$\cite{Hewett:1988xc} but here we simply refer to them as $S_{1,2}$.  Why is this $E_6$ structure interesting from the portal matter perspective? Here we have an anomaly-free setup which already contains vector-like fermions which transform under the SM in a manner analogously 
to (some of) the SM fermions thus being able to play the role of portal matter and which are related to the SM by the action of the $SU(2)_I$ group, \ie, the value of $T_{3I}$ 
differentiates the SM from exotic fields. Such a possibility was already foreseen in {\bf I}. 
Furthermore, if we were to (mistakenly as we will soon see) identify $Q_D$ with this diagonal generator then the above requirement, the condition $Tr YQ_D=0$ would be automatically 
satisfied so that $\epsilon$ would indeed be finite and calculable at 1-loop as desired. One could go even further and imagine that at some large scale, say $\sim 10$ TeV,  the breaking 
$SU(2)_I \to U(1)_D$ via a real $SU(2)_I$ triplet thus giving heavy masses to the non-hermitian `off-diagonal' $W_I^{(\dagger)}$ gauge bosons coupling to the non-hermitian pair of 
$SU(2)_I$ isospin raising and lowering operators and thus identifying the diagonally coupled $W_I^0$ with the dark photon which is left to get mass at the $\sim$ GeV scale or 
below. Assuming that the $SU(2)_I$ gauge coupling, $g_I$, is even remotely close in magnitude to the usual weak coupling this seemingly successful and interesting approach hits 
a significant snag, \ie, in such a setup the dark photon will have significant couplings,  $\sim g_IT_{3I}$, to some of the SM fields (since they carry $T_{3I}\neq 0$) beyond those 
usually induced by the loop-suppressed KM. This is inconsistent with the desired properties of the dark photon outlined above; thus, we see that we cannot make the identification of $Q_D$ with $T_{3I}$, and that the dark photon cannot be embedded in this simple $E_6$-inspired framework. This setup does have a number of very nice features, albeit with some important ingredients that are still missing. To keep these nice features while repairing the problems we need to go beyond the simple $E_6$ framework. 

First, let us establish some simplifying notation; we will denote the SM $SU(3)_c\times SU(2)_L\times U(1)_Y$ group structure as $3_c2_L1_Y$. Similarly, the above $E_6$ decomposition 
would then be written as $E_6 \to 2_I6\to 2_I51_6$ and the relevant gauge group above the $\sim 10$ TeV scale described above then would be $3_c2_L1_Y2_I$, since $SU(5)$ is broken 
at the very high mass scale $\gsim 10^{14-16}$ GeV. Next, we will take a bottom-up approach to discover what it is we need to add to the previous $E_6$-inspired setup to maintain 
the attractive features above while avoiding the dark photon coupling problem. For the moment, we keep the fermionic matter content of the model the same as that in the previous 
paragraph. Let us 
imagine that we augment the $2_I$ gauge group by an additional $U(1)_{I'}=U(1)_{I_Y}$ (which we will use interchangeably below) with its own gauge coupling $g_I'(=g_{I_Y})$ so 
that above $\sim 10$ TeV the relevant group is now 
$3_c2_L1_Y2_I1_{I_Y}$. In such an approach, the relevant KM is then between the usual SM hypercharge gauge boson and the corresponding gauge boson of $1_{I_Y}$ so that, \eg, 
$\epsilon \sim \sum_i Y_iY_{I_i}~ ln (m_i^2/\mu^2)$. In a sense the SM electroweak group is now seen to be `mirrored' in `I-space,' but with several important differences. 
We now further imagine that at the $\sim 10$ TeV breaking scale $2_I1_{I_Y} \to 1_D$, \ie, $U(1)_D$ with a (at this scale) massless dark photon via a $2_I$ doublet vev (which we'll call 
$v_I$ for the moment), in a manner completely analogous to the Higgs in the SM.  This is now very familiar physics: to first approximation the $W_I$ get a mass $\simeq g_Iv_I/2$ while 
the $Z_I$, which couples to a linear combination of the $T_{3I}$ and $Y_I$ generators, gets a mass $\simeq \sqrt{g_I^2+g_I'^2}v_I/2$ so that $M_{Z_I}=M_{W_I}/c_I$, \etc, (where 
$c_I=\cos \theta_I$ with $\tan \theta_I=g_{I_Y}/g_I$, again in complete analogy with the SM). Similarly, the dark photon, which couples to the combination $Q_I=T_{3I}+Y_I/2$ remains, massless down to the $\sim 1 $ GeV scale where $U(1)_D$, literally dark QED, is broken. Note that quite generally the couplings of the $Z_I$ to the 
additional $E_6$ matter fields will now be quite different than in the $E_6$ case itself due to the additional $U(1)_{I_Y}$ factor. Furthermore, for a range of $x_I=s_I^2>0.75$, the $Z_I$ can 
decay to $W_IW_I^\dagger$, which is kinematically forbidden in the ordinary $E_6$-inspired setup.

The low-energy theory, at/below $\sim 10$ TeV, in now seen to have 3 relevant, but widely separated, scales: $\sim10$ TeV where a vev $\sim v_I$ breaks $2_I1_{I_Y} \to 1_D$, the 
usual $\sim 250$ GeV electroweak scale where $2_L1_Y\to1_{QED}$ and the $\sim $ GeV scale where $1_D$ is broken and the dark photon obtains a mass. We now ask: in such a setup 
is it possible to assign the $Y_I$ charges consistently to the various $3_c2_L1_Y2_I1_{I_Y}$ matter representations so that $Q_I=0$ holds for all 
SM fields while the exotic fermions carry $Q_I\neq 0$, thus allowing us to identify $Q_I$ with $Q_D$? As we will shortly see the answer to this question is in the affirmative.

Since, as we will see, all of the fermions in the 27 of $E_6$ which transform non-trivially under $2_I$ must also have $1_{I_Y}$ charges, while the SM fermions which are singlets under $2_I$ must not receive $1_{I_Y}$ charges to avoid coupling to the dark photon, it is immediately clear that we cannot identify $1_{I_Y}$ with the $1_6$ arising from the $E_6$ breaking above. This implies that $1_{I_Y}$ is {\it not} a subgroup of $E_6$, and furthermore that the relevant GUT-like group is not the direct product $E_6 \times 1_{I_Y}$, as this would cause all fields in the 27 to have the same charge under $1_{I_Y}$. {\footnote {We note that the generator $Y_I$ does {\it not}, in 
fact, commute in general with those of $E_6$, so this is not just $E_6\times U(1)_{I'}$ and instead some other large unification group would be relevant at very high scales.}}. Hence, at 
higher scales $3_c2_L1_Y2_I1_{I_Y}$ (or more compactly $2_I1_{I_Y}5$) must be part of a larger group, \eg, $SU(8)$, which does not contain $E_6$ as a subgroup. Thus, while $E_6$ has and will continue to inspire and provide us some guidance in our model 
construction, especially in the matter sector, it will not directly feature in the final UV aspects of the present work. In later work\cite{tomorrow} we will more explicitly discuss how the 
present framework fits into a more unified GUT-like structure.

\section{Realizing the Bottom-Up Model}

\subsection{Basic Model Structure} \label{sec:bas_mod}

Although we will not make explicit use of the $E_6$ group below, comparisons with that quite familiar scenario which leads to the additional $SU(2)_I$ gauge group with the specific model 
we consider here will often prove useful. 
At the most basic level of $3_c2_L1_Y2_I1_{I'}$, the fermion fields as described above will fall into a number of distinct representations. Here, where relevant, we assign the (conjugate) SM 
fields to be the $T_{3I}=1/2(-1/2)$ `upper (lower) member' of $2_I$ left-handed doublets with the corresponding exotic (conjugate) fields then having $T_{3I}=-1/2(1/2)$. 
The fields in the familiar $SU(5)$ 10 representation: $(u,d)_L^T, u_L^c,e_L^c$ are thus singlets under $2_I1_{I'}$ in analogy with the case describe above, \ie, $(u,d)_L^T \sim (3,2,1/6,1,0)$,
$u_L^c \sim (\bar 3,1,2/3,1,0)$ and $e_L^c \sim (1,1,1,1,0)$ under $3_c2_L1_Y2_I1_{I'}$, respectively.{\footnote {Note that when discussing SM fermions we will employ first generation 
labels.}}  On the other hand, all the other exotic fields (and the remaining SM ones) are then found to transform non-trivially, \eg,  
$(E^c,N^c)_L^T \sim (1,2,1/2,1,1)$ and $h_L \sim (3,1,-1/3,1,1)$. The remaining leptonic fields are then seen to form a bi-doublet under $2_L2_I$ similar to what one encounters in the 
Left-Right Symmetric Model\cite{LRM}, $[(\nu,e)^T;(N,E)^T]_L \sim (1,2,-1/2,2,-1/2)$ with the $2_L(2_I)$ generators acting vertically(horizontally) while $(h^c,d^c,)_L$ is an $2_L$ singlet but 
an $2_I$ doublet $\sim (\bar 3,1,1/3,2,-1/2)$. Finally, we also have the SM singlets $(S_2,S_1)_L \sim (1,1,0,2,-1/2)$; as a further distinction from the pure $E_6$ model we will now make 
this representation also vector-like under $2_I$ by adding the corresponding conjugate fields, $(S_1^c,S_2^c)_L \sim (1,1,0,2,1/2)$, to the spectrum which will assist in our 
model-building efforts below. Note that this is a particular convenient {\it choice} and alternatives are possible where these fields remain 2-component fields which obtain a Majorana mass. 
A short calculation shows that indeed the fermionic contributions to $Tr YY_I=0$ so that $\epsilon$ is in fact finite and calculable in this setup (assuming a similar cancellation happens 
in the scalar sector). It is also important to note that the above assignments lead to $Q_I=T_{3I}+Y_I/2 =0$ for all SM fields (and also $S_2$) while their exotic partners have 
$Q_I(h_{L,R})=1$ and $Q_I(N,E,S_1)_{L,R}=-1$ so that the dark photon, $V=A_I$, prior to mixing does not couple to the SM but couples in a vector-like manner to all the exotic 
fermions. Note further that $Tr Q_I=0$ as one might expect from any linear combination of GUT (here $SU(6)$-like) generators. In general, it may be that the exotics pair up only with 
one of the SM generations and not the others or that there are exotic partners for all three SM generations; we will keep these possibilities open in the discussion below but for convenience 
continue to employ first generation labels, constructing the model in a single generation language. 

An observation that we will {\it not} make use of here to keep the discussion general is that the fermion and Higgs fields that we introduce naturally fit into the embedding 
$2_I1_{I'}\to 3_I$, with $SU(3)_I$ broken at an even larger mass scale. We will return to this observation in later work as it provides one of the natural next steps in the constructing of 
a more UV-complete model. 

Given the augmentation of both the low-energy ($\sim 10$ TeV) gauge group and the particle content one can ask if the gauge and mixed gauge-gravity anomalies still cancel in the case 
of $3_c2_L1_Y2_I1_{I'}$ as they did automatically in the previously described 
$E_6$ group; the vector-like nature of much of the above structure is helpful here, but we note that fermion fields are generally 
{\it not} vector-like under $2_I1_{I'}$ itself.{\footnote {We note that the fermions in analogous $SU(2)_I$ situation are also not vector-like under $2_I$, but that did not prevent the cancellation of all of the anomalies.}}
As might be expected, all of the dozen or so potential anomaly contributions involving only SM gauge fields or admixtures of SM gauge fields with those of $2_I1_{I'}$ or gravity are found to 
cancel completely as do those arising from $SU(2)_I^3$, \ie, $Tr ~T_{3I}^3(=0)$, and the mixed gauge-gravity case $Tr~Y_I (=0)$. However, two of the necessary anomaly-free 
conditions (assuming only a single generation of exotic fermions), $Tr~ T_{3I}^2(Y_I/2)=-5/4$ and $Tr~(Y_I/2)^3=15/4$, are clearly non-zero so that further additional fermions must be added to 
the spectrum. Provided that they are SM singlets, which is certainly the simplest possibility, we need only ensure that the condition $Tr~Y_I=0$ remains valid when these new fermions are 
included. There are several possible solutions, with the simplest being the addition of a left-handed $2_I$ triplet, $T_L$, with $Y_I/2=1$ plus a left-handed $2_I$ doublet, $D_L$, with $Y_I/2=-3/2$; the specific 
choice of such new fields will not be of direct interest in the discussion below. Of course, in a bottom-up approach we can always add further vector-like fermions as long as they don't 
spoil this anomaly cancellation. These fields will play very little role in the phenomenological discussions below. The fermionic content of the model discussed in above, including the minimal content required for anomaly cancellation, is summarized in Table \ref{fermtab}.

\begin{table}[h]
\caption{Fermionic Field Content}\label{fermtab} 
The fermionic content of the theory, including the additional $SU(2)_I$ doublet and triplet fermions $D_L$ and $T_L$ which are necessary to cancel anomalies involving $U(1)_I$. Note that $(d_L~h_L)^c$ and $(L_L~H_L)$ are doublets of $SU(2)_I$, the latter being a bidoublet under $SU(2)_L\times SU(2)_I$. 
\begin{center}
\begin{tabular}{ l  c  c  c  c  c  c  c }
\hline		
SU(5) & & SU(3)$_C$ & $T_{3L}$  & $Y$/2 & $T_{3I}$  & $Y_I$/2 & $Q_D$  \\ 
\hline
{\bf 10} & $Q \equiv \begin{pmatrix} u \\ d\\\end{pmatrix}_L $  & {\bf 3} & $\begin{pmatrix} 1/2 \\ -1/2 \\ \end{pmatrix}$ & 1/6 & 0 & 0 & 0 \vspace{.1cm} \\
 & $u^c_L$ & $\bf{ \bar3}$ & 0 & -2/3 & 0 & 0 & 0 \vspace{.1cm}\\
 & $e^c_L$ & $\bf{ 1 }$ & 0 & 1 & 0 & 0 & 0 \vspace{.1cm}\\
$\bf{\bar5}$& $L \equiv \begin{pmatrix} \nu \\ e\\\end{pmatrix}_L $ & $\bf{ 1 }$ & $\begin{pmatrix} 1/2 \\ -1/2 \\ \end{pmatrix}$ & -1/2 & 1/2 & -1/2 & 0 \vspace{.1cm}\\
 & $d^c_L$ & $\bf{ \bar3 }$ & 0 & 1/3 & 1/2 & -1/2 & 0 \vspace{.1cm}\\
$\bf{\bar5}$ & $H \equiv \begin{pmatrix} N \\ E\\\end{pmatrix}_L $ & $\bf{ 1 }$ & $\begin{pmatrix} 1/2 \\ -1/2 \\ \end{pmatrix}$  & -1/2 & -1/2 & -1/2 & -1 \vspace{.1cm}\\
 & $h^c_L$ & $\bf{\bar3 }$ & 0 & 1/3 & -1/2 & -1/2 & -1 \vspace{.1cm}\\
{\bf 5} & $H^c \equiv \begin{pmatrix} E \\ N\\\end{pmatrix}^c_L $ & $\bf{ 1 }$ & $\begin{pmatrix} 1/2 \\ -1/2 \\ \end{pmatrix}$ & 1/2 & 0 & 1 & 1 \vspace{.1cm}\\
 & $h_L$ & $\bf{ 3 }$ & 0 & -1/3 & 0 & 1 & 1 \vspace{.1cm}\\
{\bf 1} & $\begin{pmatrix} S_2 \\ S_1 \\ \end{pmatrix}_{L,R}$ & $\bf{ 1 }$ & 0 & 0 & $\begin{pmatrix} 1/2 \\ -1/2 \\ \end{pmatrix}$ & -1/2 & $\begin{pmatrix} 0 \\ -1 \\ \end{pmatrix}$  \vspace{.05cm}\\
{\bf 1} & $D_{L}\equiv \begin{pmatrix} D^- \\ D^{--}\\\end{pmatrix}_L$ & $\bf{ 1 }$ & 0 & 0 & $\begin{pmatrix} 1/2 \\ -1/2 \\ \end{pmatrix}$ & -3/2 & $\begin{pmatrix} -1 \\ -2 \\ \end{pmatrix}$  \vspace{.05cm}\\
{\bf 1} & $T_{L}\equiv \begin{pmatrix} T^{++} \\ T^{+}\\ T^{0} \\ \end{pmatrix}_L$ & $\bf{ 1 }$ & 0 & 0 & $\begin{pmatrix} 1 \\ 0 \\ -1 \\ \end{pmatrix}$ & 1 & $\begin{pmatrix} 2 \\ 1 \\ 0 \\\end{pmatrix}$  \vspace{.05cm}\\
\hline
\end{tabular}
\end{center}
\end{table}

In order to generate all of the gauge and fermion masses as well as the appropriate hierarchy of mass scales, $\sim 10$ TeV, $\sim 250$ GeV and $\sim$ at most a few GeV, several 
different Higgs fields are required which will transform non-trivially under the SM, $2_I1_I$, or both. This requires a scalar sector only slightly larger than in the usual $SU(2)_I$ model  
with some parameter tuning necessary to generate the hierarchal vev structure. The easiest way to uncover these Higgs fields is to require that all of the previously 
discussed fermions obtain masses at the appropriate scales while simultaneously maintaining the desired hierarchy of gauge symmetry breakings. This will, of course, require siome 
fine-tuning in the Higgs potential. 
The simplest situation is the case of the $u$-quark, as neither of its chiral components carries any $2_I1_{I'}$ quantum numbers. A conventional SM-like Higgs doublet $H_1 =(H^+,H^0)^T \sim (1,2,1/2,1,0)$ can thus generate a mass term via the coupling 
\begin{equation}
\lambda_u \bar u_R \begin{pmatrix}u_L \\ d_L \\ \end{pmatrix}_i \begin{pmatrix} H^+ \\ H^0 \\ \end{pmatrix}_j \epsilon^{ij} +h.c.\,,
\end{equation}
when $H^0$ obtains a vev $\left<H^0 \right> =v/\sqrt 2$ with $v\sim 100$ GeV. For the $d$-quark, a different multiplet, $H_2$, is required, with a Yukawa coupling of the form
\begin{equation}
\lambda_d \begin{pmatrix} \bar h_R & \bar d_R \\ \end{pmatrix}_J \begin{pmatrix}u_L \\ d_L \\ \end{pmatrix}_i \begin{pmatrix} h_1^0 & h_2^0 \\ h_1^- & h_2^- \\  \end{pmatrix}_{jJ} \epsilon^{ij} +h.c.\,,
\end{equation}
from which it is clear that $H_2$ is a $2_L2_I$ bidoublet, $H_2$ $\sim (1,2,-1/2,2,1/2)$. We have adopted the convention that $SU(2)_L$ indices $i$ label rows while $SU(2)_I$ indices $I$ label columns. Note that since both of the $T_{3L}=1/2$ entries in $H_2$ 
are electrically neutral, $h_{1,2}^0$, both of these fields may obtain vevs, $v_{1,2}/\sqrt 2$, with $v_2 \sim 100$ GeV. Since $Q_I(h_{1,2}^0)=1,0$, $v_2$ (in combination with $v$ above) breaks the SM gauge group but has no impact on $U(1)_D$. On the other hand, since $h_1^0$ carries a dark charge its vev generates a mass for the dark photon and so 
$v_1 \sim$ a few GeV or less. Thus we see that while $v_2$ generates the $d$ mass, $v_1\neq 0$ generates mass-mixing between $d$ and $h$, a role filled by the dark Higgs in {\bf I}. Furthermore, as we will see below, since $h_1^0$ has {\it both} $T_{3L}$ and $Q_I \neq 0$  it also generates a mass mixing between the dark photon and the SM $Z$ which is not 
suppressed by a KM factor but (essentially) only the vev hierarchy ratio squared, a feature 
{\it not} present in {\bf I}. The corresponding Yukawa coupling 
\begin{equation}
\lambda_e \bar e_R \begin{pmatrix} \nu_L & N_L \\ e_L & E_L \\ \end{pmatrix}_{iI} \begin{pmatrix} h_1^0 & h_2^0 \\ h_1^- & h_2^- \\  \end{pmatrix}_{jJ} \epsilon^{ij} \epsilon^{IJ}+h.c.\,,
\end{equation} 
is also seen to generate the electron mass as well as $e-E$ mass mixing via this vev $v_1$.

A different Higgs field is needed to generate the exotic fermion masses via the couplings
\begin{equation} 
\lambda_h\begin{pmatrix} \bar h_R & \bar d_R \\ \end{pmatrix}_I h_L \begin{pmatrix} h_3^0 & h_4^0 \\  \end{pmatrix}_I+h.c.\,, 
\end{equation}
and 
\begin{equation}
\lambda_E \begin{pmatrix} \bar N_R \\ \bar E_R \\ \end{pmatrix}_i \begin{pmatrix} \nu_L & N_L \\ e_L & E_L \\ \end{pmatrix}_{iI} \begin{pmatrix} h_3^0 & h_4^0 \\  \end{pmatrix}_J \epsilon^{IJ} +h.c.\,,
\end{equation}
where $H_3 = (h_3^0,h_4^0) \sim (1,1,0,2,-1/2)$, with both neutral members, $h_{3,4}^0$, generally obtaining vevs, $v_{3,4}/\sqrt 2$. Here we see that $Q_I(h_3^0)=0$ so that 
$v_3 \sim 10$ TeV generates  
the large $N,E$ and $h$ masses while simultaneously breaking $2_I1_{I'} \to 1_D$ giving masses to the $W_I^{(\dagger)}$ and $Z_I$ gauge bosons. Since $Q_I(h_4^0)\neq 0$ the vev 
$v_4 \sim $ a few GeV or less contributes to the breaking of $1_D$ (while {\it not} generating any additional $Z$-dark photon mixing) and leads to a further contribution to 
$e-E$ and $d-h$ mass mixing (of the opposite helicity) at the $\sim $ GeV scale.  Finally, we note that at this level of discussion $(S_2,S_1)$ can have a bare Dirac mass term, 
$M$, consistent with all the $3_c2_L1_Y2_I1_{I'}$ gauge symmetries. For now it will be assumed that $M$ can be either quite large, of order the scale of scale of $2_I$ breaking, {\it or} 
might be small as $\lsim 1$ GeV, where the DP gets a mass, although this would likely be a very highly tuned 
value. As mentioned earlier, we could instead choose the potentially more interesting path to make the $S_i$ Majorana fields employing an $SU(2)_I$ triplet in the case when the exotic 
fields come in two or more generations. Furthermore, perhaps even more interestingly, we could imagine even more complex scenarios where other mass terms exist due linking the 
$S_i$ with the other neutral fields $\nu,N$ so that we can make interesting models of neutrino masses; this will require additional Higgs fields. However, neither of these paths will 
be followed here and we leave this for later work. We emphasize that we are free to add additional (vector-like) fermion fields, so long as they don't violate the `finiteness' condition 
$Tr YY_I=0$. 
Finally, note that we will also add one new $SU(2)$ singlet, electrically charged scalar, $H_4 \sim (1,1,1,1,1)$, which (clearly) does 
not get a vev, and is introduced solely to ensure that the scalar contributions to $TrYY_I$ vanishes in this bottom-up construction, playing no role in the phenomenology below. The various Higgs fields, and their vevs, are summarized in Table \ref{higgstab}. Note that the Higgs sector structure just described is somewhat more complex than seen in the traditional $SU(2)_I$ scenario, though they both have many of the same features. 

Now that we have elucidated the necessary set of Higgs fields needed to generate all the fermion masses and the breakings of the necessary gauge symmetries we can 
write the corresponding potential in the form 

\begin{equation}
\begin{aligned}
V = &\mu_1^2 H_1^\dagger H_1+\mu_2^2 \textrm{Tr}(H_2^\dagger H_2) +\mu_3^2 H_3^\dagger H_3 + \mu_4^2 |H_4|^2 + \lambda_1  (H_1^\dagger H_1)^2 + \lambda_2 [ \textrm{Tr}(H_2^\dagger H_2)]^2 \\ & + \alpha_1 \textrm{Tr}(H_2^\dagger H_2 H_2^\dagger H_2) + \alpha_2 \textrm{Tr}(H_2^\dagger H_2 \tilde H_2^\dagger \tilde H_2) + \lambda_3 (H_3^\dagger H_3)^2 + \lambda_4 |H_4|^4 \\ &+ \lambda_5 H_1^\dagger H_1 \textrm{Tr}(H_2^\dagger H_2) + \lambda_6  H_1^\dagger H_1  H_3^\dagger H_3 + \lambda_7  H_1^\dagger H_1 |H_4|^2 + \lambda_8 H_3^\dagger H_3 \textrm{Tr}(H_2^\dagger H_2) \\ &+ \lambda_9 |H_4|^2 \textrm{Tr}(H_2^\dagger H_2) + \lambda_{10} H_3^\dagger H_3 |H_4|^2 + \rho H_3 \tilde H_2^\dagger H_1 + \rho^* H_1^\dagger \tilde H_2 H_3^\dagger,
\end{aligned}
\end{equation}
where we have maintained the convention that $H_3 = (h_3^0 ~ h_4^0)$ and $H_1 = (H^+ ~ H^0)^T$, and we have defined $\tilde H_2 \equiv \epsilon H_2^* \epsilon$. The phase of the coupling $\rho$ may be absorbed into the relative phases of $H_1$ and $H_3$, leaving us with 17 real parameters. From the form of this potential it is obvious that after spontaneous symmetry breaking several fine-tunings in the various parameters are necessary to generate the required hierarchy of the vevs discussed above.

\begin{table}
\caption{Higgs Sector Content}\label{higgstab}
The Higgs content required to generate masses for the fermionic content of the theory. The charges under relevant gauge groups are summarized, and the vev arrangement and approximate scales of the vevs are listed. The vevs are assumed to be real. 
\begin{center}
\begin{tabular}{ l c c c c c }
\hline
$\Phi$ & SU(2)$_L$ & $Y/2$ & SU(2)$_I$ & $Y_I/2$ & $\left< \Phi \right>$ \\
\hline
$H_1$ & {\bf 2} & 1/2 & {\bf 1} & 0 & $\frac{1}{\sqrt{2}}\begin{pmatrix} 0 \\ v \\ \end{pmatrix} \sim \begin{pmatrix} 0 \\ 100 \textrm{ GeV} \\ \end{pmatrix}$  \vspace{0.1cm}\\
$H_2$ & {\bf 2} & -1/2 & {\bf 2} & 1/2 &  $\frac{1}{\sqrt{2}}\begin{pmatrix} v_1 & v_2 \\ 0 & 0 \\ \end{pmatrix} \sim \begin{pmatrix} 1 \textrm{ GeV} & 100 \textrm{ GeV} \\ 0 & 0 \\ \end{pmatrix}$ \vspace{0.1cm}\\
$H_3$ & {\bf 1} & 0 & {\bf 2} & -1/2 & $\frac{1}{\sqrt{2}}\begin{pmatrix} v_3 & v_4 \\ \end{pmatrix} \sim \begin{pmatrix} 10 \textrm{ TeV} & 1 \textrm{ GeV} \\ \end{pmatrix}$ \vspace{0.1cm}\\
$H_4$ & {\bf 1} & 1 & {\bf 1} & 1 & 0 \\
\hline 
\end{tabular}
\end{center}
\end{table}

From this discussion, and as we will see below, it is clear that the dark photon will eventually couple to the SM fields in several ways: ($i$) via the usual KM $\sim e\epsilon Q$, ($ii$) via 
$\epsilon$-unsuppressed mass mixing with the SM $Z$ boson induced by $v_1\neq 0$ and ($iii$) via the $v_{1,4} \neq 0$ induced mass mixing of the exotic fermions with their 
SM partners. Note that the latter two possibilities automatically lead to parity-violating interactions of the dark photon with (some of) the SM fields. If all these effects are of a similar 
magnitude the nature of the dark photon interactions with the SM fields may be quite different than is usually anticipated. 

\subsection{Gauge Boson Masses and KM} \label{sec:gauge_sec}

The couplings of the various physical gauge fields to the previously introduced fermions and scalars is determined by both the presence of KM as well as mass mixing among the various 
weak eigenstates. 
Based upon the Higgs sector described in the previous subsection, it is straightforward to determine to leading order the masses of the non-hermitian gauge fields, \ie, the SM $W$ 
and the $W_I$ which couple to the two sets of raising and lowering operators of the two $SU(2)$'s; these (diagonal) mass terms are given by 
\begin{equation}
M_W^2=  \frac{g^2}{4} (v^2+v_1^2+v_2^2)\simeq \frac{g^2}{4} (v^2+v_2^2)  ~~~~~~   M_{W_I}^2 =\frac{g_I^2}{4} (v_3^2+v_1^2+v_2^2+v_4^2) \simeq \frac{g_I^2}{4} v_3^2\,,
\end{equation}
where in the second step we have noted and made use of the large vev-squared hierarchies $v_3^2>>v^2,v_2^2>>v_{1,4}^2$ based on the suggestive numerical values discussed 
above. Here we can define the sum $v_{SM}^2=v^2+v_2^2$ for later use below. As in the type-II Two Higgs Doublet Model\cite{Branco:2011iw}, we see that $v$ gives mass 
to the $u$ quarks 
while $v_2$ provides mass to the SM $d$ quarks; by analogy, we may write $v=v_{SM} \sin \beta$ and $v_2=v_{SM} \cos \beta$ in familiar notation. 
Note that $W_I$ can have sub-leading mixings with the neutral hermitian gauge bosons, primarily with the dark photon, via the $v_{1,4}$ vevs that we will discuss further below.

In the case of the real neutral, diagonally-coupled gauge bosons, the relevant part of the covariant derivative (suppressing Lorentz indices here) is given in the weak basis by the 
combination
\begin{equation}
~~~~ gT_{3L}W_3 +g_Y \frac{Y}{2}\hat B +g_I T_{3I} W_{3I} +g_{Y_I} \frac{Y_I}{2}  \hat B_I \,.
\end{equation}
Before we can analyze the masses and couplings of the three massive neutral hermitian gauge bosons in the present setup, we first must remove the effects of KM that arise from a Lagrangian term, now following conventional normalization, of the form (keeping the $`w\leftrightarrow I' $symmetry)
\begin{equation}
{\cal L}_{KM}=  ~\frac{\epsilon}{2c_wc_I} \hat B_{\mu\nu} \hat B_I^{\mu\nu} \,,  
\end{equation}
where $\hat B_{\mu\nu}$ is the field strength the SM $1_Y$ gauge field, $\hat B_I^{\mu\nu}$ is the corresponding $1_{I'}$ field strength and where the finite value of the fermionic contributions to $\epsilon$ (which, to be general, we will assume are chiral) are now given in the notation above by 
\begin{equation}
\epsilon =c_w c_I\frac{g_Y g_{Y_I}}{24\pi^2} \sum_i ~\frac{Y_i}{2} \frac{Y_{I_i}}{2}~ ln \frac{m^2_i}{\mu^2}\,.
\end{equation}
An additional overall factor of 1/2 is present in the corresponding sum of potential complex scalar contributions. Here we see explicitly that in this setup the exotic vector-like fermions 
as well as their $SU(2)_I$ (bi)doublet SM partners are {\it both} playing the role of portal matter fields. In the case of a single generation of exotic fermions we can 
already evaluate this sum explicitly to find that (treating the Dirac mass of the SM neutrino to be $\sim m_e$ for now)
\begin{equation}
\epsilon = 3.87\times 10^{-4}~\Big(\frac{g_Is_I}{gs_w}\Big)~\Bigg[ 3\log \frac{m_E}{m_h}+\log \frac{m_e}{m_d} +\log \frac{m_{H_4}}{m_{H_2}}\Bigg]\,,
\end{equation}
where the term in the square bracket is $O(1)$ (and likely to be {\it negative}) since there are no large hierarchies anticipated. Since $\epsilon \sim 10^{-(3-4)}$ we can safely work, 
most of the time, to 
leading order in this parameter, removing the KM by the mapping $\hat B\to B+\epsilon B_I/(c_wc_I)$ and $\hat B_I \to B_I$ and dropping terms of $O(\epsilon^2)$ or smaller. Following 
this step, it is 
useful to make the following familiar rotations as the mass eigenstates are not far from the usual SM expectations, with $A,Z$ being the usual SM fields:
\begin{equation}
~~~~~~~~~~~W_3=c_wZ+s_wA~~~~~~~~~~~~B=c_wA-s_wZ\,, 
\end{equation}
and correspondingly for the $2_I1_{I'}$ gauge fields with $s_w \to s_I$, \etc. Then the above piece of the covariant derivative, after removal of KM, can be written as
\begin{equation}
~~~~~eQA+\frac{g}{c_w}(T_{3L}-s_w^2Q)Z+e_I(Q_I+\eta_2 \frac{Y}{2})A_I+\frac{g_I}{c_I}(T_{3I}-s_I^2Q_I-\eta_1 \frac{Y}{2})Z_I\,,
\end{equation}
where here we have used the usual relation $e=gs_w$ (similarly $e_I=g_Is_I$) and defined the $\eta_i$ parameters to be the combinations of couplings and mixing angles
\begin{equation}
~~~~~\eta_1=\frac{\epsilon gs_ws_I}{g_Ic_w^2}~~~~~~~~~\eta_2=\frac{\epsilon gs_w}{g_Is_Ic_w^2}=\frac{\eta_1}{s_I^2}\,.
\end{equation}
To determine the gauge boson masses we recall that the above covariant derivative 
expression acts on a set of electrically neutral Higgs multiplet members so that $T_{3L}|vevs>=-\frac{Y}{2}|vevs>$. In 
the case of $Z_I$, the $\eta_1$ term only acts on Higgs fields which have SM couplings and which have vevs much smaller than $v_3$. Since this term is already $\epsilon$ suppressed 
it can be dropped. Thus, to leading order in the small parameters and hierarchical squared vev ratios, the relevant piece of the covariant derivative acting on Higgs vevs is simply (recalling 
that some Higgs carry $Q_I\neq 0$)
\begin{equation}
~~~~~~~~~\frac{g}{c_w}T_{3L}Z+(e_IQ_I-\frac{g}{c_w}\epsilon t_wT_{3L})A_I+\frac{g_I}{c_I}(T_{3I}-s_I^2Q_I)Z_I\,.
\end{equation}
Note the familiar form of the dark photon, $A_I$, coupling to SM matter before mass mixing. The $3\times 3$ symmetric $A_I-Z-Z_I$ mass (squared) matrix can now be written as 
\begin{eqnarray}
{\cal M}^2 & =  \left( \begin{array}{ccc}
                           (g_Is_I)^2(v_1^2+v_4^2)+\frac{g^2}{4c_w^2}\epsilon^2 t_w^2(v^2+v_2^2) & \frac{gg_Is_I}{2c_w}v_1^2-\frac{g^2}{4c_w^2}\epsilon t_w(v^2+v_2^2) & M_{13}^2\\                    
                          - & \frac{g^2}{4c_w^2}(v^2+v_1^2+v_2^2)& \frac{gg_I}{4c_wc_I}[(1-2s_I^2)v_1^2-v_2^2]\\
                          - & -& \frac{g_I^2}{4c_I^2}[v_3^2+v_2^2+(1-2s_I^2)^2(v_1^2+v_4^2)]\\
                         \end{array}\right) \,,
\end{eqnarray}
and where 
\begin{equation}
M_{13}^2=\frac{g_I^2s_I}{2c_I}[v_1^2+v_4^2](1-2s_I^2)+\frac{gg_I}{4c_wc_I}\epsilon t_wv_2^2\,.
\end{equation}
Further note that, \eg,  since $v_4^2<<v_3^2$ that, up to corrections of order $(v^2,v_2^2)/v_3^2$, one finds $c_IM_{Z_I}=M_{W_I}$ as expected. While $A_I-Z_I$ mixing is seen to be very 
highly suppressed by the ratios $v_{1,4}^2/v_3^2 \sim 10^{-(7-8)}$, $Z-Z_I$ and $Z-A_I$ mixing may be of some phenomenological importance. To leading order in the vev ratios one finds
\begin{equation}
\theta_{ZZ_I} \simeq \frac{g_I/c_I}{g/c_w}~\frac{M_Z^2}{M_{Z_I}^2}~ \frac{v_2^2}{v^2+v_2^2}\,,
\end{equation}
where the last ratio of vevs is $\cos^2 \beta$ in the language above and in the 2HDM, and is $O(1)$, and similarly
\begin{equation}
\theta_{ZA_I} \simeq -\epsilon t_w +\frac{gg_Is_I}{2c_w} ~\frac{v_1^2}{M_Z^2}\equiv -\epsilon t_w+\sigma\,.
\end{equation}
Thus we see that the SM $Z$ picks up additional suppressed couplings to the `dark' fields via its mixing with both $A_I$ and $Z_I$ and that the $Z_I$ also picks up an additional coupling to 
the SM. These couplings are given by the terms
\begin{equation}
~~~~ \big[g_Is_I Q_I\theta_{ZA_I}+\frac{g_I}{c_I}(T_{3I}-s_I^2Q_I) \theta_{ZZ_I}\big]Z-\frac{g}{c_w}(T_{3L}-s_w^2Q) \theta_{ZZ_I}Z_I\,. 
\end{equation}
These highly suppressed mixing-induced modifications to the SM couplings and the corresponding new couplings to previously `invisible' matter are too small to be presently observable. 
The corresponding $Z$ mass shift is then to leading order 
\begin{equation}
\frac{\delta M_Z^2}{M_Z^2}\simeq ~- \theta_{ZZ_I}^2 \frac{M_{Z_I}^2}{M_Z^2}  =-\gamma~\frac{M_Z^2}{M_{Z_I}^2}\,.
\end{equation}
where $\gamma>0$ is an $\lsim O(1)$ parameter. 
On the other hand,  $A_I$ itself picks up new interactions with the SM fields through its mixing with the $Z$ which, after some algebra, implies that $A_I$ now would couple 
to the combination
\begin{equation}
~~~~ g_Is_IQ_I+e\epsilon Q-\sigma \frac{g}{c_w}(T_{3L}-s_w^2Q)\,.
\end{equation}
Recall that $e_I=g_Is_I$ is what one usually calls the `dark' coupling, $g_D$, in an EFT approach. 
Note that the $Z$-like coupling term, proportional to $\sigma \sim 10^{-4}$ for typical parameter values, is absent in most treatments but arises here due to a Higgs field 
in the bi-doublet carrying both weak isospin and, effectively, a non-zero value of $Q_I$. Such a term clearly produces parity violation, which can lead to important 
phenomenological implications. Also, importantly, $A_I$ via $\sigma \neq 0$ now necessarily couples to the SM neutrinos in a generation-independent manner, leading to potential 
impact in sensitive neutrino experiments such as DUNE.  
Correspondingly, due to this mixing, the physical $A_I$ mass to NLO in the small parameters from the above considerations is
\begin{equation}
M_{A_I}^2 =g_I^2s_I^2(v_1^2+v_4^2)+\sigma(2\epsilon t_w-\sigma)M_Z^2\,,
\end{equation}
where the second term, which can sometimes be numerically important for lighter dark photons and can be of either sign, is also new and thus can be of some general significance. 
To see that both signs are possible in principle, consider the ratio $r_0=\sigma/(\epsilon t_w)$ so that the coefficient of the $M_Z^2$ term is just $r_0(2-r_0)(\epsilon t_w)^2$. 
Numerically, we indeed find that $r_0$ can easily be $O(1)$: 
\begin{equation}
r_0\simeq 0.256 ~\Big(\frac{10^{-4}}{\epsilon}\Big)~\Big(\frac{v_1}{1 \rm GeV}\Big)^2 ~\Big(\frac{g_Is_I}{gs_w}\Big)\,,
\end{equation}
so that the additional term can be negative for some parameter space regions. We note, however, that for dark photon masses $\gsim 50-100$ MeV and $\epsilon \lsim 10^{-(3-4)}$ the first term 
in the expression above is likely to be the far dominant one.

Thus far we have not analyzed the impact of the light vevs $v_{1,4}\neq 0$. Since these lead to the breaking of $U(1)_D$, thereby generating the $A_I$ mass, the non-hermitian fields 
$W_I^{(\dagger)}$ may also mix (slightly) with both the $Z_I$ and $A_I$ by terms such as 
\begin{equation}
~~~~ \frac{g_I^2s_Iv_3v_4}{2\sqrt 2} (W_I+W_I^\dagger)(A_I-t_IZ_I)\,.
\end{equation}
While the $Z_I$ physics we are concerned with here is not at all significantly influenced by this effect, for $A_I$ this mixing induces an additional new coupling term given by
\begin{equation}
~~~ - g_Is_I\frac{v_4}{v_3}\big(T_I^++T_I^-\big)\,,
\end{equation}
where $T_I^\pm$ are the $SU(2)_I$ raising and lowering operators. Here we see that although $A_I$ mixes with the non-hermitian $W_I$ and $W_I^\dagger$, it does so in such a 
way that $A_I$ remains real, as it should. Note that this new interaction is only suppressed by a single power of $v_4/v_3 \sim 10^{-4} \sim \epsilon$, so it can be of some 
numerical consequence and leads to an off-diagonal coupling between the exotic fermions and their SM partners for the dark photon. Further note that a second term of this type of order $v_1v_2/v_3^2$ is also induced, but it is numerically negligible in that the further shift in $A_I$ couplings is only of the order $\sim 10^{-6}$. 
When both of these $W_I$ and $_I^\dagger$ terms are {\it combined} they lead to a {\it negative} shift in $A_I$ mass squared $\delta m_{A_I}^2 \simeq -g_I^2s_I^2v_4^2$ which is of the 
same order as discussed already above, resulting in the final leading order result for the physical $A_I$ mass squared given by
\begin{equation}
M_{A_I}^2 =g_I^2s_I^2v_1^2+\sigma(2\epsilon t_w-\sigma)M_Z^2\,. 
\end{equation}
As noted above, for the parameter ranges of interest to us we expect that the first term will generally be numerically dominant in this expression.

Without studying the details of the rather complex Higgs potential of this setup described above (which is beyond the scope of the present work) we can make some simple observations with 
respect to the numbers of degrees of freedom that remain after spontaneous symmetry breaking; the most straightforward case is the number of charged states. $H_1$ contains 
one charged state $H^+$, while the bi-doublet $H_2$ contains two, $h_{1,2}^-$, and one linear combination of these three states must be eaten to supply the longitudinal components of $W_L^\pm$ implying that 
(including also the state $H_4$) three charged states remain in the physical spectrum. In the neutral Higgs sector there are a total of 10 real fields: 5 CP-even states and 5 CP-odd states. In the case of original five CP-odd neutral fields, we need to supply longitudinal components 
to the $Z_I,Z$, and $A_I$, which implies that three linear combinations of the CP-odd states become Goldstone bosons. The $A_I$ likely eats some combination of the $Q_I \neq 0$ CP-odd $a^0_{1,4}$ states. The final gauge boson in need of a longitudinal component is $W_I^{(\dagger)}$. Since it carries a non-zero $Q_I$, $W_I^{(\dagger)}$ must eat both a real CP-even state, likely some linear combination of $h^0_4$ and $h^0_1$, as well as a CP-odd state, likely the linear combination of $a^0_1$ and $a^0_4$ not eaten by the $A_I$. Thus, we find that four CP-odd states and one CP-even state become goldstone bosons, leaving 5 neutral real scalars remaining in the spectrum, made of some linear combinations of the 4 uneaten CP-even states and 1 uneaten CP-odd state. We identify one of these linear combinations, most likely an admixture of the CP-even isodoublet fields within $H^0$ and $h^0_2$, with the SM Higgs.

\subsection{Fermion Mixing} \label{sec:ferm_mix}

As was seen in the previous section, the exotic fermions naturally mix with their SM $SU(2)_I$ partners via the same set of Yukawa couplings that generate all of the `diagonal' fermion 
masses themselves. There are several model-building possibilities here, but for simplicity of our discussion we will assume that this mixing only occurs with the first generation. It is easy 
to mutate this into the 
case where the mixing is dominated by a different generational choice, where multiple families of exotic states exist each mixing primarily with a single SM generation, and the more 
general case where the mixing can be quite complex. In this simple single generation case that we 
consider, \eg, the $2\times2$ $d-h$ mass matrix in the weak eigenstate basis $\bar{\cal D}_R^0 {\cal M}_d {\cal D}_L^0$, where ${\cal D}^0=(d^0,h^0)^T$, is given by 
\begin{eqnarray}
{\cal M}_d & =  \left( \begin{array}{cc}
                         m_d^0 & m_h^0 \frac{v_4}{v_3}  \\
                          m_d^0\frac{v_1}{v_2}  & m_h^0 \\
                          \end{array}\right) \,,
\end{eqnarray}
where we have defined $m_{d,h}^0=\lambda_{d,h} v_{2,3}/\sqrt 2$ employing the notation above. Clearly since the off-diagonal elements are `small', straightforward algebra leads to 
$m_{d,h}^0\simeq m_{d,h}$, which are the corresponding physical particle masses, to a very good approximation. This mass matrix is easily diagonalized, as usual, via the 
bi-unitary transformation  
$M_D=U_R{\cal M}_dU_L^\dagger$, where $M_D$ is diagonal so that ${\cal D}_{L,R}=U_{L,R}{\cal D}^0_{L,R}$ are the mass eigenstates.  Note that $U_L$ is determined via the 
relation $M_D^2=U_L{\cal M}_d^\dagger {\cal M}_d U_L^\dagger$ while $U_R$ is similarly determined via $M_D^2=U_R{\cal M}_d {\cal M}_d^\dagger U_R^\dagger$. To leading 
order in the small parameters corresponding to vev ratios from above we then find that
\begin{eqnarray}
 U_L & \simeq  \left( \begin{array}{cc}
                         1 & -\frac{m_d^0}{m_h^0} \frac{v_1}{v_2}  \\
                          \frac{m_d^0}{m_h^0} \frac{v_1}{v_2} & 1 \\
                          \end{array}\right) \,,
\end{eqnarray}
implying that $U_L$ is numerically very close to the unit matrix (which we will assume from now on) since the off-diagonal term is $\sim O(10^{-8})$ or less, while for $U_R$ we find 
instead to this same order in the small parameters 
\begin{eqnarray}
 U_R & \simeq  \left( \begin{array}{cc}
                         1 & -\frac{v_4}{v_3}  \\
                          \frac{v_4}{v_3} & 1 \\
                          \end{array}\right) \,,
\end{eqnarray}
where, as we saw above, $\theta_R \simeq -v_4/v_3$ is of order $\epsilon$, and can be phenomenologically important. Note that these same conclusions would hold if $h$ mixed with 
any of the three SM generations. A very similar result is obtained in the case of $e-E$ mixing with identical results as above, but with the interchange $U_L\leftrightarrow U_R$. These 
results are qualitatively similar to what was obtained in {\bf I} and lead to similar phenomenological implications as we will discuss below.

We now see that the combined effect of the $A_I$ coupling term $\sim (T_I^++T_I^-)$ discussed in the last subsection above and the fermion mixing seen here now yields an effective 
exotic-SM {\it off-diagonal} coupling for the dark photon that is not explicitly $\epsilon$ suppressed and is given by 
\begin{equation}
~~ -2g_Is_I \frac{v_4}{v_3} (\bar h\gamma_\mu d_R+\bar d\gamma_\mu h_R)A_I^\mu\,,
\end{equation}
with a similar result holding in the case of, \eg, $e-E$ mixing with $m_h \to m_E$ and with $R\to L$ in the expressions above. Here we see that this coupling is roughly of order $\epsilon$,  
as will generally be assumed in the phenomenological analysis below and as was anticipated in {\bf I}. From this expression we can determine the induced coupling of the longitudinal component 
of $A_I$ to this current structure as well as that of the associated Goldstone boson by employing the Equivalence Theorem\cite{GBET}. Approximating $m_{A_I}=g_Is_Iv_1$ in our 
mass range of interest (given the arguments above), we find that the overall strength of this interaction, $\lambda$, is given by
\begin{equation}
~ \lambda =-\Big(\frac{v_4}{v_1}\Big)~\Big(\frac{2m_h}{v_3}\Big)=- 0.3 ~\Big(\frac{v_4}{v_1}\Big)~\Big(\frac{m_h}{1.5 \rm TeV}\Big) ~\Big(\frac{10 \rm TeV}{v_3}\Big)~\sim O(1)\,,
\end{equation}
especially if $v_4>v_1$ which can easily happen. This is semi-quantitatively similar to what we obtained in {\bf I} for the toy model we constructed there. As was found there, this implies 
that decays such as $h\to dA_I$ or $E\to e A_I$ via the $A_I$'s longitudinal component experience no $\epsilon$-like suppression and will be completely dominant over other anticipated 
processes such as $h\to dZ,H$ and $h\to uW$. The fact $\lambda$ is $O(1)$ leads to numerous phenomenological implications, some of which were hinted at in {\bf I}. Explicitly, the 
Goldstone boson associated with $A_I$, \ie, $G_{A_I}$, has an off-diagonal coupling given by 
\begin{equation}
~~ i\lambda (\bar hd_R-\bar dh_L)G_{A_I}\,,
\end{equation}
with an analogous term, \eg, for $E-e$-type interactions with the opposite helicity structure. 
It is easy to see that the physical CP-even scalar state(s) containing a significant $Re h_4^0$ component will also have off-diagonal, SM-exotic fermion couplings of very similar 
strength, \ie, $\sim \lambda (\bar hd_R+\bar dh_L)S$, as was found in the toy model presented in {\bf I}. This state would essentially be the real, CP-even partner to the CP-odd $A_I$'s  
associated Goldstone boson above. In the limit where the $2_I1_{I_Y}$ breaking scale becomes large, as in the toy 
model case, we would expect that this real scalar state, $S$, is light, $\lsim$ a few GeV, and with a mass set by the $U(1)_D$ breaking scale. Without a detailed study of the Higgs 
potential of the present setup shown above (which is comparable in complexity in our case compared to, \eg,  that of the Left-Right Symmetric model with Dirac neutrinos\cite{Bolton:2019bou}, 
though here there is no 
L-R symmetry relating the 17 free parameters in the potential), which is beyond the scope of the present work, we will assume that the vacuum structure of potential allows this 
state to be present in the spectrum with the anticipated mass and couplings as in {\bf I}.

The introduction of vector-like fermions which mix with SM fermions via Yukawa couplings, as outlined above, may induce new flavor changing neutral currents through, \eg, the SM Z boson coupling. FCNCs in this context have been studied in the literature \cite{ishiwata2015new,bobeth2017patterns,bobeth2017yukawa}, and are proportional to products of the VLF-SM Yukawa couplings and the ratio of the SM vev to the VLF mass, $\lambda_i \lambda_j^* v^2/M^2$. The case here is slightly different, however, as the mixing between the VLF and the SM fermions  are only non-negligible for the SM fermion which transforms the same way under the SM gauge groups as the VLF, $\ie$ above we have $d_R-h_R$ and $e_L-E_L$ mixing but no $d_L-h_L$ or $e_R-E_R$ mixing at order $\epsilon$. Since the fermions which experience mass mixing couple identically to the $Z$, there are no $Z$-mediated FCNCs in this model at order $\epsilon$ {\footnote{We note that $Z$ mediated FCNCs do occur due to mixing in the left-handed down-like quark sector, but as noted in the above analysis of $U_L$, we expect these terms to be $O(10^{-(10-16)})$, depending on the mass of the down-like quark, and we will neglect these couplings.}}. Nevertheless, we see there are FCNCs also mediated by the dark photon $A_I$ which we will briefly discuss, using the down-like quark sector as an example.

The most general flavor structure of this model has an additional set of VLFs for each generation, so that the relationship between the mass eigenstates and the gauge eigenstates, ${\cal D}_{R}=U_{R}{\cal D}^0_{R}$, becomes a mixing between six states which may be written in block form as
\begin{eqnarray}
 U_R & \simeq  \left( \begin{array}{cc}
                         D_R & -\frac{v_4}{v_3} U_{R,12} \\
                          \frac{v_4}{v_3} U_{R,21} & H_R \\
                          \end{array}\right) \,,
\end{eqnarray}
where $D_R$ mixes the SM down-like quarks with themselves, $H_R$ mixes the VLFs with themselves, and $U_{R,12}$ and $U_{R,21}$ mix the SM and VLF components. We have defined the SM-VLF mixing blocks with a prefactor of $v_4/v_3 \simeq 10^{-4}$ to emphasize the small mixing between the SM and VLF states, so $U_{R,12}$ and $U_{R,21}$ may have components which are order unity. This mass mixing induces tree-level FCNC couplings between the down-like quarks and the $A_I$, given by
\begin{equation}
g_I s_I A_I^\mu \bar{{\cal D}}_{R i}\gamma_\mu V^I_{ij} {\cal D}_{R j},
\end{equation}
where $V^I$ is a $6 \times 6$ matrix which may be written in block form as

\begin{eqnarray}
 V^I & \simeq  \left( \begin{array}{cc}
                         \frac{v_4^2}{v_3^2} U_{R,12} U_{R,12}^\dagger & -\frac{v_4}{v_3} U_{R,12}H_R^\dagger \\
                          -\frac{v_4}{v_3} H_R U_{R,12}^\dagger & H_R H_R^\dagger \\
                          \end{array}\right) \,.
\end{eqnarray}

Since the $A_I$ is much lighter than the SM $Z$, it may be produced on-shell in meson decays, which could provide constraints on the flavor sector of these models. As an example, we consider the tree-level processes $b\rightarrow s + A_I$, $b \rightarrow d + A_I$, and $s\rightarrow d + A_I$, with a long-lived $A_I$ which escapes the detector. These decays mimic the SM processes $B \rightarrow K \nu \bar\nu$, $B \rightarrow \pi \nu \bar\nu$, and $K \rightarrow \pi \nu \bar\nu$, respectively, and thus may be constrained by limits on the branching fractions of these decay channels. We estimate the branching fractions for the corresponding new physics processes $B \rightarrow K A_I$, $B \rightarrow \pi A_I$, and $K \rightarrow \pi A_I$ using the hadronic form factors of refs. \cite{leutwyler1984determination,ball2005new}, taking $m_{A_I} = m_{light}$, the mass of the light meson in the decay product, and assuming $g_I^2 s_I^2 \simeq 0.1$. Our estimates of these branching fractions and the current limits are summarized in Table \ref{branchtab}. 

From these estimates we see that the observed branching fraction of $K^+ \rightarrow \pi^+ \nu \bar\nu$ is quite constraining on the flavor structure of the model, requiring either a flavor structure such that $|(U_{R,12} U_{R,12}^\dagger)_{sd}|^2 \lesssim 10^{-8}$, which can happen if, \eg, $U_{R,12} \lsim 10^{-2}$, 
or that the decay be kinematically inaccessible with $m_{A_I} > 350$ MeV in order to trivially avoid the constraint. The constraints from $B$ decays are comparatively weaker, with a relatively mild flavor suppression factor $|(U_{R,12} U_{R,12}^\dagger)_{bd} |^2 \lesssim 0.1$ proving sufficient to evade the bound on ${\cal B}(B^+ \rightarrow \pi^+ \nu \bar\nu)$, and the bound on ${\cal B}(B^+ \rightarrow K^+ \nu \bar\nu)$ providing little constraint, even for $|(U_{R,12} U_{R,12}^\dagger)_{bs} |^2 \sim O(1)$. 

We note that FCNCs may also impact neutral meson oscillations, but due to the light mass of the $A_I$ these effects are difficult to estimate, and an analysis of the impact of weakly coupled FCNCs with light mediators on neutral meson oscillations is beyond the scope of the present work. Further, we note that a similar set of $A_I$-mediated FCNCs will be present in the lepton sector, which could be probed by, \eg, $\mu \rightarrow e$ measurements, though the rate would be suppressed by a factor of $\epsilon^2 v_4^4/v_3^4 \sim 10^{-24}$ at least. Decays forbidden in the SM, such as $K^+ \rightarrow \pi^+ \mu^+ e^-$, could also provide a very clean probe of the FCNCs discussed here, though this rate would be further suppressed by a factor of $O(10^{-16})$ relative to the $K^+ \rightarrow \pi^+ A_I$ discussed above. 

\begin{table}
\caption{FCNC Branching Fraction Estimates vs. SM Bounds} \label{branchtab}
Tree-level estimates of $A_I$-mediated FCNC branching fractions were estimated using hadronic form factors of refs. \cite{leutwyler1984determination,ball2005new}, assuming $m_{A_I} = m_{light}$, the mass of the light meson in the decay product, and $g_I^2 s_I^2 \simeq 0.1$.
\begin{center}
\begin{tabular}{ | c | c | }
\hline
Tree-Level Estimate & Current SM Limit \\
\hline
${\cal B}(K^+ \rightarrow \pi^+ A_I) \sim 3\times10^{-3} \left|(U_{R,12} U_{R,12}^\dagger)_{sd} \right|^2$ & ${\cal B}(K^+ \rightarrow \pi^+ \nu \bar\nu) = (1.7 \pm 1.1) \times 10^{-10}$ \cite{tanabashi2018review} \\
${\cal B}(B^+ \rightarrow K^+ A_I) \sim  8 \times10^{-6} \left|(U_{R,12} U_{R,12}^\dagger)_{bs} \right|^2$ & ${\cal B}(B^+ \rightarrow K^+ \nu \bar\nu) < 1.6 \times 10^{-5}$ \cite{babar2013search} \\
${\cal B}(B^+ \rightarrow \pi^+ \nu \bar\nu) \sim  6\times10^{-5} \left|(U_{R,12} U_{R,12}^\dagger)_{bd} \right|^2$ & ${\cal B}(B^+ \rightarrow \pi^+ \nu \bar\nu) < 1.4 \times 10^{-5}$ \cite{babar2013search} \\
\hline 
\end{tabular}
\end{center}
\end{table}

\section{Survey of Some Bottom-Up Phenomenology}

Given the rich structure we have introduced above, we might expect that this setup will have a complex phenomenology which has partial overlap with both the $SU(2)_I$  
models and that of the conventional dark photon picture, perhaps augmented by some of what we discussed previously in {\bf I}. The first and most obvious topic to address is how the physics of the new heavy gauge bosons and exotic fermions differs from the more familiar and well-studied $SU(2)_I$ from $E_6$-inspired models. One 
certain issue we need to address is the nature 
of the interplay between the exotic and SM sectors: is there a single set of exotic states that primarily mixes with the corresponding fields in a single SM generation (and which one 
is it, \eg, do we have dominant $h-d, h-s$ or $h-b$ mixing?) or is there a set of exotic fields corresponding to each generation? In traditional $E_6$-inspired models the answer is clear, while 
here we see that there are several different possibilities which lead to somewhat different phenomenology. In the discussion that follows we will ignore issues related 
to flavor-changing processes and consign such discussions to later work.

These model-dependent differences make themselves felt in even the simplest production process, \ie, that 
of $Z_I$ production in the Drell-Yan channel when the $Z_I$ can only decay to SM final states with the most trivial difference being the overall $Z_I$ coupling strength which in 
$E_6$ models is proportional to $\sim \sqrt{\frac{5s_w^2}{3}}$ due to GUT coupling requirements. Here, the overall coupling is set by $g_I/c_I$ so that $Z_I$ production cross sections 
can be determined up to an overall factor of $r=(g_I/c_I)/(g/c_w)$ which we expect to be $O(1)${\footnote {In what follows we will employ the Narrow Width Approximation for our $Z_I$ 
analyses.}}. Of course, a potentially more significant difference is whether the $Z_I$ couples only to $d\bar d$, $s\bar s$ or $b\bar b$ initial states (or all three) and whether it will  
appear in only the $e^+e^-$, $\mu^+\mu^-$, or $\tau^+\tau^-$ channels (or, again, all three). For example, we are reminded that in the more well-studied case, there is a universality 
of interactions among the generations so that all three initial states as well as all three final state will contribute to the potential $Z_I$ Drell-Yan cross section and corresponding 
decay signatures at the LHC. In the present situation, all of these scenarios need to be explored independently.  

The simplest situation, and the one that allows us to make most direct contact with previous studies, is the production of $Z_I$ with identical couplings to all three SM generations but 
which is 
kinematically forbidden to decay into any of the exotic vector-like fermions or into $W_IW_I^\dagger$. In such a case, $\sigma B_\ell$ in the NWA is shown in Fig.~\ref{ZI} assuming that 
$r=\frac{g_I/c_I}{g/c_w}=1$; note that the overall cross section in this case is independent of the value of $x_I$ (since SM states all have $Q_I=0$) 
and that it is simply proportional to $r^2$ so that other cases are easily 
obtained by a simple overall rescaling. As can be seen from this 
Figure, the present ATLAS null searches\cite{Aad:2019fac} employing 139 fb$^{-1}$ of 13 TeV integrated luminosity excludes $Z_I$ masses below $\simeq 5.2$ TeV under these set of 
assumptions. A similar null search performed at the 14 TeV HL-LHC with 3 ab$^{-1}$ of luminosity\cite{ATLASNote} would increase the exclusion limit on a $Z_I$ to masses 
below $\simeq 5.9$ TeV under the same set of assumptions.
\begin{figure}[htbp]
\centerline{\includegraphics[width=5.0in,angle=0]{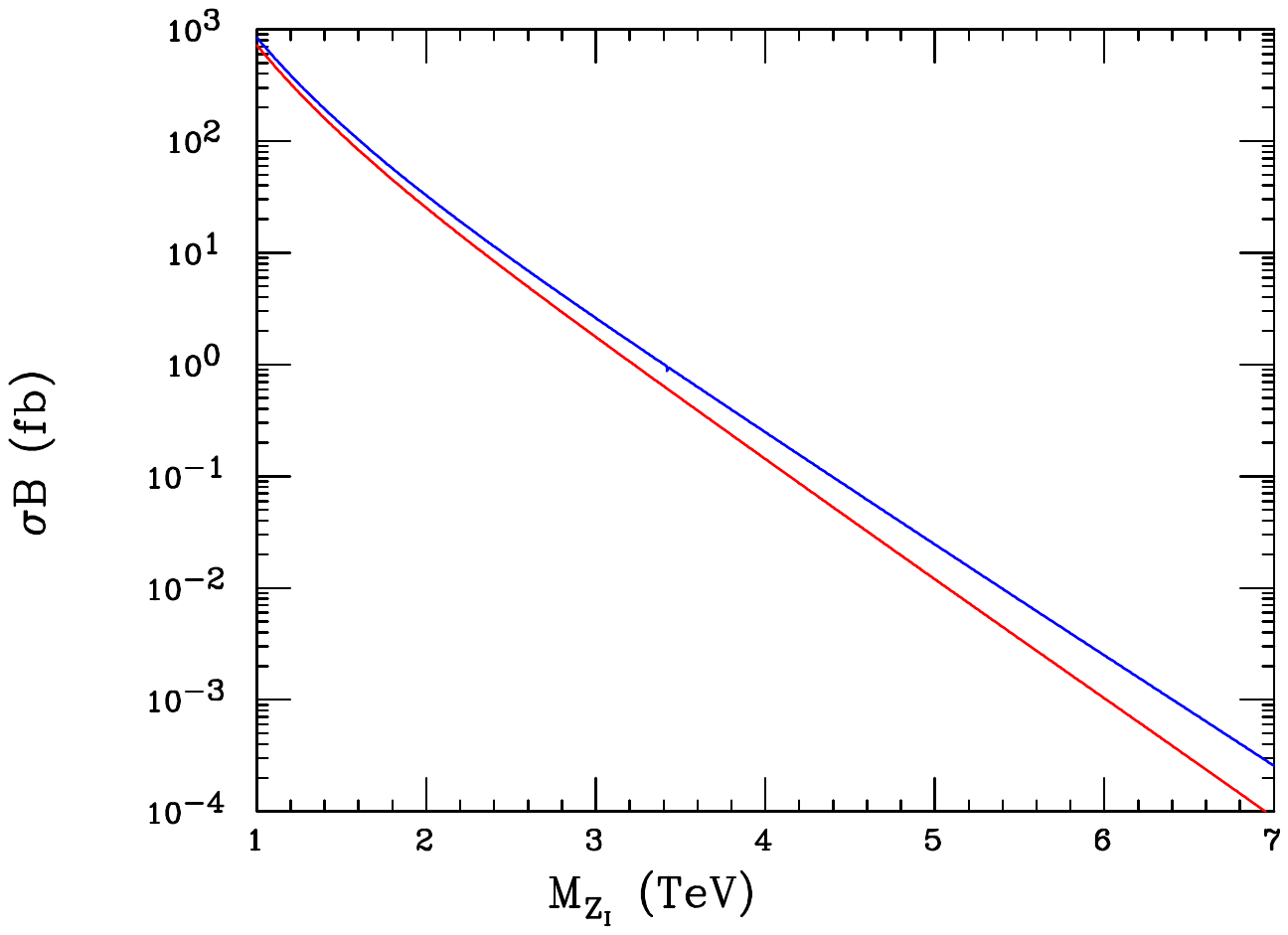}}
\vspace*{-1.50cm}
\caption{$\sigma B_\ell$ for $Z_I$ production at the $\sqrt s=13$ (red) and 14 (blue) TeV LHC where only decays to SM fields are assumed to be kinematically 
allowed and with $r=\frac{g_I/c_I}{g/c_w}=1$ being assumed. Here the $Z_I$ is assumed to couple universally as in the original most well-studied scenario. }
\label{ZI}
\end{figure}

A second possibility, as discussed above, is that only a single set of exotic fermions exist which mix with a particular SM generation. In such a case only a single SM generation couples 
to $Z_I$ and so only the searches in a particular dilepton channel are applicable for setting constraints. Thus, \eg, if only the third generation carries non-zero $2_I1_{I'}$ quantum 
numbers, the relevant process to examine is then $b\bar b\to Z_I \to \tau^+\tau^-$. We first consider the simplest case where the $Z_I$ still cannot decay to any of the exotic states so 
that the cross section remains $x_I$ independent with only an overall sensitivity to the coupling ratio $r$ as above. The production rates for this case at the $\sqrt s=$13 and 14 TeV 
LHC are shown in Fig.~\ref{Dylan2} {\footnote {Here, and in what follows, we will ignore the possible effects of small inter-generational fermion mixings which could result in 
flavor-changing dark currents. Note that in the quark case this 
involves possible inter-generational mixings among the right-handed fields which are not well-probed by SM measurements and may even be absent in some scenarios.}}. 

\begin{figure}[htbp]
\vspace*{0.5cm}
\centerline{\includegraphics[width=4.5in,angle=0]{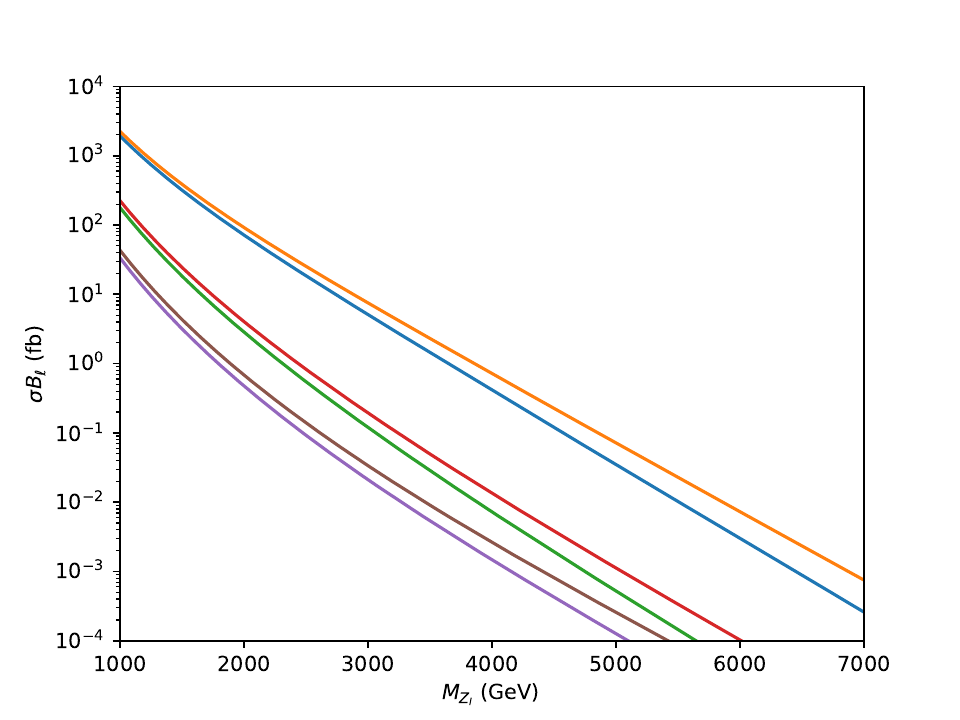}}
\vspace*{0.30cm}
\caption{$\sigma B_\ell$ for $Z_I$ production at the $\sqrt s=13$ (lower) and 14 TeV (upper) LHC for each pair of curves assuming that the $Z_I$ only couples to a single SM 
generation and decays to exotic partners are forbidden, taking $r=1$. From top to bottom, the $Z_I$ couples to the first, second, and third generation of the SM, respectively.}
\label{Dylan2}
\end{figure}

Fig.~\ref{Dylan1} then shows how these $Z_I$ production cross sections in the various dilepton channels translate into current LHC search limits and the expectations for 
the HL-LHC assuming either universal couplings or to only one of the SM generations. In order to obtain these results in the case of the $\tau^+\tau^-$ third generation couplings, we have 
recasted the results from an ATLAS $b\bar b\to H\to \tau^+\tau^-$ study\cite{ATLAStau36,ATLASNote2} making corrections for the acceptance differences between spin-0 and spin-1 resonances. 

\begin{figure}[htbp]
\vspace*{0.5cm}
\centerline{\includegraphics[width=3.7in,angle=0]{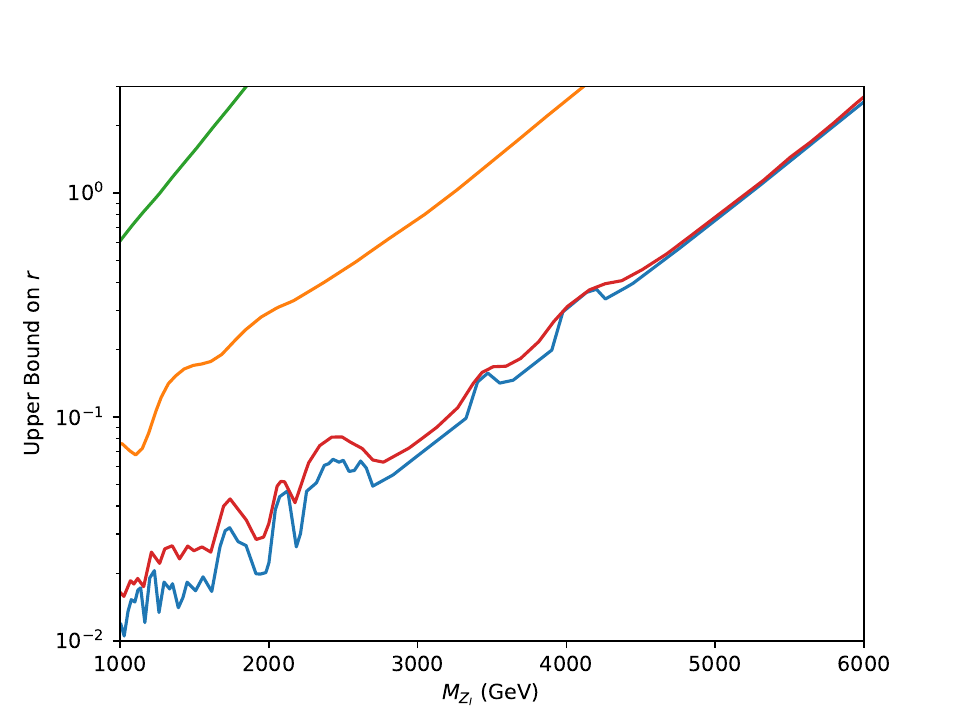}
\hspace*{-0.7cm}
\includegraphics[width=3.7in,angle=0]{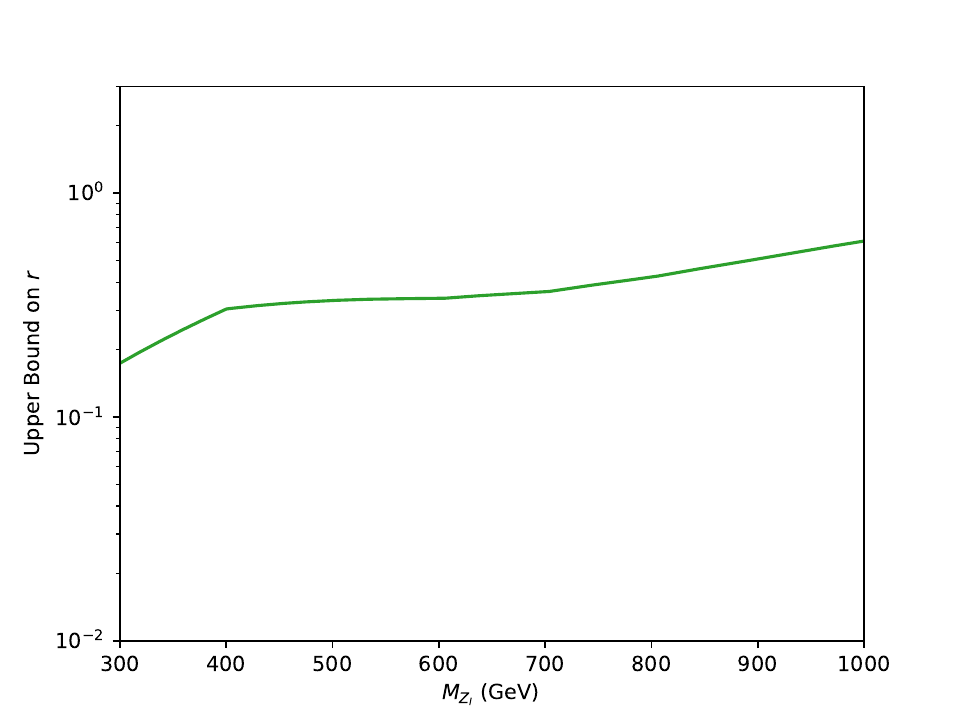}}
\vspace*{1.0cm}
\centerline{\includegraphics[width=3.7in,angle=0]{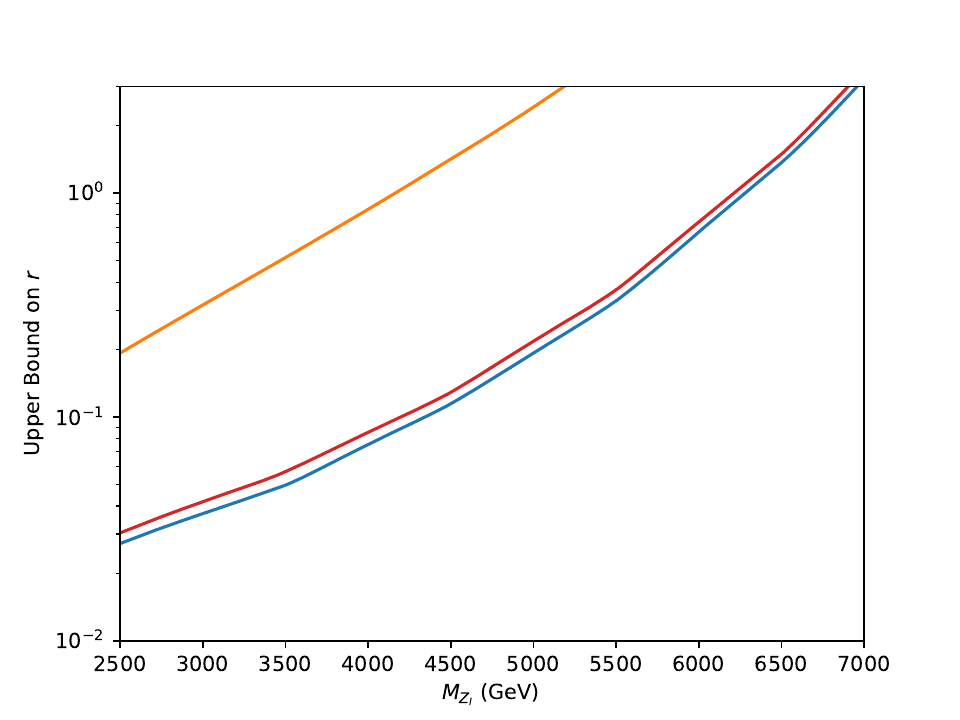}
\hspace*{-0.7cm}
\includegraphics[width=3.7in,angle=0]{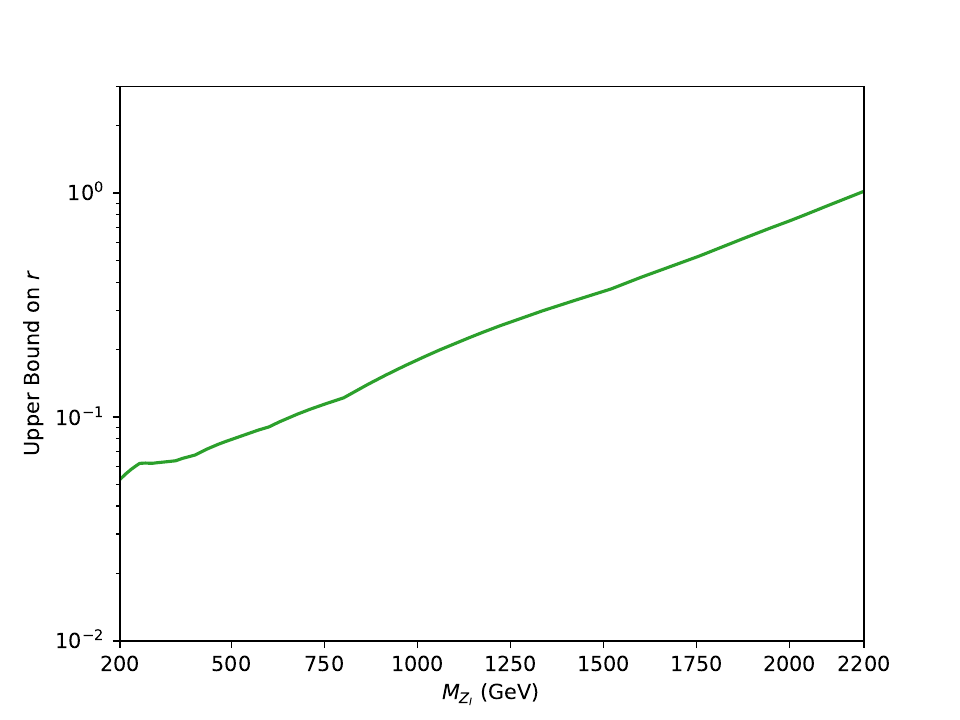}}
\vspace*{0.50cm}
\caption{(Top Left) Limits from the $\sqrt{s} = 13$ TeV LHC on the values of the parameter $r$ as described in the text, as a function of the mass of the $Z_I$, employing the results from ATLAS searching for dilepton decays \cite{Aad:2019fac,ATLAStau36}. From left to right the curves correspond to only third generation couplings (green), only second generation couplings (gold), universal couplings (red), and only first generation couplings (blue). (Top Right) Extension 
of the results in the previous panel for the case of third generation couplings to lower $Z_I$ masses. (Bottom Left) Same as the top left panel, but 
now employing an ATLAS analysis assuming a null result at the HL-LHC with $\sqrt s=14$ TeV and L=3 ab$^{-1}$ \cite{ATLASNote}. (Bottom Right) Corresponding limit in the $\tau^+\tau^-$ case employing the ATLAS heavy  
Higgs study\cite{ATLASNote2} with acceptance corrections included for a spin-1 state.}
\label{Dylan1}
\end{figure}

It is interesting to explore how these results change as we move further away from these rather vanilla scenarios. In addition to variations in the overall coupling, \ie, $r\neq 1$, 
the $Z_I$ may 
also decay into one or more exotic fermion final states and also into $W_IW_I^\dagger$ which would modify the branching fraction for the leptonic decay mode used in the search. 
Furthermore, the $Z_I$ may {\it not} couple universally, perhaps only to $d\bar d$, $s\bar s$ or $b\bar b$ initial states and consequently only leptonically decay to $e^+e^-$, 
$\mu^+\mu^-$ or $\tau^+\tau^-$ final states, respectively, as we saw above.  In such cases one also needs to employ the individual constraints as applicable to each of these final 
states as previously considered\cite{Aad:2019fac}.

When the $Z_I$ is sufficiently massive it can also decay into some if not all of the exotic non-SM states; the partial widths for these decays, unlike those to SM final states, will 
explicitly depend upon the value of $x_I$. Furthermore, once $x_I>0.75$ the $Z_I\to W_IW_I^\dagger$ channel also opens up via the usual non-Abelian trilinear coupling. 
The main effect of these new decay modes of the 
$Z_I$ for our discussion is to reduce the value of the relevant leptonic branching fraction, $B_\ell$, and thus a correspondingly reduced signal rate resulting is a suppression of 
the search/exclusion reach. This effect, however, is expected to be relatively mild over most of parameter space as $B_\ell$ is reduced by at most a factor of $\simeq 6.8$, even 
in the extreme case when all three generations of exotic states (without much phase space suppression) as well as $W_IW_I^\dagger$ are allowed to contribute to the total 
$Z_I$ width, and this maximum reduction only occurs when $x_I\simeq 1$. Fig.~\ref{exo} gives us a 
feel for how large a reduction in $B_\ell$ may occur where, for purposes of demonstration in the universal case, where we have included decays into 3 generations of degenerate 
$N,E,h,S_{1,2}$ states as well as $W_IW_I^\dagger$ for $x_I>0.75$. Very similar reduction factors will also occur in the cases where there is only a single generation of 
exotics which mix with only a single SM generation. From Fig. \ref{exo} we see that the typical leptonic branching fraction seen here will have very little influence on the $Z_I$ search 
reach at the HL-LHC.

\begin{figure}[htbp]
\centerline{\includegraphics[width=5.0in,angle=0]{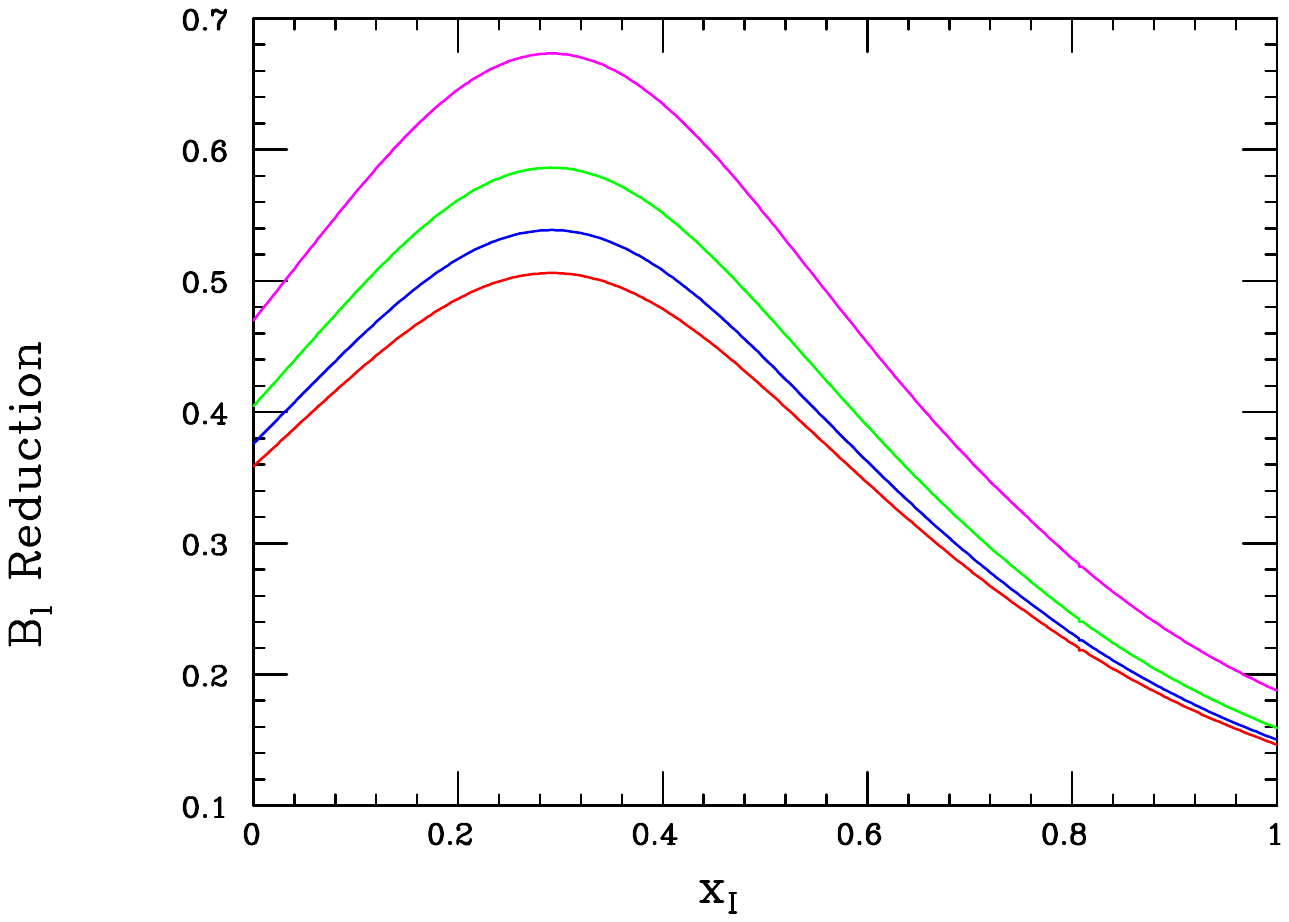}}
\vspace*{-1.50cm}
\caption{Multiplicative $B_\ell$ suppression factor for the $Z_I$ as a function of $x_I$ due to the additional non-SM decays into 3 generations of degenerate exotic fermions 
plus $W_IW_I^\dagger$ (for $x_I>0.75$) as discussed in the text. From bottom to top the curves assume $m_{exotic}/m_{Z_I}=0, 0.2, 0.3$ and 0.4, respectively.}
\label{exo}
\end{figure}

\begin{figure}[htbp]
\centerline{\includegraphics[width=5.0in,angle=0]{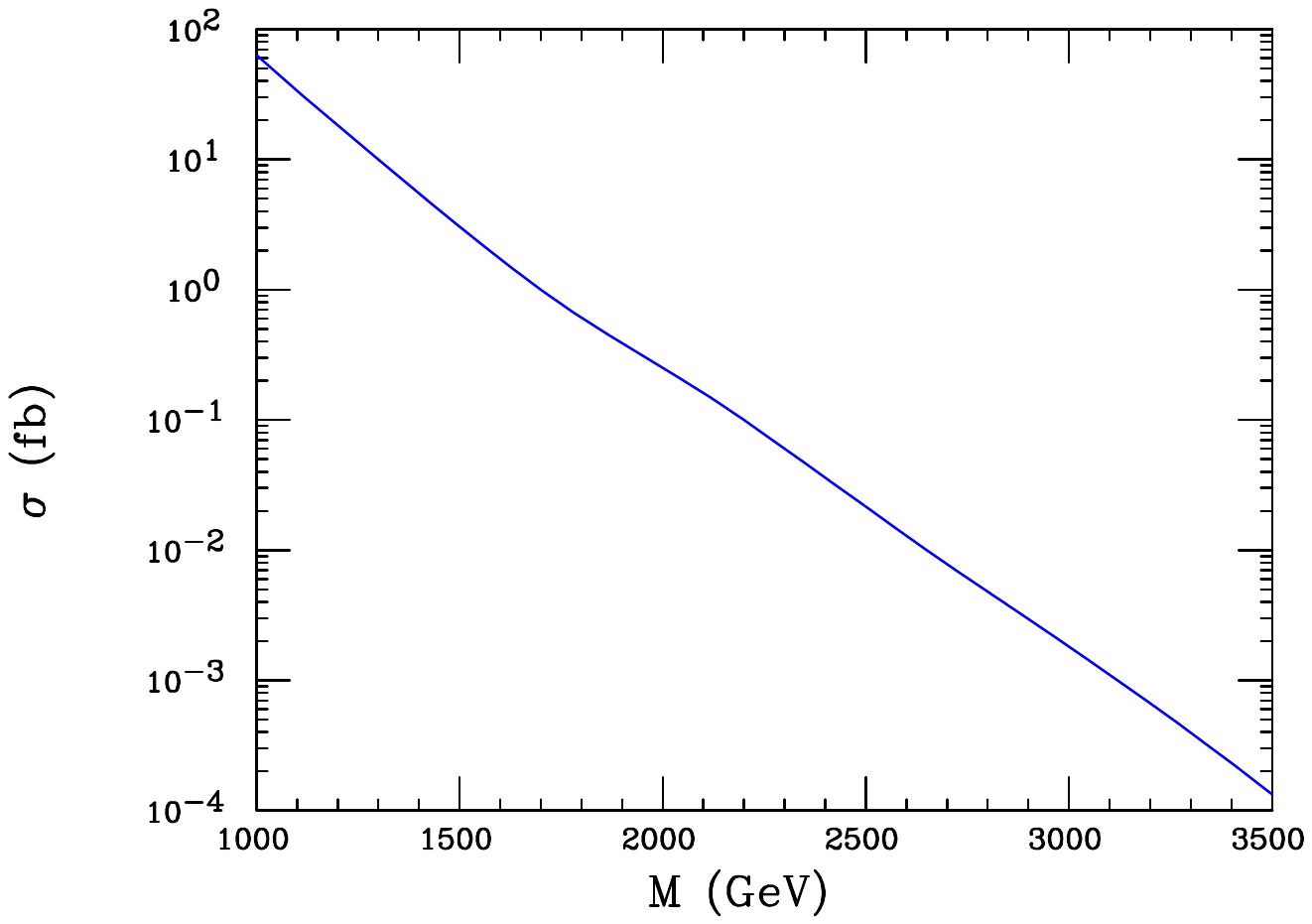}}
\vspace*{-1.50cm}
\caption{Pure QCD NNLO $h\bar h$ production cross section at the $\sqrt s=14$ TeV LHC as a function of $m_h$ using HATHOR following Ref.~\cite{hpair}.}
\label{vlq14}
\end{figure}

Perhaps at a similar level of relative `simplicity' is the direct pair production of the exotic fermions $h,E,N$, \etc, themselves which may be accessible to LHC experiments. As in {\bf I}, 
once produced $h$ will (far) dominantly decay into its associated SM partner $q=d,s$ or $b$, which will appear as a jet, plus $A_I$ or $S$ which produce either MET or lepton-jet signatures. 
$h$, being a color triplet, is certainly pair produced via QCD from, at LO, 
$q\bar q, gg$ annihilation, as is the case of the top quark as well as other more familiar vector-like quarks. For the $\sqrt s=13$ TeV LHC the cross section due to these processes 
is given in {\bf I} while for the case of $\sqrt s=14$ TeV\cite{hpair,LHCTop} the slightly larger cross section is shown in Fig.~\ref{vlq14} with the caveat here that no additional, non-electroweak 
interactions are present; this caveat needs some explanation. As discussed in {\bf I}, the decay of $h\bar h$ are not those of conventional vector-like states but instead 
will lead to two, non-back-to-back jets plus MET, if $A_I/S$ are very long-lived, or lepton-jets\cite{ljet,mores,ljet2} if the decays occur inside the detector. Decays of $A_I/S$ far from 
the detector may be captured by specialized experiments looking for long-lived states\cite{Alpigiani:2018fgd,Ariga:2018zuc} such as FASER and MATHUSLA.

While the $gg\to h\bar h$ process is left unaltered by the additional interactions described above, the $q\bar q \to h\bar h$ process may be significantly 
modified when $q=d,s$ or $b$ via 
a $t-$channel longitudinal $A_I$ exchange (or equivalently, the $G_{A_I}$ Goldstone boson) as well as that due to the corresponding light CP-even scalar, $S$, provided that 
the parameter $\lambda$ is sufficiently large. In a similar vein, the $t$-channel $W_I$ exchange may also produce a potentially important contribution but it is suppressed by both 
the relatively small $SU(2)_I$ gauge coupling as well as the large mass of the $W_I$ itself; the neglect of this contribution at this level of discussion is similar to the neglect of $Z$ exchange 
in the case of top pair production within the SM context. Clearly, the numerical 
impact of these new exchanges will differ significantly depending upon which SM quark mixes with the exotic partner $h$. To quantify this possibility, consider the modification to the 
$q\bar q \to h\bar h$ differential cross section at LO, where we will neglect the masses, $\lsim 1$ GeV, of the exchanges in the $t$-channel compared to other mass scales in the 
process. Defining the coupling ratio $\chi=\lambda^2/(4\pi \alpha_s)$, with $\lambda$ as given above, and defining $z=\cos \theta^*$, with $\theta^*$ being the partonic center of mass frame 
scattering angle, we find, for the quark that mixes with $h$, that
\begin{equation}
\frac{d\sigma}{dz}=\frac{\pi \alpha_s^2\beta}{9\hat s}~\Big(B_1+2\chi B_2+ \frac{9}{2} \chi^2 B_3\Big)\,,
\end{equation}
where $\beta^2=1-4m_h^2/\hat s$, and where 
\begin{equation}
B_1=2-\beta^2(1-z^2),~~~B_2=\frac{(1-\beta z)^2+1-\beta^2}{(1+\beta^2)/2-\beta z},~ ~~B_3=\frac{(1-\beta z)^2}{[(1+\beta^2)/2-\beta z]^2}\,,
\end{equation}
with $B_1$ being the conventional LO pure QCD result.  The impact of this exchange will not only depend on the value of $\lambda$, but even more importantly on which of the 
$q=d,s,b$ initial state quarks participates in the mixing with $h$ since their parton densities are all quite different. As we will see, not only is $h(\bar h)$ production pushed 
more forward(backward) due to this $t-$channel exchange as one might expect, but the overall total $h\bar h$ production cross section also increases, in some cases significantly. 

\begin{figure}[htbp]
\centerline{\includegraphics[width=4.0in,angle=0]{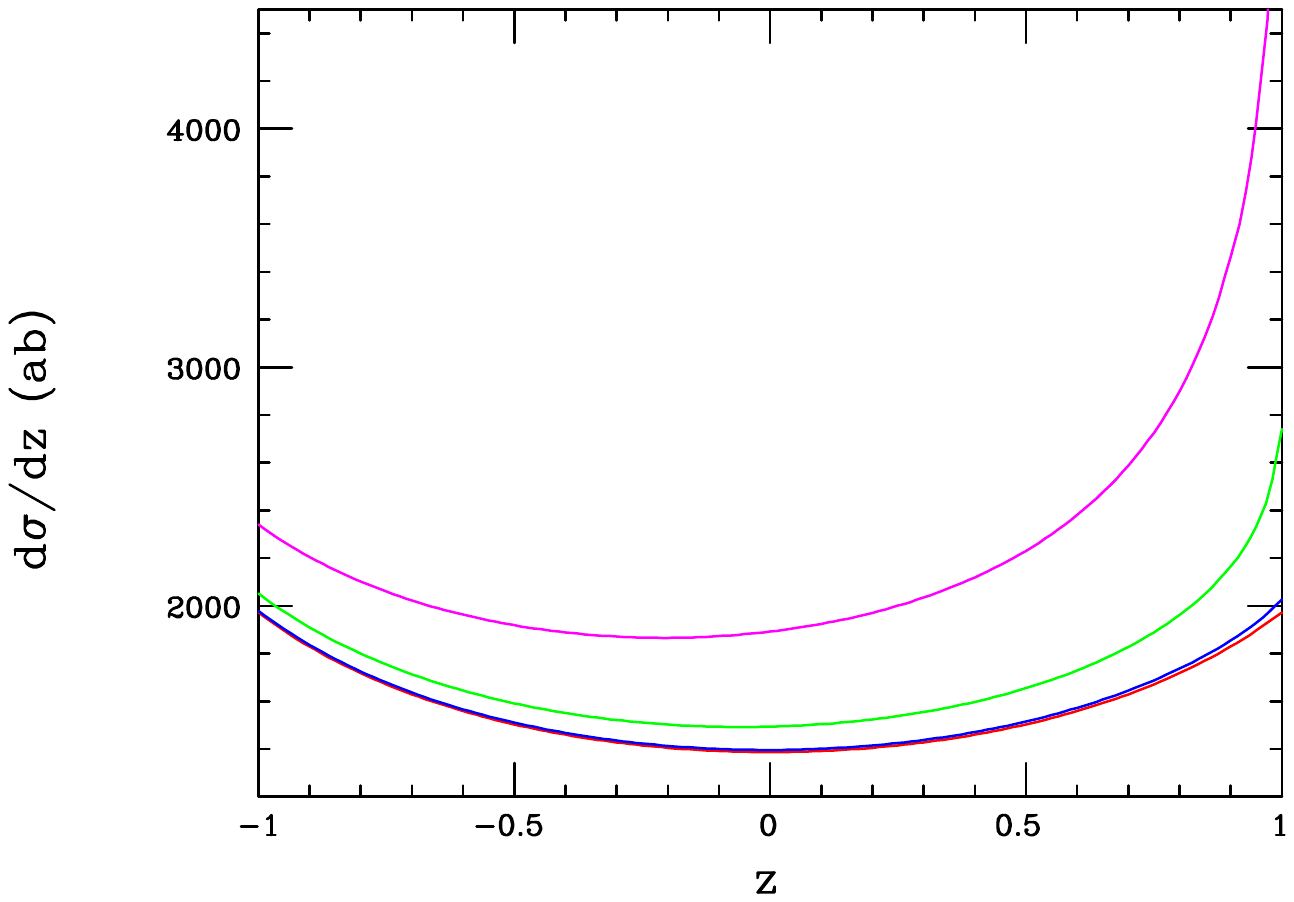}
\hspace*{-1.8cm}
\includegraphics[width=4.0in,angle=0]{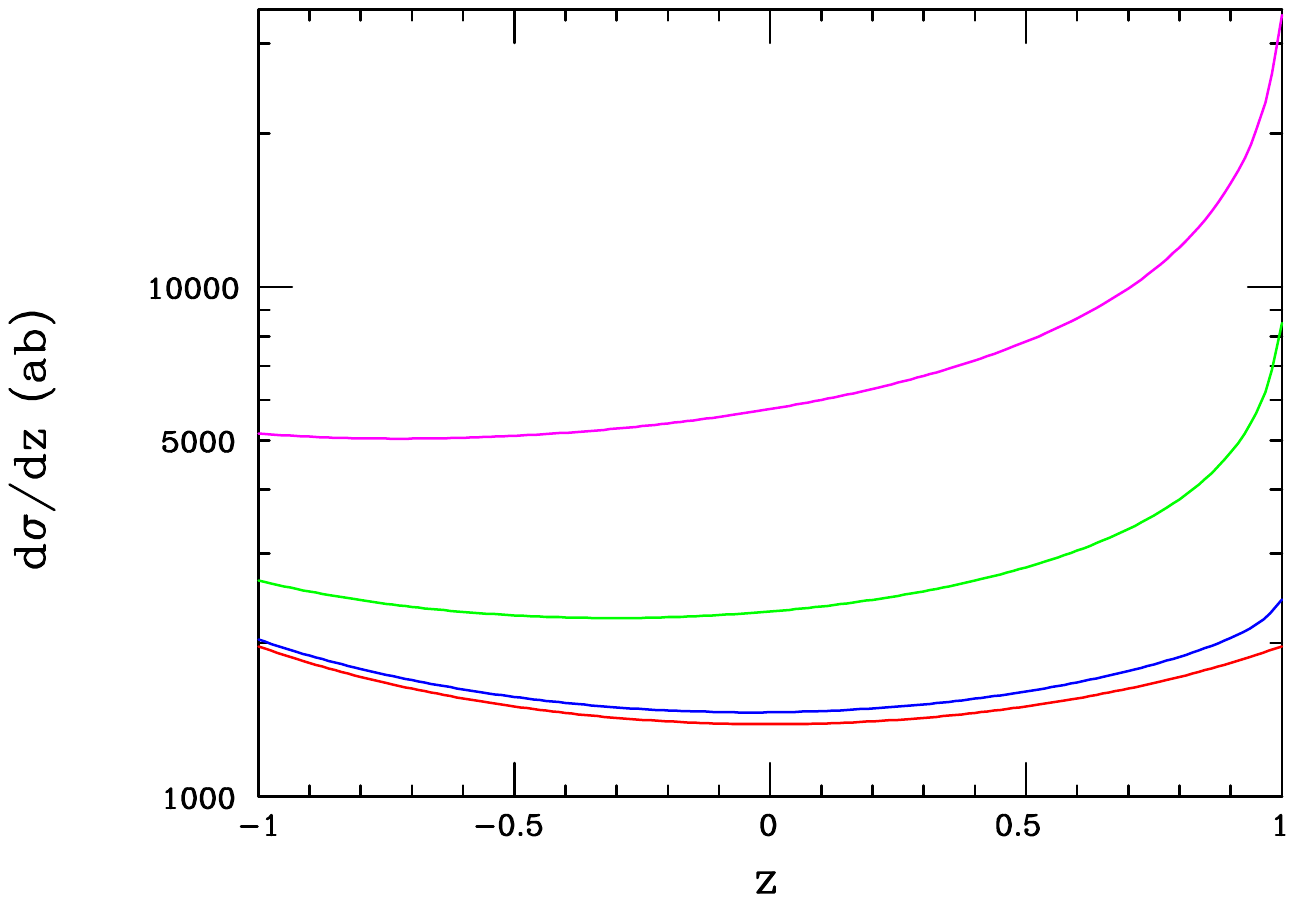}}
\vspace*{-1.5cm}
\centerline{\includegraphics[width=4.0in,angle=0]{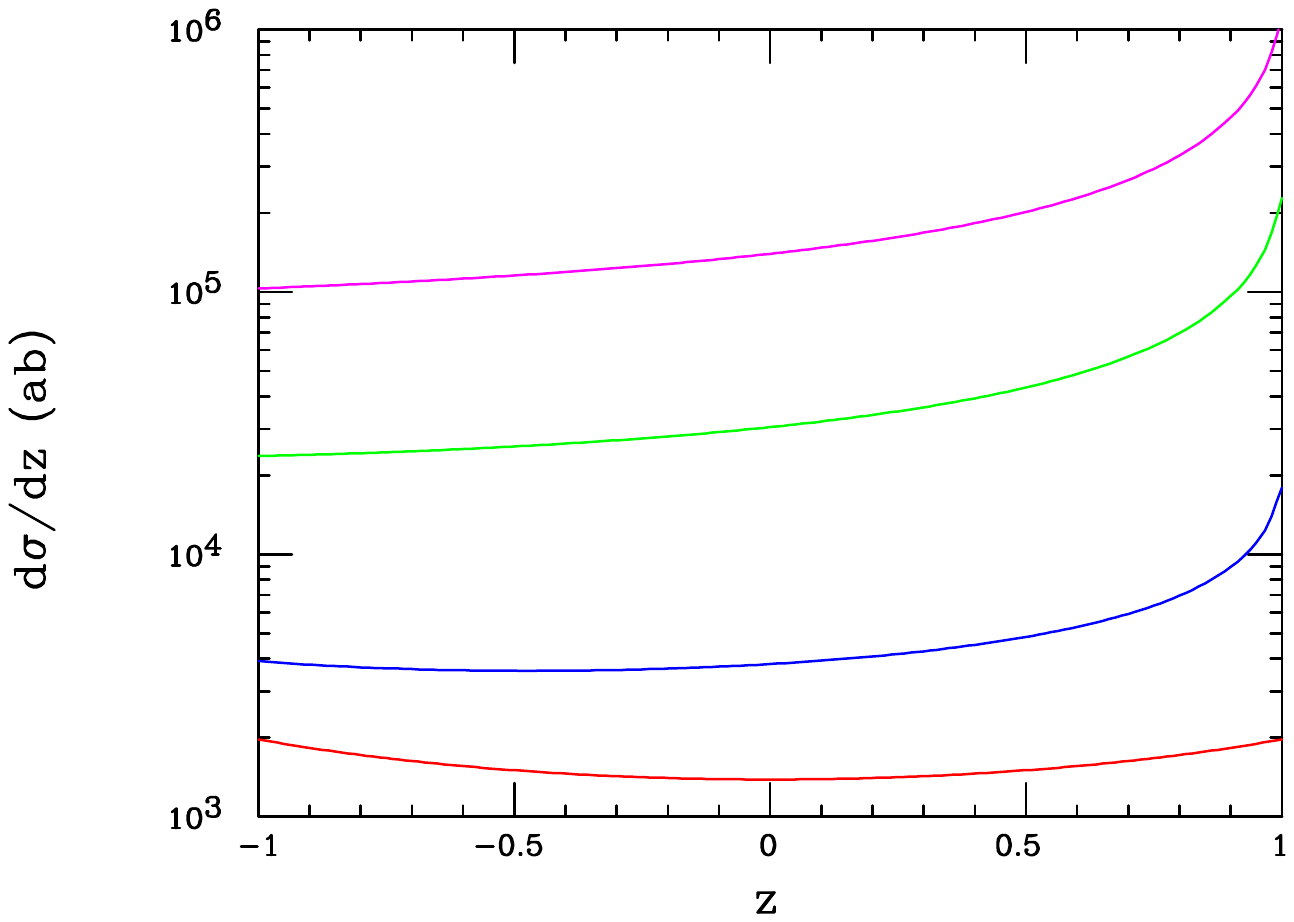}}
\vspace*{-1.20cm}
\caption{(Top Left) Angular distribution for $h\bar h$ production at the 14 TeV LHC assuming $m_h=1.5$ TeV and where the $h$'s SM partner is the $b$ quark. From bottom to top 
the curves assume that $\lambda=0,1,2,3$, respectively. (Top Right) Same as the previous panel but now taking $h$'s SM partner to be the $s$ quark. (Bottom) Same as the 
previous panel but now taking $h$'s SM partner to be the $d$ quark. Very similar distributions are obtained when, \eg, $m_h=2$ TeV is assumed. }
\label{hhdif}
\end{figure}

To get an idea of the impact of $\lambda \neq 0$ for the various non-universal $q=d,s,b$ choices, we have taken the LO $q\bar q$ and $gg$ processes and re-weighted them by 
K-factors to recover the corresponding NNLO SM total cross section result when they are combined and $\lambda=0$ is assumed\cite{hpair,LHCTop}.  Fig.~\ref{hhdif} shows 
these modifications  
to the $h\bar h$ angular distributions for the three non-universal choices assuming $m_h=1.5$ TeV at the 14 TeV LHC, and taking rather large values of $\lambda$  for purposes 
of demonstration. Here we see the obvious result that for any fixed value of $\lambda$ the overall impact of the $t-$channel exchange increases dramatically as the choice of $q$ goes 
from $b$ to $s$ to $d$. For example, in the case $q=b$, a substantial impact is only found when $\lambda$ is quite large $\simeq 3$. However, for $q=d$, we see that a reasonable 
impact is seen even for values of $\lambda$ less than unity. In all cases, as expected, we see that the impact is largest in the forward direction due to the $t-$ channel nature of the 
exchange. Since the QCD aspects of heavy vector-like quark production are well-known\cite{hpair}, a measurement of the {\it total} cross section for $h\bar h$ production would perhaps 
give us a rough indication of the value of $\lambda$ in some cases. Fig.~\ref{hhtot} shows the total $h\bar h$ cross section at 14 TeV as a function of $\lambda$ for the three 
choices $q=b,s,d$. Here we see that noticeable effects may be visible for $\lambda \simeq 3,1$ and 0.3, respectively, for these three cases.

\begin{figure}[htbp]
\centerline{\includegraphics[width=4.0in,angle=0]{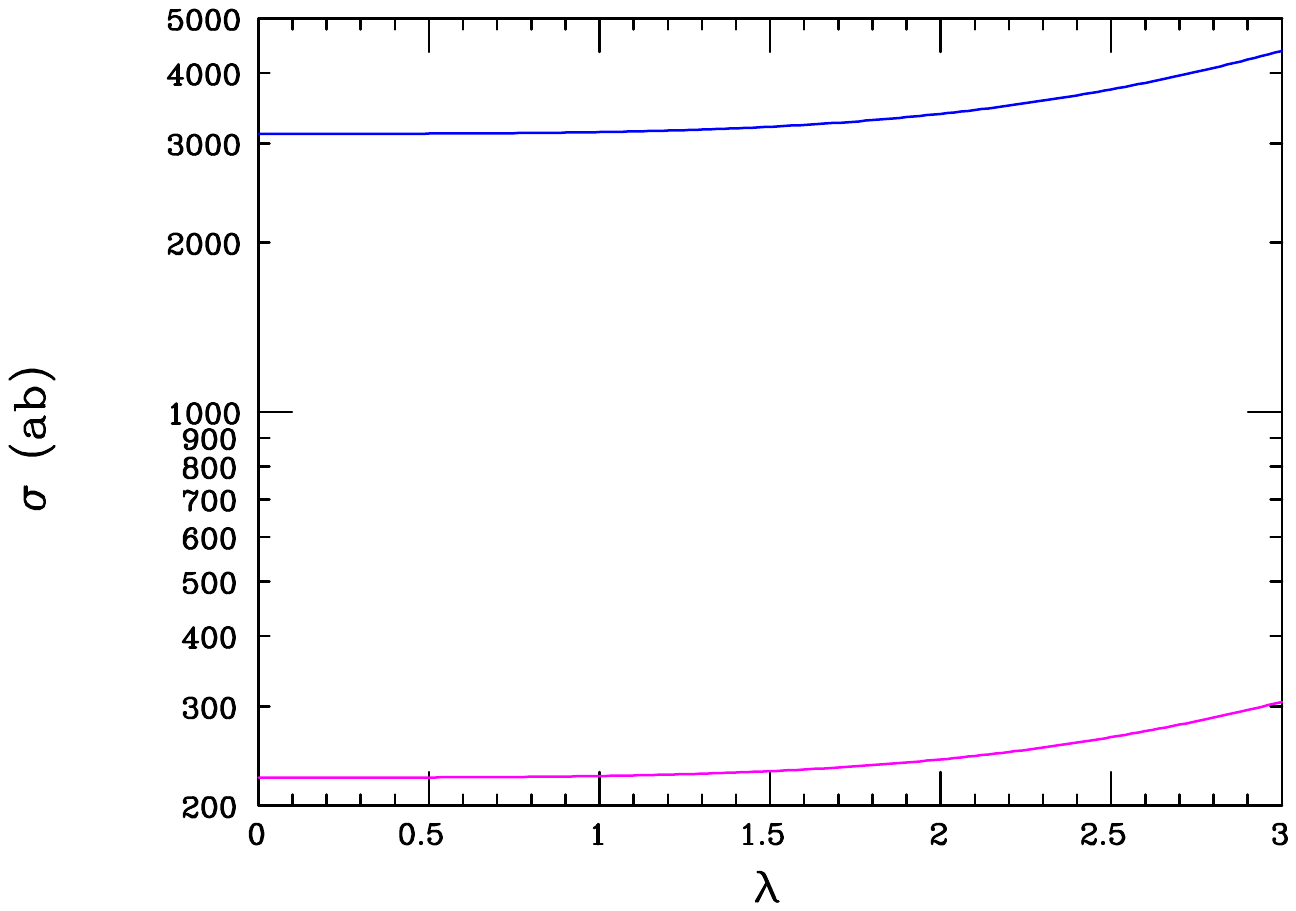}
\hspace*{-1.8cm}
\includegraphics[width=4.0in,angle=0]{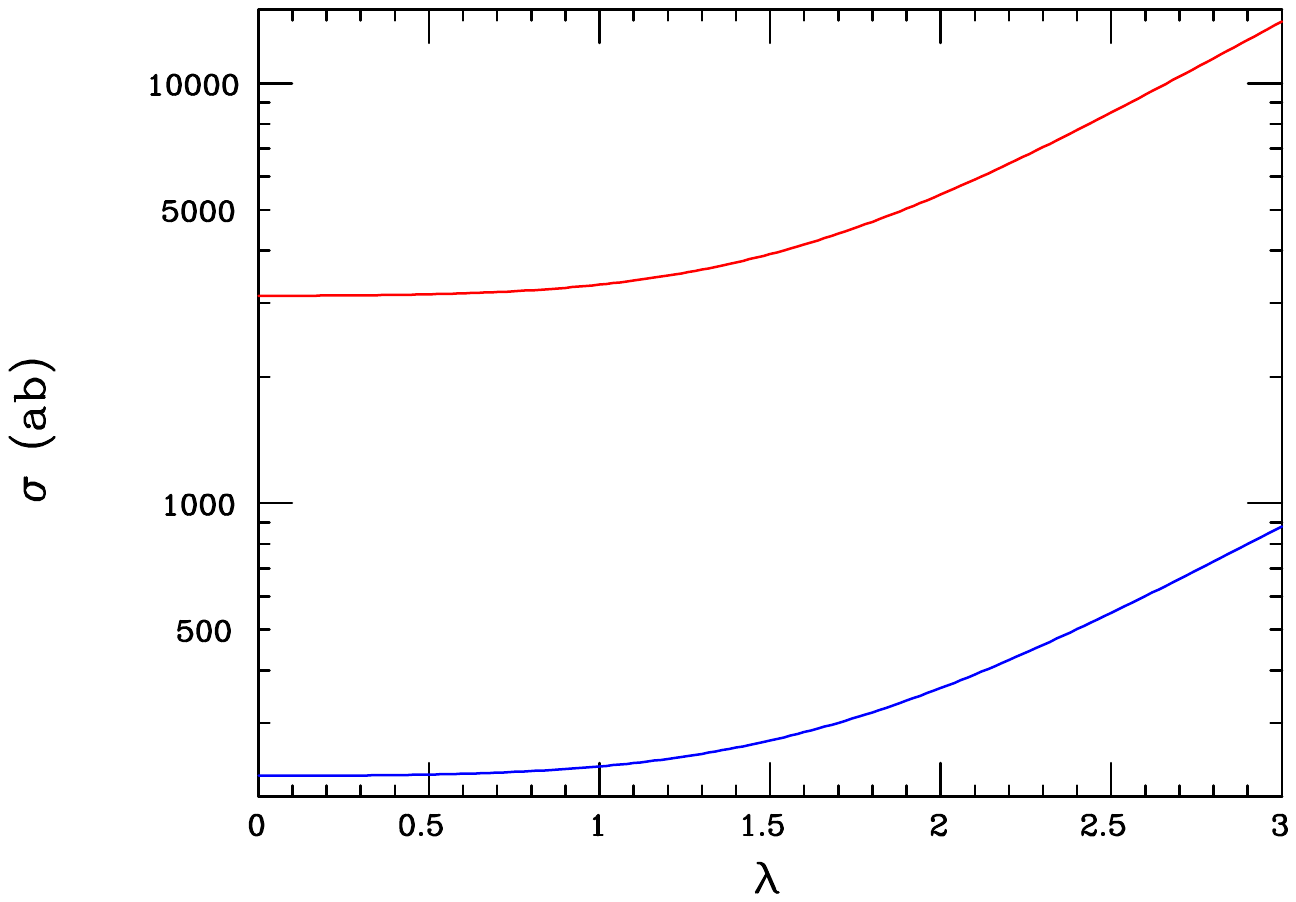}}
\vspace*{-1.5cm}
\centerline{\includegraphics[width=4.0in,angle=0]{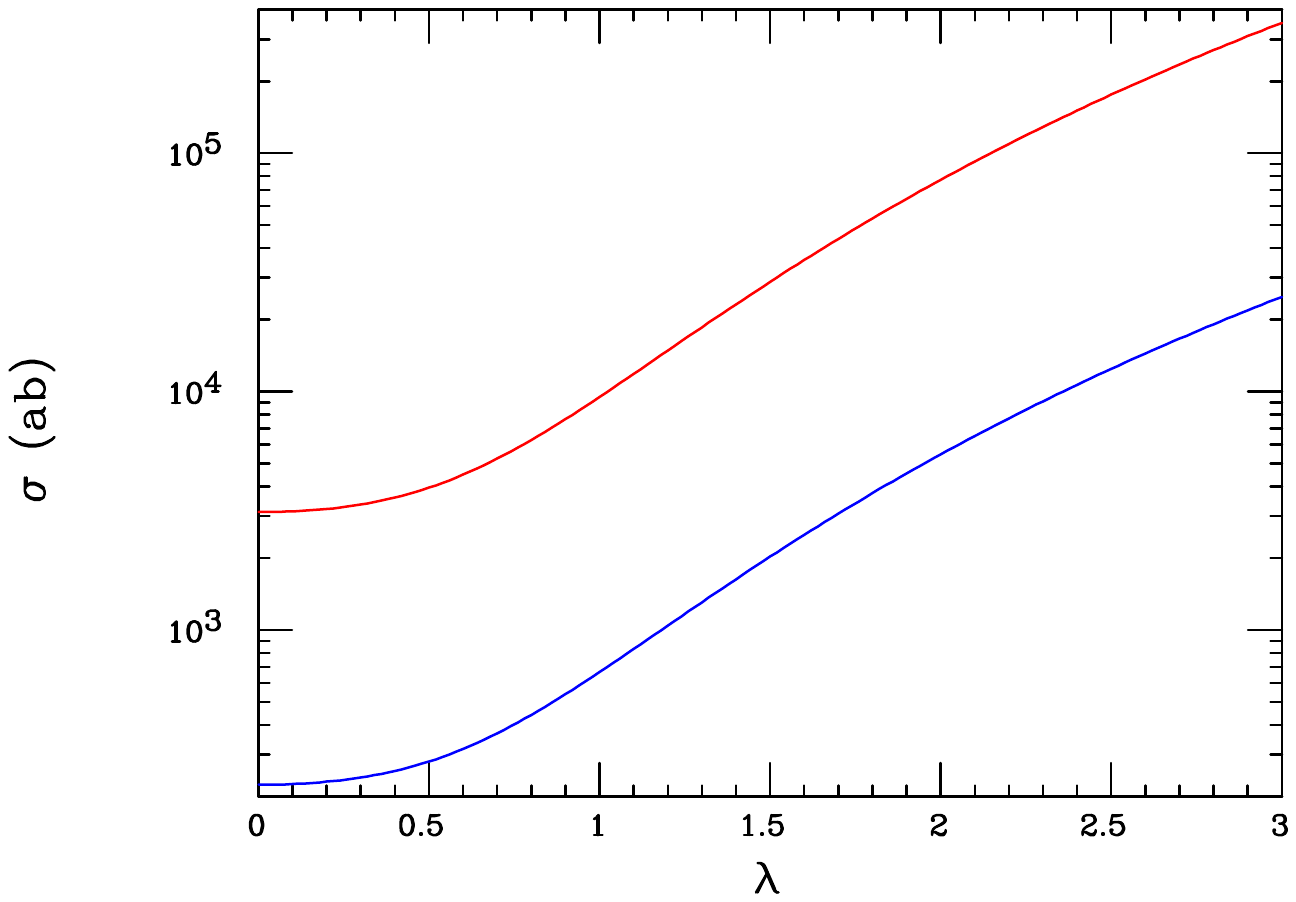}}
\vspace*{-1.20cm}
\caption{(Top Left) Total cross sections corresponding to the angular distributions for $h\bar h$ production at 14 TeV shown in the previous Figure but now as a function of 
$\lambda$ and assuming that $m_h$=1.5(2) TeV for the top(bottom) curve. Here $h-b$ mixing is assumed. (Top Right)  Same as the previous panel but now taking $h$'s SM 
partner to be the $s$ quark. (Bottom) Same as the previous panel but now taking $h$'s SM partner to be the $d$ quark.}
\label{hhtot}
\end{figure}

Unlike for the QCD color triplet exotic fermion $h$, $N$ and $E$ form a vector-like SM weak isodoublet so they can only be produced via the $2_L1_Y2_I1_{I'}$ interactions. Apart 
from possible resonant production in the decay of the $Z_I$ and $W_I$ gauge bosons (as was discussed above and will be further below), SM $W^\pm$ exchange provides 
the largely dominant, model-independent production mechanism at the LHC; $Z$-mediated production is smaller by roughly an order of magnitude.{\footnote {For some range of 
masses $gg$-induced off-shell $Z,Z_I$ contributions may also be significant\cite{Willenbrock:1985tj} here but we will neglect this possibility in this brief discussion.}} 
The rate for this process as a function of $m_E(=m_N)$ for both the $\sqrt s=13$ and 14 TeV LHC is shown 
in Fig.~\ref{leptons}, where the possible effects of $\lambda \neq 0$ have been ignored. Here we see that such states may be visible out to masses 
$\sim 1.5$ TeV, depending on their decays and the relevant SM backgrounds. As noted above and in {\bf I}, these differ from the vector-like leptons usually discussed, \ie, $E\to eA_I,S$ rather than $E\to eH,eZ,\nu W$. However, present limits from the LHC on vector-like leptons with these `conventional' decay paths are, and may likely continue to be, rather poor due 
to large SM backgrounds except in the case where they mix with the $\tau$\cite{Bhattiprolu:2019vdu}.  

\begin{figure}[htbp]
\centerline{\includegraphics[width=5.0in,angle=0]{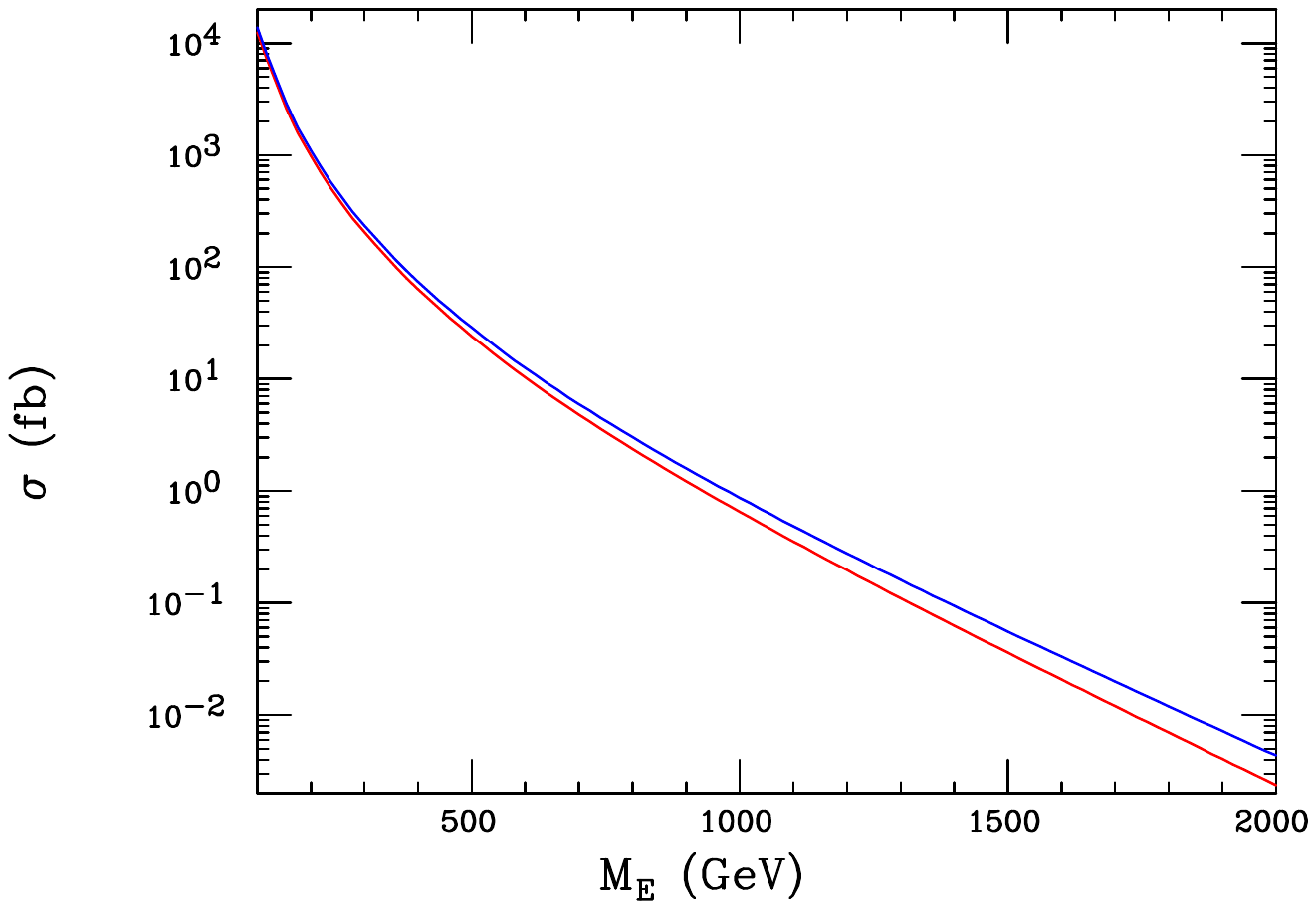}}
\vspace*{-1.50cm}
\caption{$q\bar q \to W^{\pm*}_{SM}\to E\bar N+N\bar E$ production cross section at the $\sqrt s=13$ (red) and 14 (blue) LHC as a function of $m_N=m_E$.}
\label{leptons}
\end{figure}

We now turn to the production of the $W_I$ gauge boson; there are several possible mechanisms for this in the present setup, one of which does not occur in any form in the 
usual traditional treatment. In the original $SU(2)_I$ scenario\cite{Hewett:1988xc}, it was noted that since $W_I$ coupled an exotic vector-like fermion to a SM one it could not be 
singly produced at 
colliders in the usual Drell-Yan fashion, the only options then being pair production via $q\bar q$ annihilation, which we will further discuss below, or in association with an $h$ 
via gluon-quark fusion, \ie, $qg\to hW_I$ where $q=d,s,b$.  In the present setup, this represents three distinct possibilities. The LO subprocess differential cross section for this 
reaction can be easily extracted  from the result given long ago (in a somewhat different context) by Ref.\cite{Gunion:1987xi} with a few obvious alterations, $u\to d$, 
assuming coupling to the first SM generation, $W_R \to W_I$, \etc,  including a rescaling by overall factor of $g_I^2/g^2$. The resulting cross sections as a function of the $W_I$ 
mass for various choices of $m_h$ are shown in Fig.~\ref{mhmwi}. The two upper panels correspond to the cases where $q=d$ for $\sqrt s=13$ and 14 TeV, respectively, while 
the lower panels are for 14 TeV with $q=s$ or $b$, respectively; in all cases one sees that a large part of the model parameter space is potentially accessible for all choices 
of $q$ at $\sqrt s=14$ TeV.  Once $W_Ih$ is produced, $h\to qA_I,qS$ (with $q=d,s$ or $b$) as discussed above and, if $m_{W_I}>m_{h,E}$, then $W_I\to hq,Ee$ with $h,E$ then 
decaying as previously described.{\footnote {If $W_I$ is less massive than $h$ (or $E,N$ \etc) it could possibly decay into a 3-body final state as will be discussed below.}} For $W_I h$, 
this final state is somewhat similar to 
that for $h\bar h$ production as discussed above (which provides a significant background) but with an extra jet which if not b-tagged could be easily mimicked by QCD ISR. Within 
that region of parameter space where $V,S$ decay inside the detector so that the $h$'s can be reconstructed, the corresponding reconstruction of the $W_I$ mass peak using the extra 
$q$ jet would substantially reduce the QCD background. This production process requires further study. 

Another possibility, which is more model-dependent but one we briefly mention here, is that the two SM singlet 
states $S_{1,2}$ might be relatively light with the $Q_I=-1$ state, $S_1$, split from 
and slightly heavier than the $Q_I=0$ state, $S_2$, by $2_I1_{I'}$ gauge boson loops. In this scenario, if these states are indeed light, $W_I$ can always 
decay into them. The $S_2-S_1$ mass splitting, being radiatively generated, is rather small so that if both states are light $S_1$ will generally be relatively long lived  
due to the 3-body nature of the decay over most of the parameter space. Hence, it may likely appear that $S_1$ is stable on detector length scales given the significant boost from 
the large $W_I$ mass. In such cases this final state appears as $W_I \to$ MET at a collider detector, implying that $h-W_I$ associated production 
may likely produce a monojet signature due to the $h\to qA_I,S$ decay. Of course, this scenario may be altered once any significant mixings among all the neutral fields is taken into 
account.

\begin{figure}[htbp]
\centerline{\includegraphics[width=4.0in,angle=0]{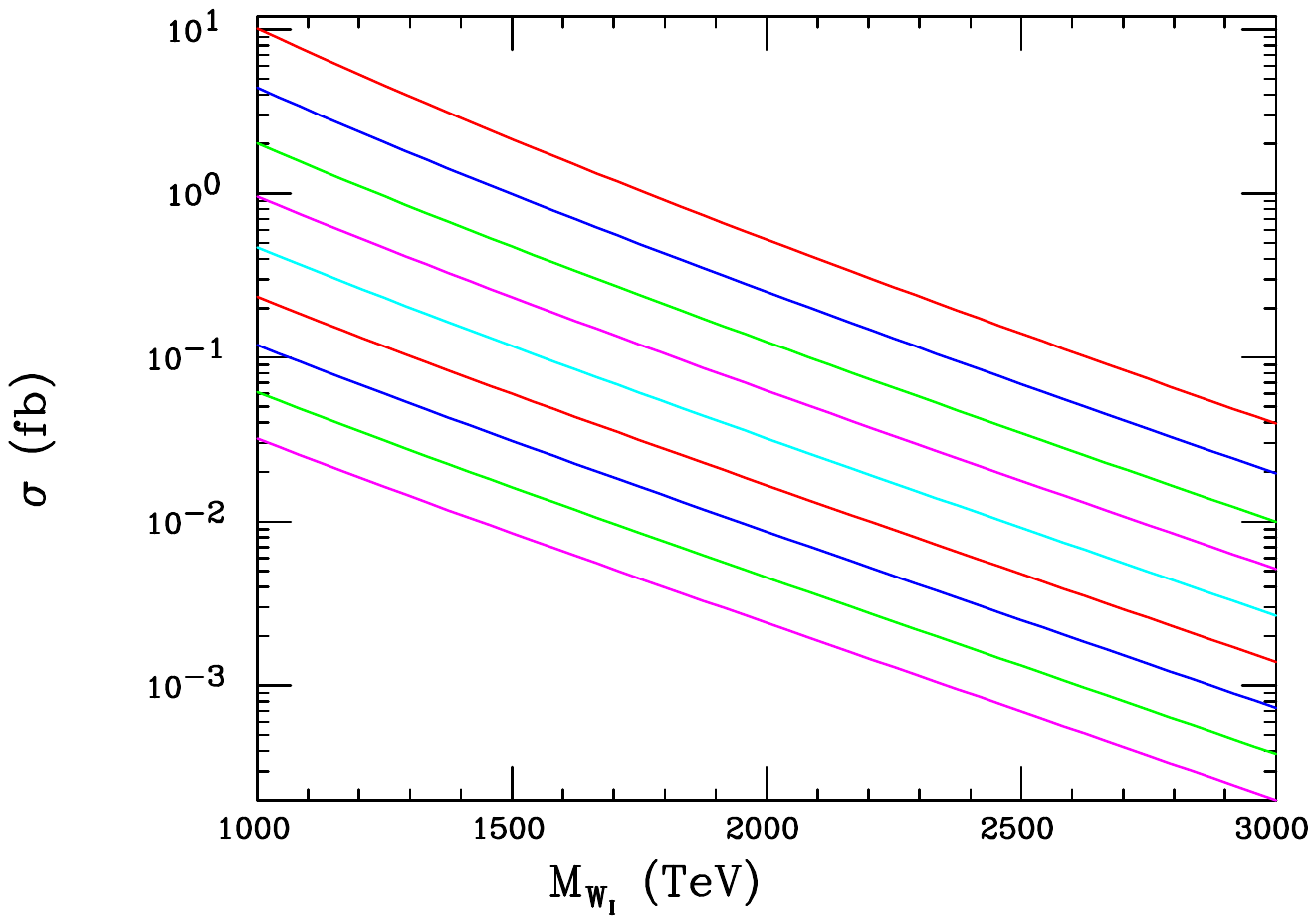}
\hspace*{-1.8cm}
\includegraphics[width=4.0in,angle=0]{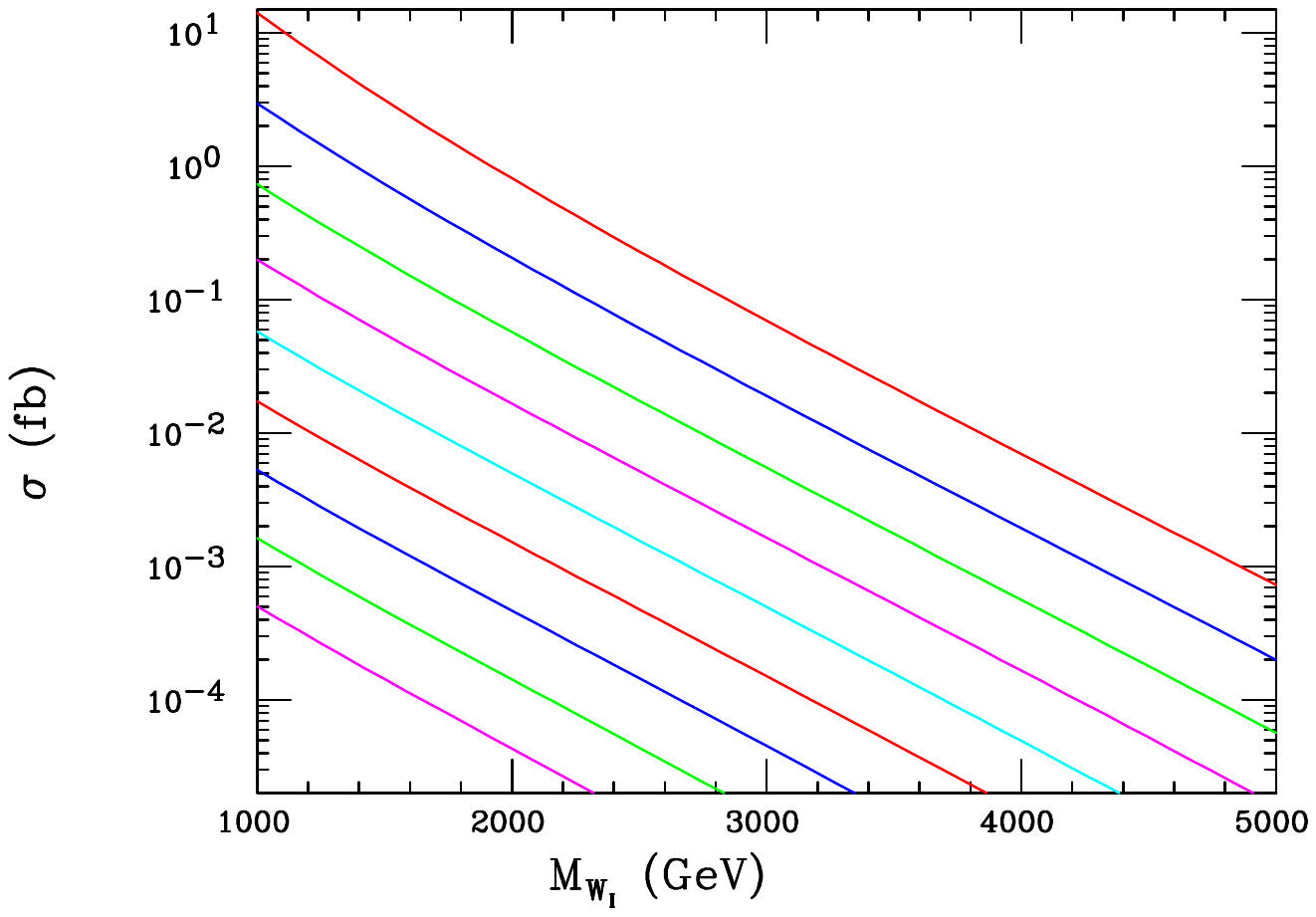}}
\vspace*{-1.5cm}
\centerline{\includegraphics[width=4.0in,angle=0]{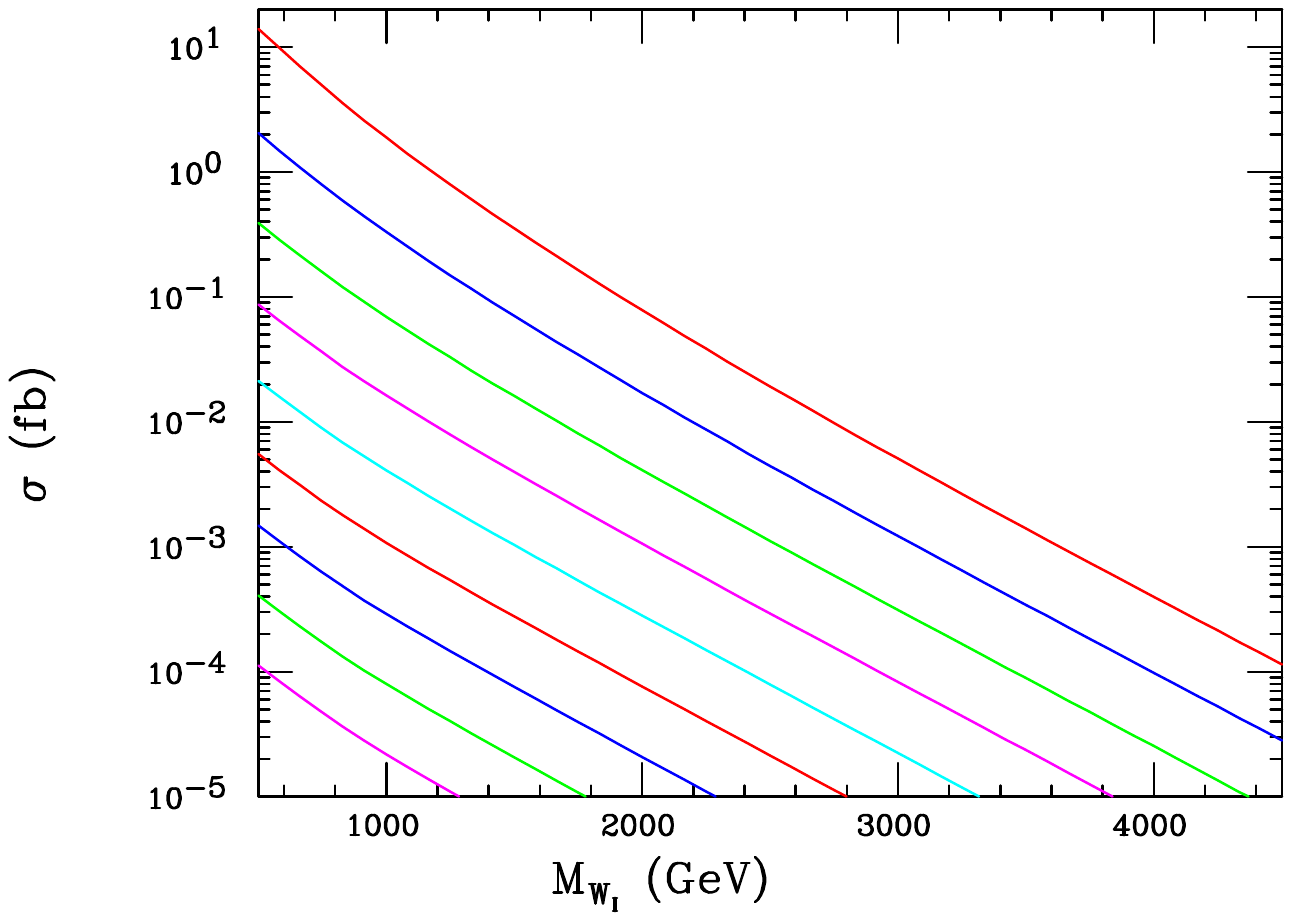}
\hspace*{-1.8cm}
\includegraphics[width=4.0in,angle=0]{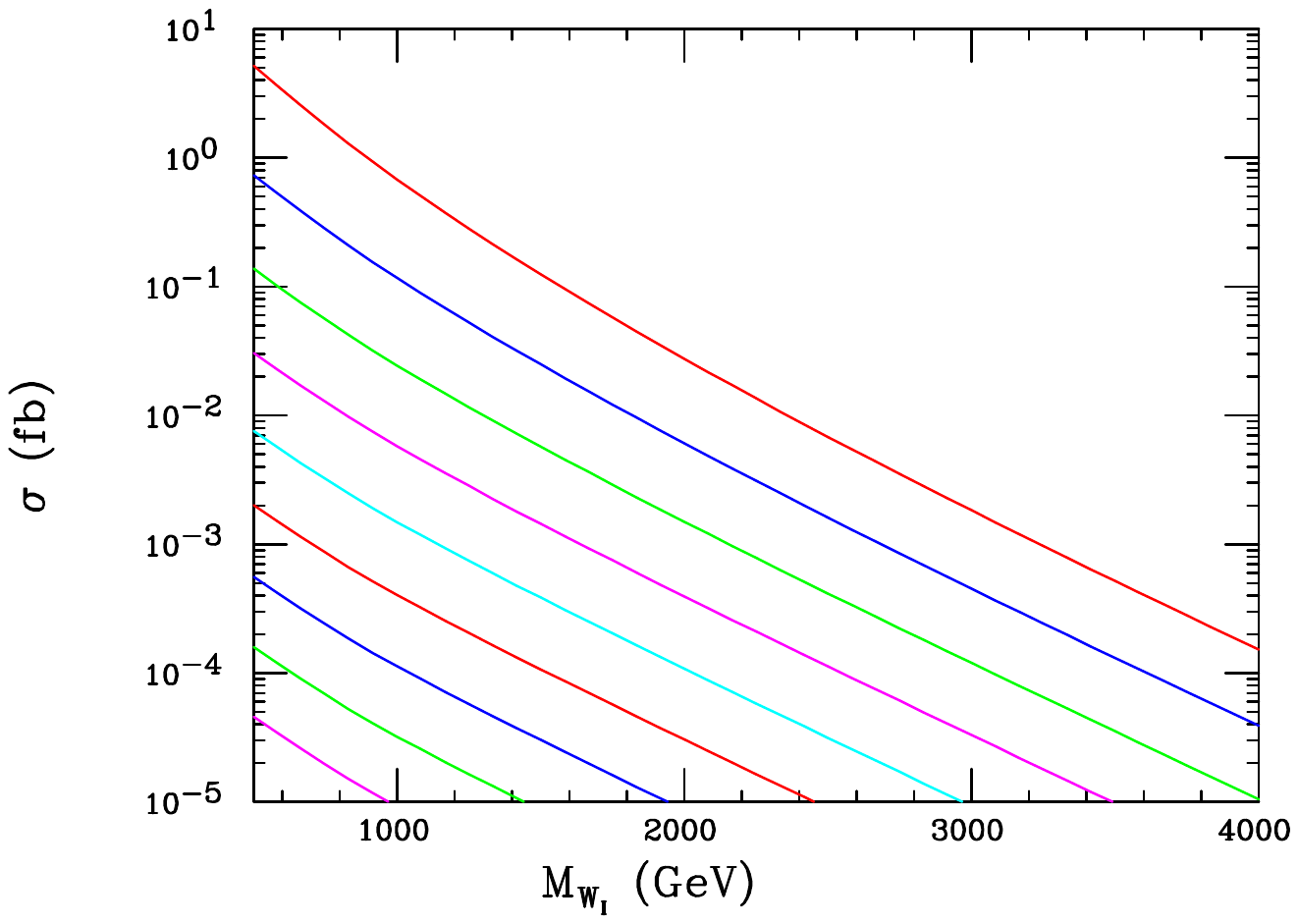}}
\vspace*{-1.20cm}
\caption{(Top Left) $gd \to hW_I+h.c.$ associated production cross section, taking $(g_I/g)^2=1$, as a function of the $W_I$ mass assuming, from top to bottom, that 
$m_h=1,1.25,..,3$ TeV, respectively, at the 13 TeV LHC with $d$ being $h$'s SM partner. (Top Right) Same as the previous panel but now for 14 TeV LHC and for 
$m_h=1,1.5,..5$ TeV. (Bottom Left) 
and (Bottom Right) Same as the previous panel but now assuming $s(b)$ is the SM partner coupling to $h$, respectively. }
\label{mhmwi}
\end{figure}
\begin{figure}[htbp]
\centerline{\includegraphics[width=4.0in,angle=0]{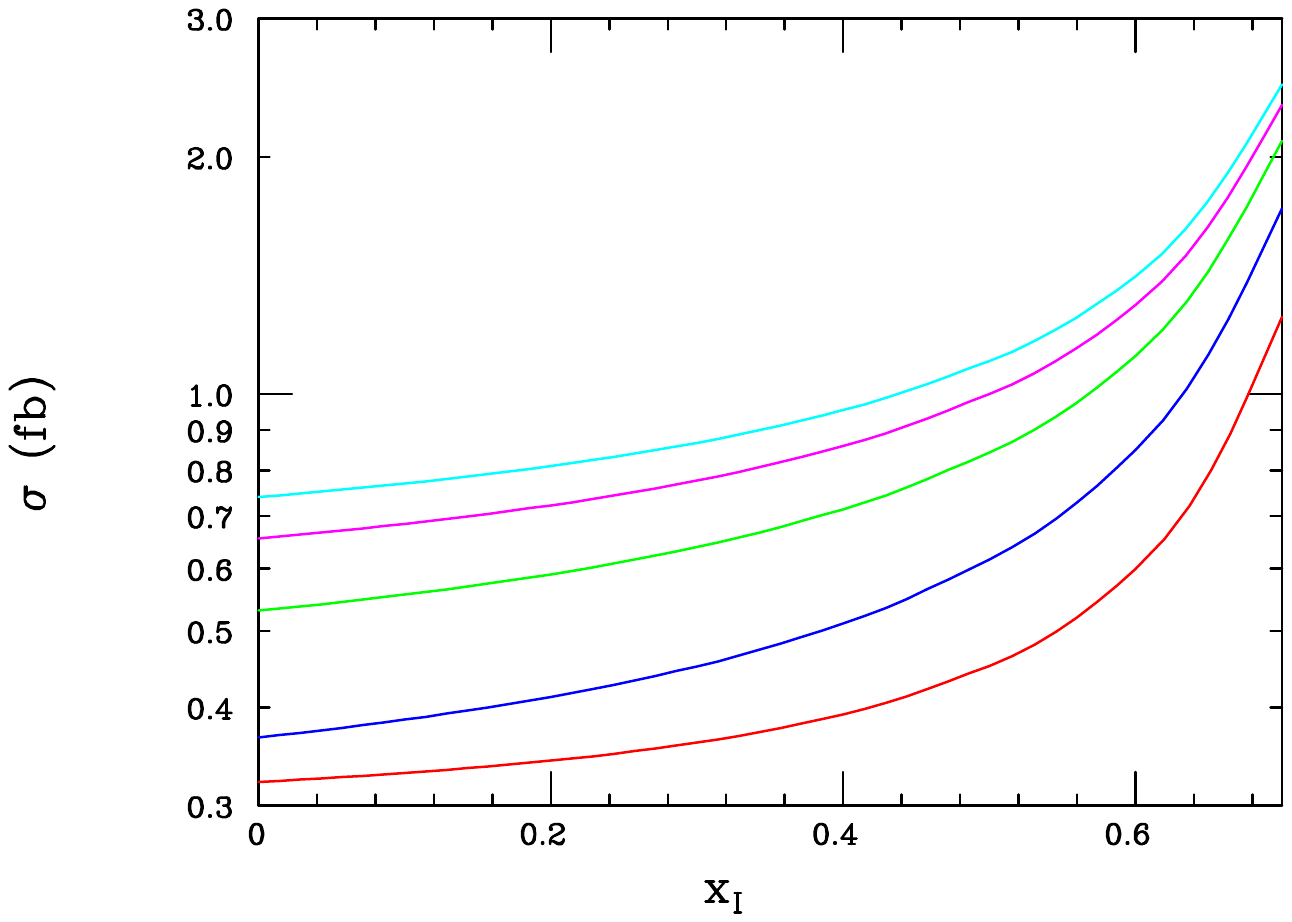}
\hspace*{-1.8cm}
\includegraphics[width=4.0in,angle=0]{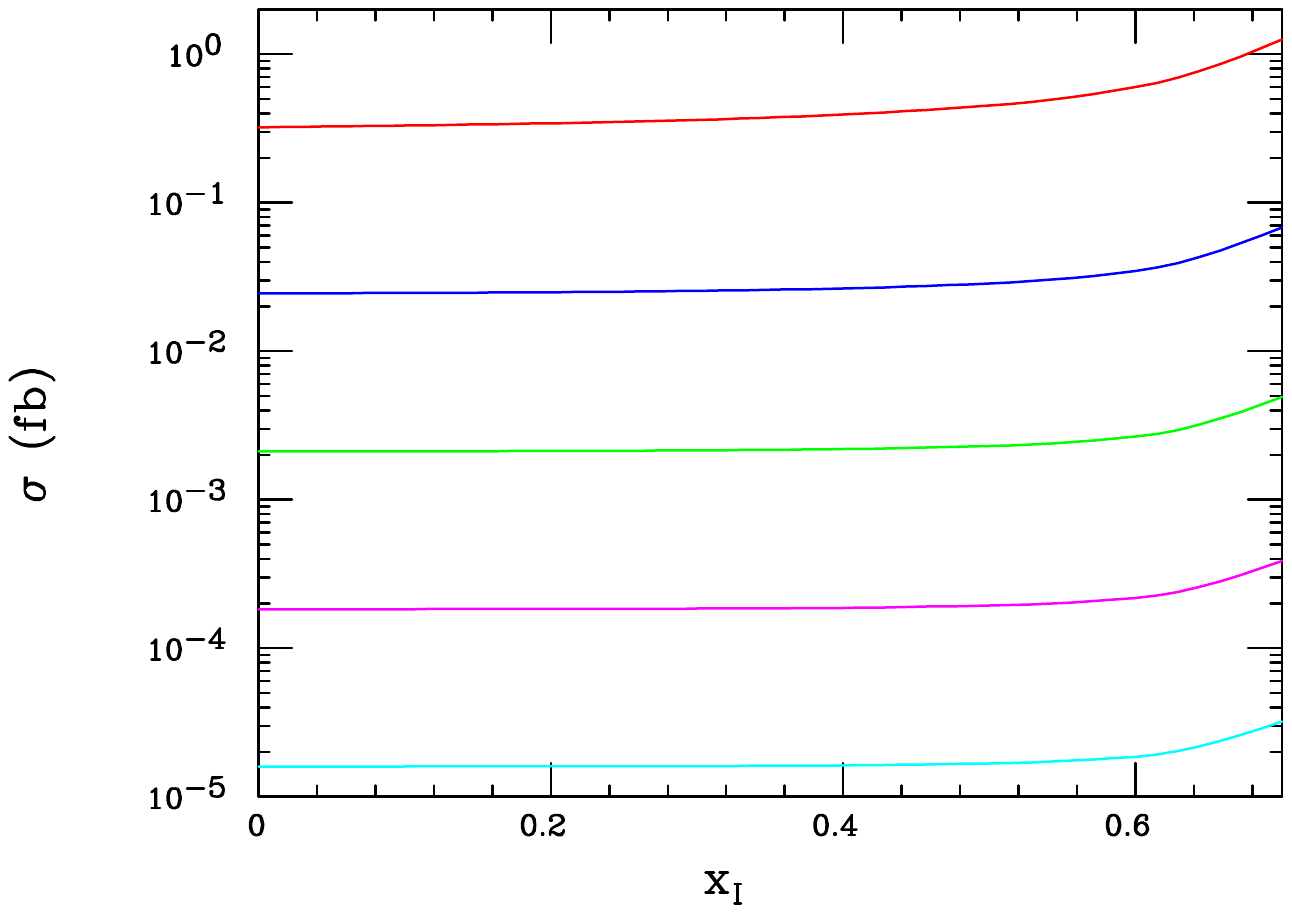}}
\vspace*{-1.5cm}
\centerline{\includegraphics[width=4.0in,angle=0]{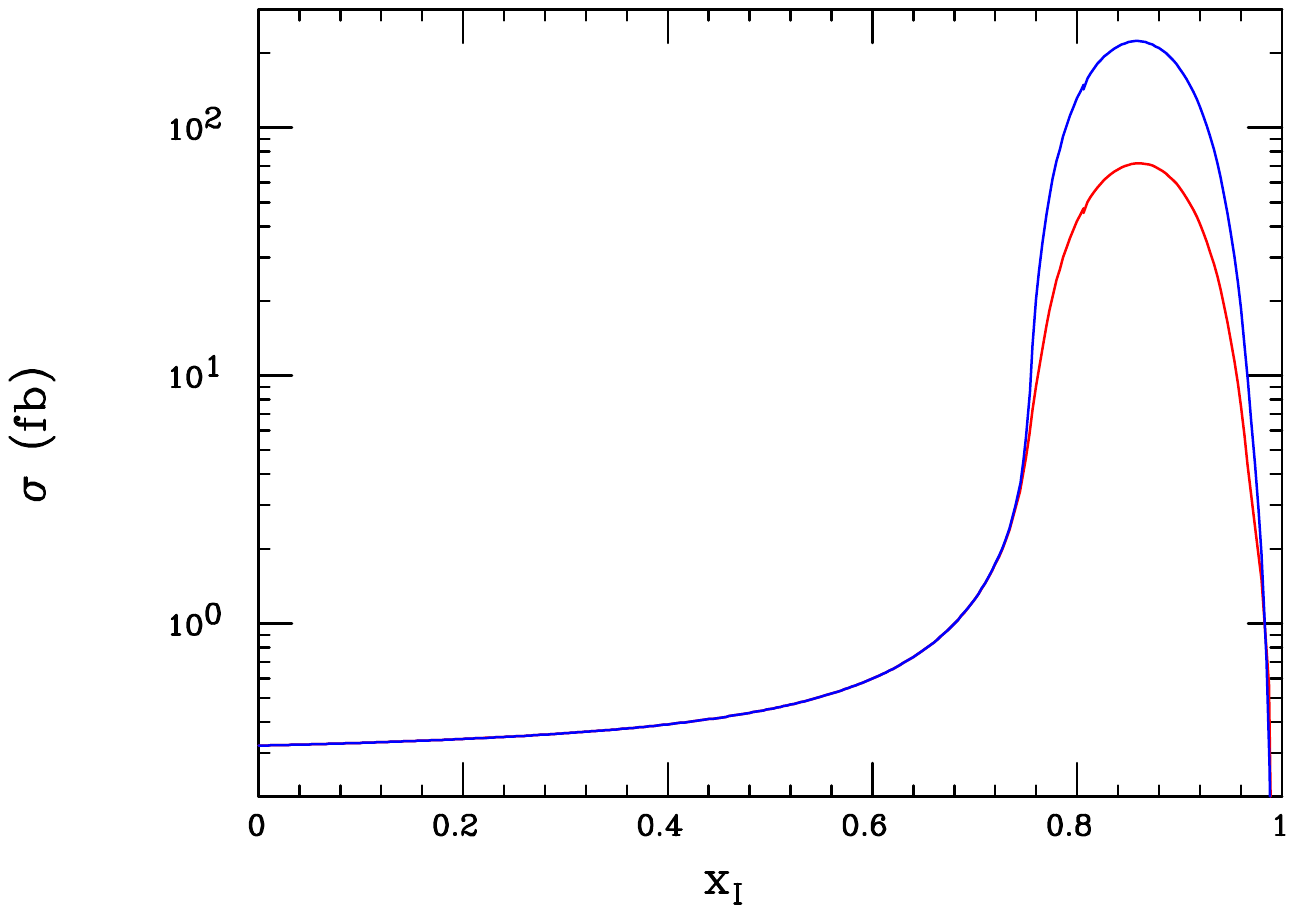}}
\vspace*{-1.20cm}
\caption{(Top Left) $q\bar q \to W_IW_I^\dagger$ total cross section at the 14 TeV as a function of $x_I=s_I^2$ assuming $M_{W_I}=1$ TeV and, from bottom to top  
$m_h=1,2,..,5$ TeV, respectively,  with $d$ being $h$'s SM partner. Here an overall scaling by $(g_I/g)^4$ is still required as in the case of associated production. (Top 
Right) Same as the previous panel but now with $m_h=1$ TeV and, from top to bottom, $m_{W_I}=1,1.5,..,3$ TeV, respectively. (Bottom) Same as the previous panel but 
now assuming $m_{W_I}=m_h=1$ TeV and including the $Z_I$ resonance region at large $x_I$ assuming that $\Gamma_{Z_I}/m_{Z_I}=0.01(0.03)$ as the top 
blue(bottom red) curve.} 
\label{wipair}
\end{figure}

$W_I$ may also be produced in pairs at the LHC. 
Since the $W_I$ is electrically neutral and carries no weak SM charges and also $Q_I(d)=0$, the dominant pair-production process $q\bar q \to W_IW_I^\dagger$, proceeds via 
$s$-channel $Z_I$ exchange as well as $h$ exchange in the $t$-channel (whose amplitudes destructively interfere to satisfy unitarity as $s\to \infty$). This is 
similar to the SM $W^+W^-$ process, but without the photon contribution and with a massive neutrino. We recall that only two 
independent mass parameters, $m_h$ and $m_{W_I}$, enter into the cross section for this process since $M_{W_I}=M_{Z_I}c_I$ at tree level in this setup. However, we note that 
both the values of (the overall factor of) $(g_I/g)^4$ and of $x_I=s_I^2$ are both undetermined in the bottom-up approach that we are following here. Recall 
that the traditional pure $SU(2)_I$ limit of the current setup is achieved in the $x_I\to 0$ limit so that the additional $U(1)$ is then decoupled.
Furthermore, in the present setup, this cross section will be highly sensitive to the choice of $q=d,s,b$ as we saw in the case of several other production processes above. 
The full differential subprocess cross section for this reaction can be extracted, with some care, from the detailed analysis presented in Ref.~\cite{Coutinho:1991pd}. A final `variable' in 
this calculation is the width of $Z_I$ which enters into the $s$-channel exchange and is particularly important for the range $x_I>3/4$ where $Z_I\to W_IW_I^\dagger$ can occur 
on-shell as discussed above. For a reasonable set of choices of exotic fermion mass variations, we generally find that $\Gamma_{Z_I}/M_{Z_I}$ usually lies in the range 
$\sim 0.01-0.03$ when $g_I/g$ is not far from unity, which we will assume in the numerical analysis that follows.

Defining as above $\beta^2=1-4M_{W_I}^2/{\hat s}$ and with $z=\cos \theta^*$, one finds the subprocess cross section to be given by 
\begin{equation}
\frac{d\sigma}{dz}=\frac{g_I^4}{g^4} ~\frac{G_F^2M_W^4}{12\pi}\frac{\beta}{\hat s}\Bigg( E_2 + \frac{-2\hat s(\hat s-m_{Z_I}^2)I+A{\hat s}^2}{(\hat s-M_{Z_I}^2)^2 +(m_{Z_I}\Gamma_{Z_I})^2}\Bigg)
\,,
\end{equation}
where the functions $E_2,I,A(\hat s, \hat t,\hat u)$ are given in Ref.\cite{Coutinho:1991pd} with 
\begin{equation}
~~~~ \hat t,\hat u =m_{W_I}^2 -\frac{\hat s}{2} (1\mp \beta z)\,.
\end{equation}
Cross sections for this process at the 14 TeV LHC are shown in Figs. \ref{wipair} and \ref{wipair2}, which show some of the detailed model parameter dependence for this process. As 
noted, these can depend quite strongly on the value of $x_I$ as it determines the $Z_I-W_I$ mass relationship and thus whether or not the $Z_I\to W_IW_I^\dagger$ process can 
happen on-shell and is thus resonantly enhanced. Since on-shell $Z_I$ production has already been discussed above, here we will {\it mostly} be interested in the case where 
$x_I<0.75$. In Fig.~\ref{wipair} we see that for values of $x_I \lsim 0.6-0.7$, away from the potential $Z_I$ 
contribution, the cross section is only weakly dependent on $x_I$. We see in the top left panel that, not unexpectedly, as we increase $m_h$ we essentially turn off the destructive 
$s/t$-channel interference and the cross section rises, reaching an asymptotic value due to the $Z_I$ exchange diagram alone.{\footnote {The asymptotic cross section value as 
$m_h\to \infty$ with $m_{W_I}=1(5)$ TeV is roughly $\sim 25\%$ larger than that when $m_h=5(10)$ TeV}} For this same range of $x_I$, with $m_h$ held fixed, the upper right panel 
shows that the production rate falls rapidly with increasing $m_{W_I}$ in a manner which is, again, relatively insensitive to the specific value of $x_I$. The lower panel shows the 
rather strong $x_I$ sensitivity to the properties of the $Z_I$ resonance once the on-shell decay process $Z_I\to W_I W_I^\dagger$ becomes allowed.

Similarly to the other production processes considered, this cross section depends upon the choice of $q=d,s,b$. Fig.~\ref{wipair2} shows this cross section as a function of $m_{W_I}$ for various values of $m_h$ while holding 
$x_I=0.25$ fixed but varying the choice of $q=d,s,b$ as the 14 TeV LHC. Here we see that, \eg, assuming $m_h=1$ TeV and demanding a target cross section of at least 10(1) ab for purposes 
demonstration, $m_{W_I}$ is restricted to be $\lsim$ 1.7(2.2), 1.4(1.9) and 1.1(1.5) TeV, respectively, assuming $q=d,s,b$. Clearly, we see that the associated 
production channel generally wins in providing the largest signal rate for $W_I$ production provided that the value of $m_h$ is not too large.

\begin{figure}[htbp]
\centerline{\includegraphics[width=4.0in,angle=0]{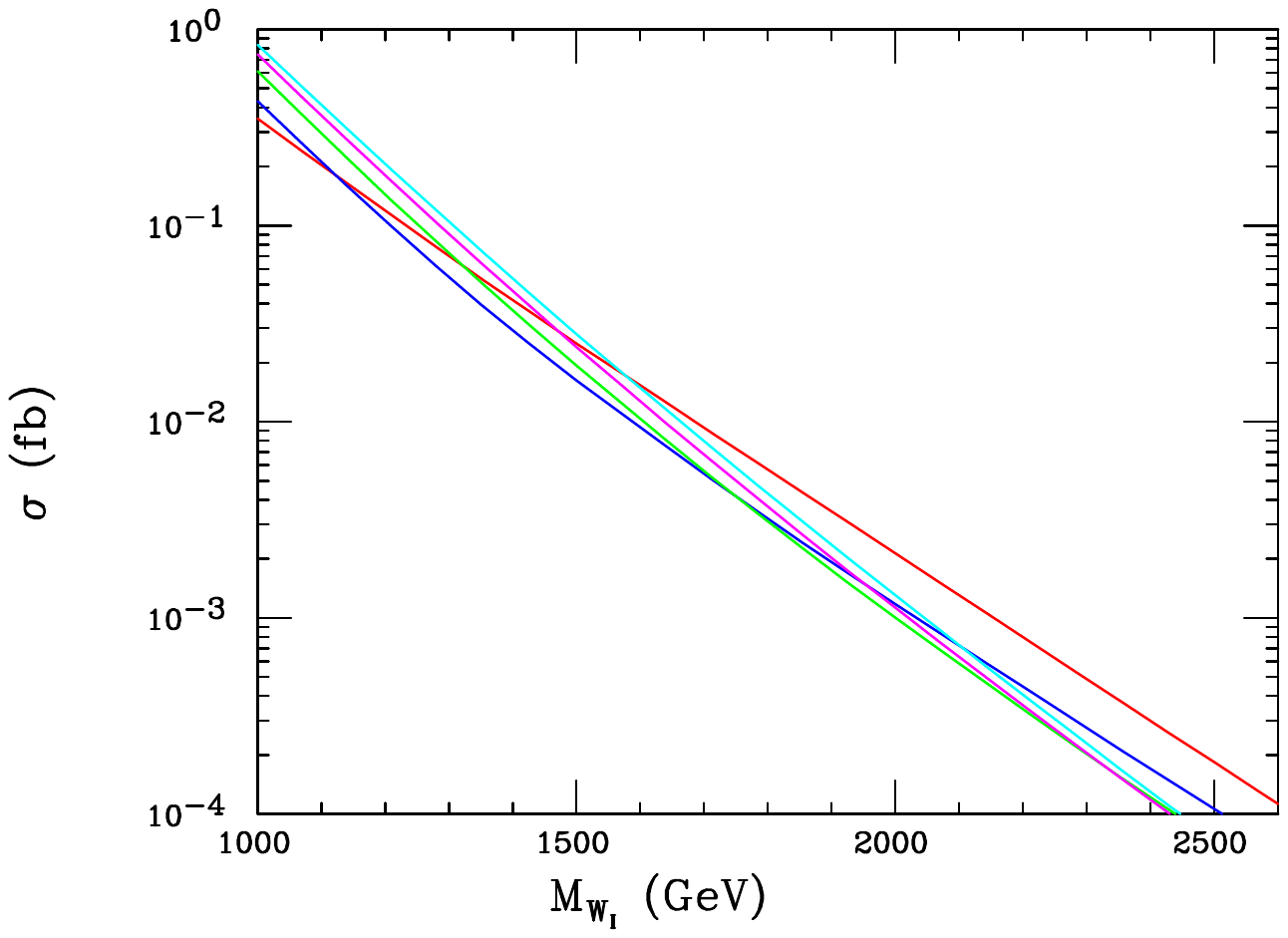}
\hspace*{-1.8cm}
\includegraphics[width=4.0in,angle=0]{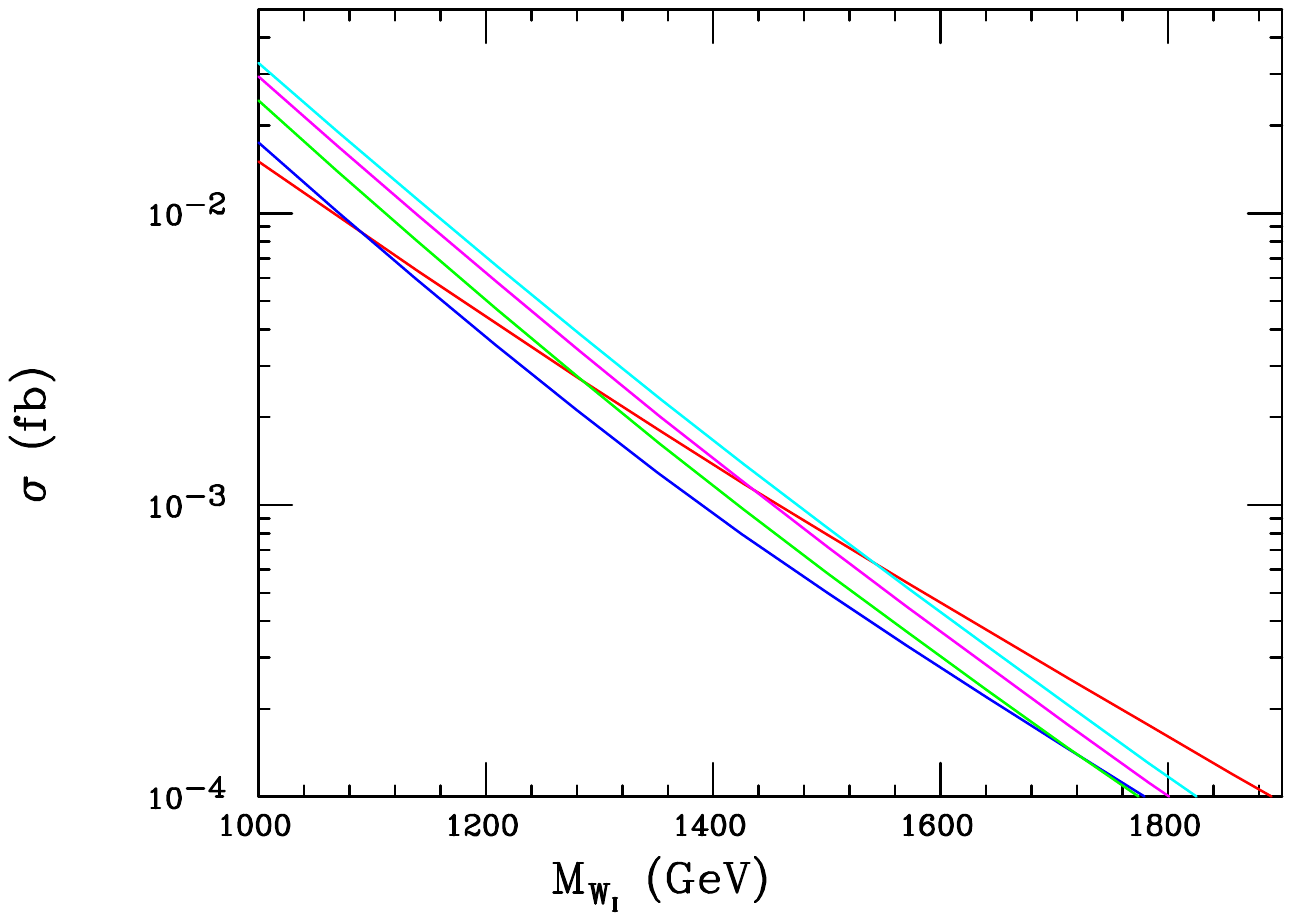}}
\vspace*{-1.5cm}
\centerline{\includegraphics[width=4.0in,angle=0]{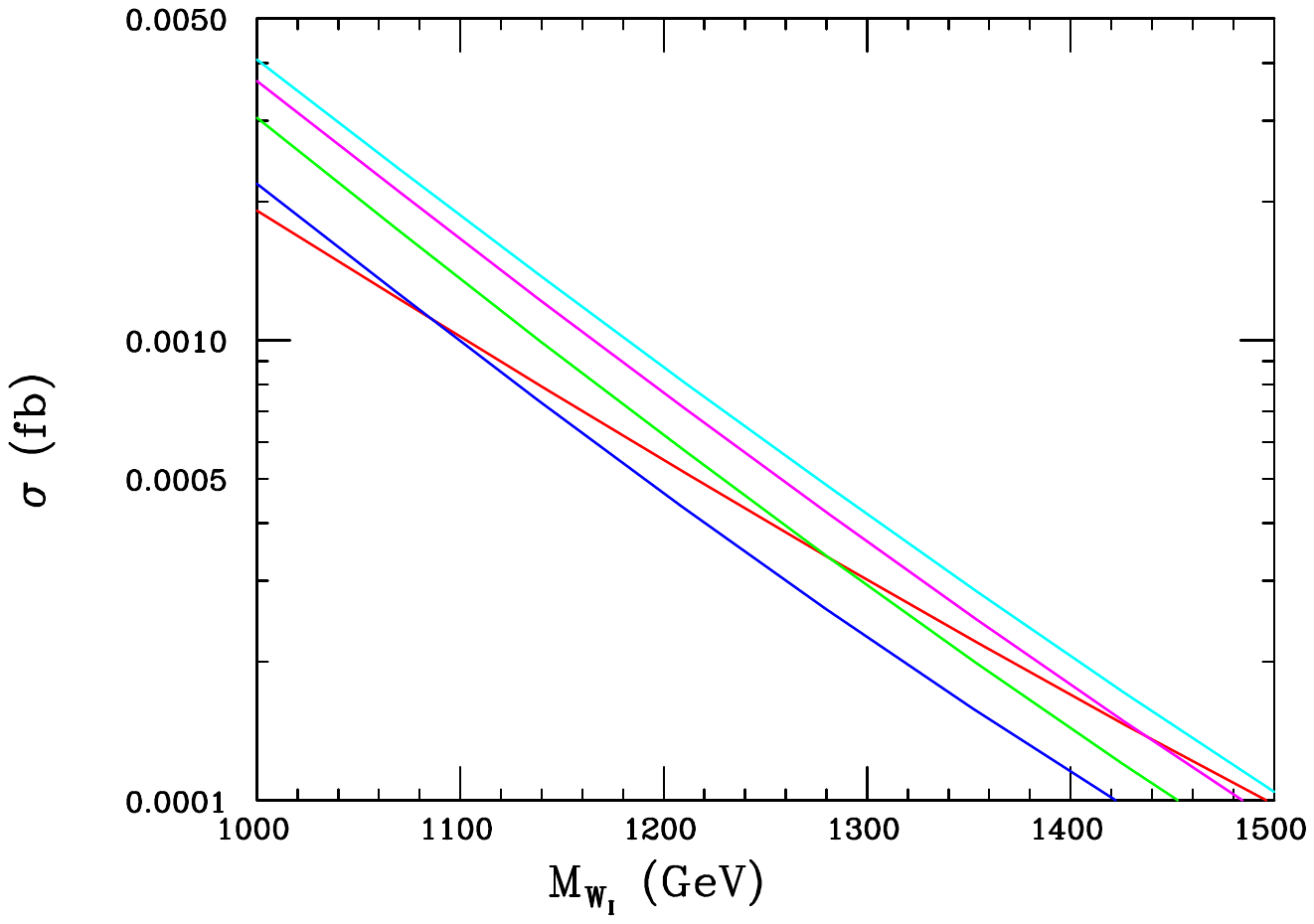}}
\vspace*{-1.20cm}
\caption{(Top Left ) Same as the in the previous Figure but now assuming that $x_I=0.25$ and displayed as a function of $m_{W_I}$ with, from bottom to top at the left axis, 
$m_h=1,2,..,5$ TeV. (Top Right, Bottom) Same as the previous panel but now assuming $s(b)$ is the SM partner to $h$, respectively.}
\label{wipair2}
\end{figure}

Once the $W_I^{(\dagger)}$ is produced its main decay paths are back into a SM-exotic fermion pair, \eg, $\bar eE(e\bar E), \bar \nu N(\nu \bar N)$ or $\bar S_2S_1(S_2\bar S_1)$ and 
$\bar dh(d\bar h)${\footnote {Here $e,\nu,d$ are being used to represent any of the corresponding fermions of the three SM generation.}}. These 2-body partial widths are completely 
fixed by the $W_I$ and $h,E/N$, \etc, masses apart from an overall factor of $g_I^2/g^2$. However, it is always possible that $m_{E,N} $ and/or $m_h$ are larger than $m_{W_I}$ so 
that such decay modes are blocked.{\footnote {However, if the bare mass $M$ is sufficiently small then the $\bar S_2S_1$ mode will always remain open.}} Of course through the 
previously discussed fermion mass mixing effects, governed by $\theta \sim 10^{-4}$, 
decays into $e^+e^-$ and/or $d\bar d$ are always allowed, but with partial widths that are highly suppressed by $\theta^2 \sim 10^{-8}$. Much larger partial widths are potentially 
possible via the off-shell, three-body modes such as $W_I \to eE^* \to e^+e^- (d\bar d) A_I/S$, where here $A_I$ is essentially the longitudinal mode, \ie, the Goldstone boson $G_{A_I}$. 
These are suppressed by relative three-body factors of $\sim \lambda^2/16\pi^3$, which prove not be too prohibitive, and asymptotically scale as $\sim (m_{W_I}/m_E)^2$. 
Fig.~\ref{wi3b} shows the reduced partial width for this decay process, in units of $g_I\lambda/g$, as a function of $\delta=m_E/M_{W_I}$; an analogous result is obtained in the 
case of $h$ exchange except for an additional multiplicative factor of $\sim 3$ for color and QCD corrections.

\begin{figure}[htbp]
\centerline{\includegraphics[width=5.0in,angle=0]{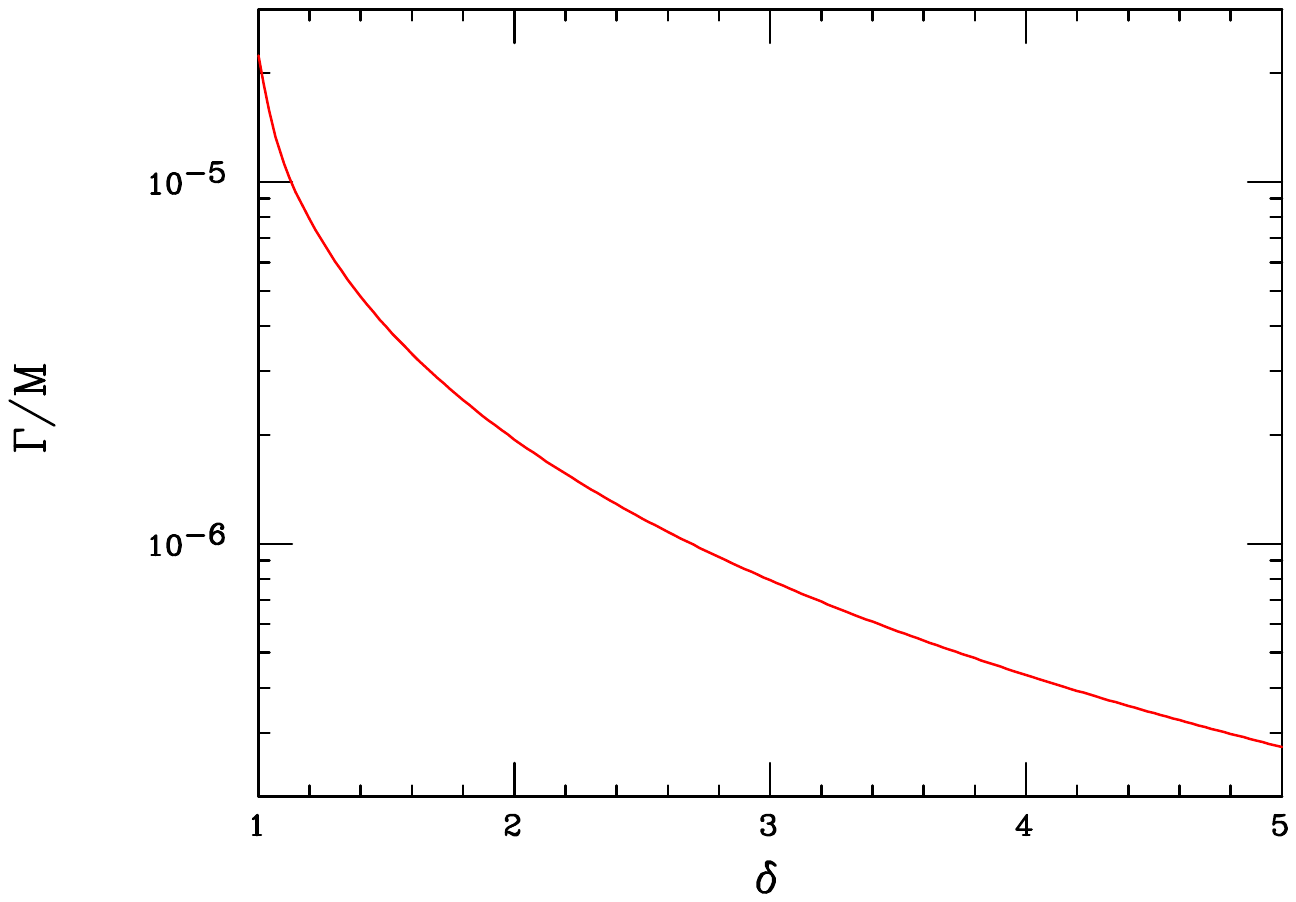}}
\vspace*{-1.50cm}
\caption{Reduced partial width, $\Gamma/m_{W_I}$, for the process $W_I \to \bar eE^* \to e^+e^-A_I/S$ as a function of $\delta=m_E/m_{W_I}$ and in units of 
$(g_I\lambda/g)^2$. Apart from a color factor of 3, a suitable redefinition of both $\lambda,\delta$ and a small correction from QCD, a similar partial decay rate would be obtained for the 
$W_I \to \bar dh^* \to d\bar dA_I/S$ process.}
\label{wi3b}
\end{figure}

In the present setup, $W_I$ can also be produced by a new mechanism not present in the classic $SU(2)_I$ scenario. Since $A_I$ contains a small admixture of $W_I+W_I^\dagger$, 
the process $q\bar q \to W_I^{(\dagger)} A_I/S+h.c.$ via $t$-channel $h$ exchange, with $q=d,s,$ or $b$ becomes possible via $\lambda\neq 0$; again, we see that it 
is the longitudinal component of $A_I$ (or, equivalently, the Goldstone boson $G_{A_I}$) that is mainly responsible for this reaction. This process has a kinematical advantage over both 
$W_IW_I^\dagger$ and $W_Ih$ production in that only a single heavy particle needs to be produced in the final state and thus can lead to the largest signal cross section for $W_I$ 
production in some parameter space regions. Also, unlike the pair production case, the amplitude is 
proportional to the product $\lambda g_I$ instead of $g_I^2$; this can be especially advantageous if $\lambda$ is indeed large.  The subprocess cross section for this reaction is given 
(in the limit where the dark photon and light scalar masses can be safely neglected) by
\begin{equation}
\frac{d\sigma}{dz}=\Big(\frac{\lambda^2 g_I^2}{g^2}\Big)~\frac{G_Fm_W^2}{96{\sqrt 2}\pi \hat s} ~\frac{m_h^2}{m_{W_I}^2}\Big(1-\frac{M_{W_I}^2}{\hat s}\Big) ~\frac{A+1-\beta^2z^2}{[1-\beta z+C]^2} ~\,,
\end{equation}
where here $\beta=(\hat s-m_{W_I}^2)/(\hat s +m_{W_I}^2)$, $z=\cos \theta^*$ as above and 
\begin{equation}
~~~A=\frac{4m_{W_I}^2\hat s}{(\hat s+M_{W_I}^2)^2}, ~~~~~~ C=\frac{2(m_h^2-m_{W_I}^2)}{\hat s +m_{W_I}^2}\,.
\end{equation}
The resulting cross section for this new associated production process as a function of $m_{W_I}$ is shown in Fig.~\ref{wis} for all three choices of initial state SM quark $q=d,s$ or $b$. 
Here we see that, \eg, if $\lambda g_I/g \simeq 1$ then cross sections of at least 10(1) ab are obtained for $m_{W_I}\lsim 3.9(4.9)$ TeV assuming $q=d$. This is a substantially larger reach than 
found in the case of pair production, but is somewhat inferior to the rates found in the case of $hW_I$ associated production under the same assumptions. For comparison, the 
corresponding results for the cases $q=s$ and $q=b$ are $m_{W_I} \lsim 2.5(3.5)$ TeV and $\lsim 2.0 (2.9)$ TeV, respectively. Depending upon the $W_I$ decay mode, this final state can also 
easily lead to MET, monojet or monolepton type signatures, although lepton-jets are possible when the $A_I$ decays sufficiently rapidly.

\begin{figure}[htbp]
\centerline{\includegraphics[width=4.0in,angle=0]{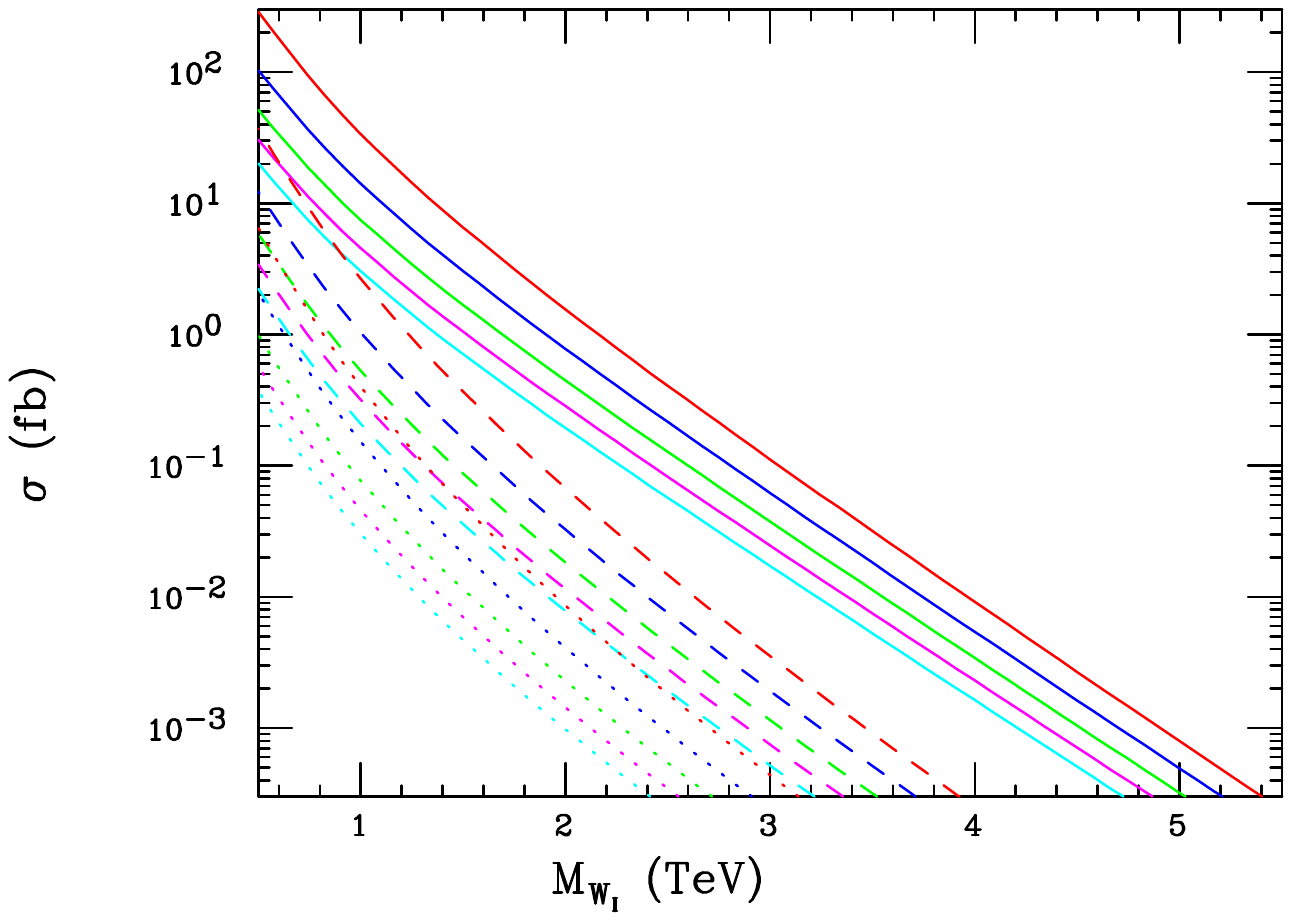}}
\vspace*{-1.5cm}
\centerline{\includegraphics[width=4.0in,angle=0]{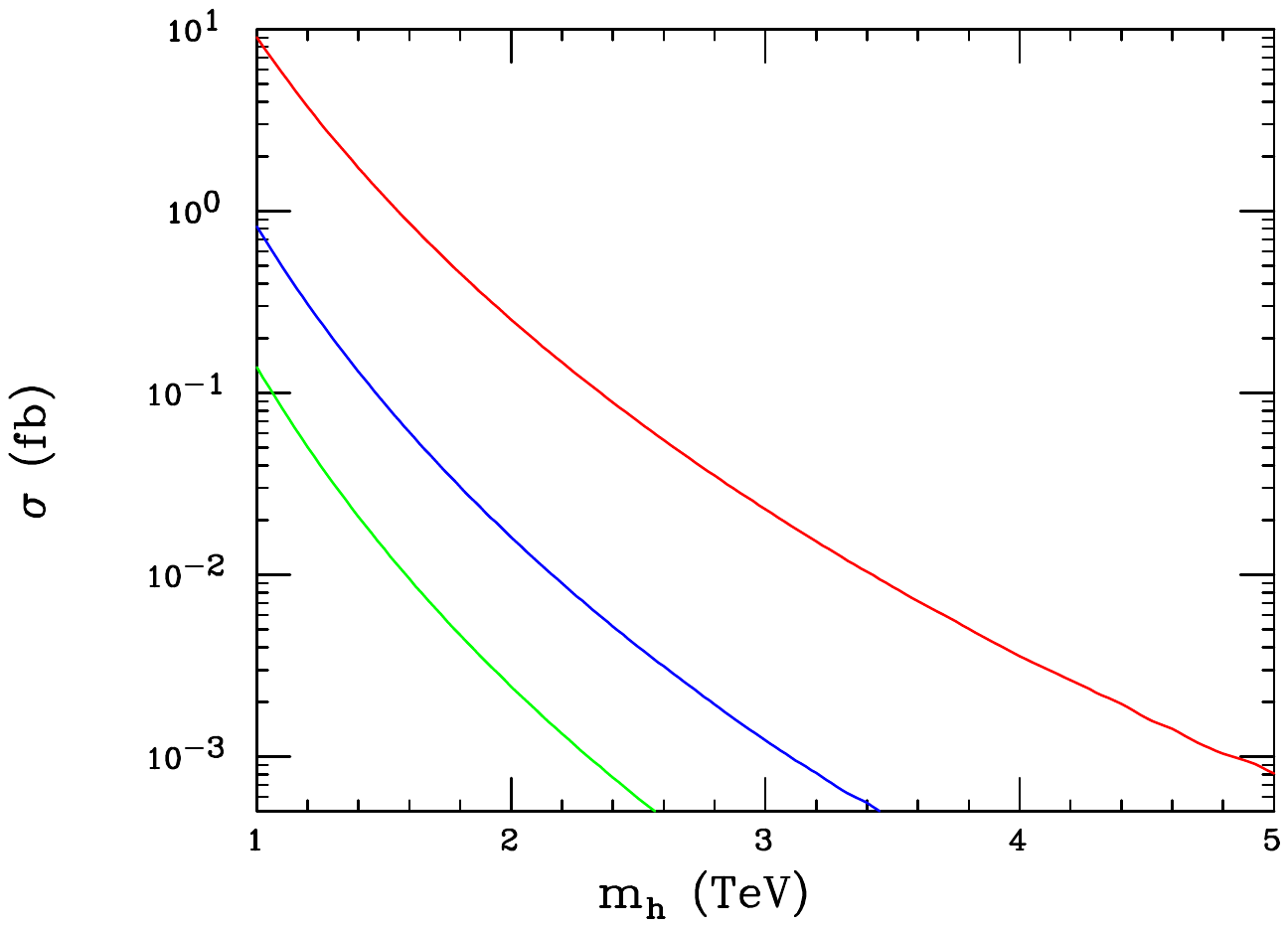}
\hspace*{-1.8cm}
\includegraphics[width=4.0in,angle=0]{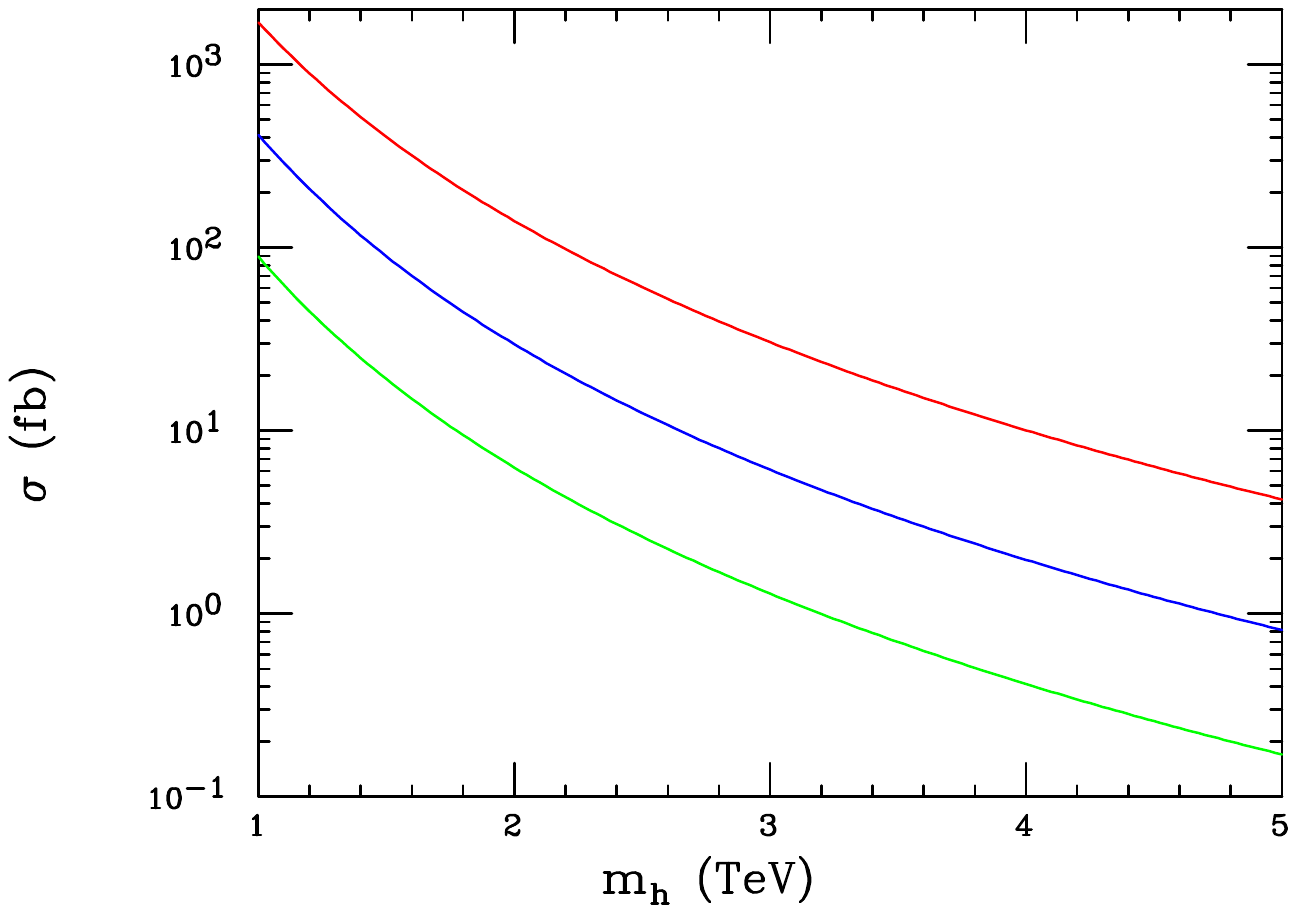}}
\vspace*{-1.20cm}
\caption{(Top) The $W_I^{(\dagger)}A_I/S$ associated production cross section as a function of $m_{W_I}$ and in units of $(g_I\lambda/g)^2$ at the $\sqrt s=14$ TeV LHC. Here the 
curves are for the choice of $d\bar d$ (solid),  $s\bar s$ (dashed) or $b\bar b$ (dotted) initial states assuming, from top to bottom in each set, that $m_h=1,..,5$ TeV, respectively. 
(Bottom Left) $A_I$ or $S$ pair production cross section in units of $\lambda^4$ as a function of $m_h$ assuming, from top to bottom, $q=d,s,b$, respectively.  
(Bottom Right) $A_IS$ associated production cross section in units of $\lambda^4$ as a function of $m_h$ assuming, from top to bottom, $q=d,s,b$, respectively. }
\label{wis}
\end{figure}

Finally, we consider the $q\bar q \to 2A_I(2S)$ production process which proceeds via $t$- and $u$-channel $h$ exchanges. In the case of the $A_I$ final state this is by far 
dominated by the longitudinal polarization mode, and so is well represented by the production of the corresponding Goldstone boson. Summing over both final states and neglecting the 
$A_I,S$ masses in comparison to other mass scales, the subprocess differential cross section for this identical particle final state process is given by
\begin{equation}
\frac{d\sigma}{dz}=\frac{\lambda^4}{48\pi \hat s} \frac{z^2(1-z^2)}{(a^2-z^2)^2} \,,
\end{equation}
where $a=1+2m_h^2/\hat s$ and $z=\cos \theta^*$. Note that the outgoing $A_I/S$ states will generally appear at large $p_T$ and not along the beam directions. 
Examining the numerator, we see that in the limit of large $m_h$ this cross section is pure $d$-wave, as we might expect 
from the production of pairs of identical spin-0 particles in the final state; this is the result of the destructive interference between the $t$-channel and $u$-channel exchange amplitudes. 
The resulting integrated cross section is shown in the lower left panel of Fig.~\ref{wis} for the different choices of $q=d,s,b$, as above. This result is numerically smaller than might be 
expected for the pair production of two very light states due to this strong destructive interference, and we see that in the large $m_h$ limit this interaction turns into an effective dim-8 
operator. For the analogous $q\bar q \to A_IS$ process, where constructive interference now occurs instead, one obtains the resulting $\sim p$-wave cross section 
\begin{equation}
\frac{d\sigma}{dz}=\frac{\lambda^4}{24\pi \hat s} \frac{a^2(1-z^2)}{(a^2-z^2)^2} \,,
\end{equation}
which is seen in the lower right panel of Fig.~\ref{wis} as a function of $m_h$ for the different choices of $q=d,s,b$. Here, unlike in the case of $2S$ or $2A_I$ production, we see a 
substantial event rate qualitatively similar to our naive expectations. As in the case of pair production, we note that the outgoing $A_I/S$ states will generally appear at large $p_T$ 
and not along the beam directions.

For both of these processes on their own, unlike the others considered previously, one sees the potential lack of obvious signal to trigger on at the LHC if the $A_I/S$ is sufficiently 
boosted so that it decays far outside the detector. In such a case one must rely on the production of (an) extra jet(s) produced by QCD to act as a trigger to produce a monojet signature. Even in the case where $A_I/S$ decay inside the detector, this additional radiation may act as a useful trigger.

In order to analyze the production of monojet events at the 13 TeV LHC, we use FeynRules \cite{alloul2014feynrules} to produce UFO files that may be passed to MadGraph5\_aMC@NLO \cite{alwall2014automated} in order to generate parton level events. We generate events for $2A_I, 2S,$ and $A_I S$ + 1-3 generator-level jets final states, as all of these subprocesses will produce large missing $E_T$, though the $2A_I$ and $2S$ final states are suppressed by the destructive interference between the $t$- and $u$-channels as discussed above. It is important to include the additional generator-level hard jets in the final states, as these allow for processes with $qg, \bar q g,$ and $gg$ initial states to contribute to the $E_T^{miss}$ + jets signal, though they do not contribute to the exclusive $2A_I,2S,A_IS$ production considered above. At the generator level, we require the leading jet have $p_{T,j1} > 150$ GeV and that $E^{miss}_T > 150$ GeV, and that all jets have $p_{T,j} > 20$ GeV. We employ the 5 flavor number scheme (5FNS), treating the $b$ quark as massless and including it in the proton PDF, and calculate the production cross section at leading order. These parton level events are then showered and hadronized within Pythia 8 \cite{sjostrand2015introduction}. We use the MLM matching scheme in order to avoid double counting jets from the parton level generation and the parton showering procedure, using the merging parameters {\bf xqcut} = 20 GeV and {\bf QCUT} = 30 GeV. In order to compare generated events to present searches, we use DELPHES 3 \cite{de2014delphes} to simulate detector effects, and make the cuts on the final states as outlined in the most recent ATLAS search \cite{aaboud2018search}. In particular, the leading jet must have $p_{T,j1} > 250$ GeV and $| \eta | < 2.4$, $E^{miss}_T$ must be at least 250 GeV (this is the requirement for the IM1 search in Ref. \cite{aaboud2018search}, the subsequent IM2-IM10 searches have successively higher cuts on $E^{miss}_T$), there must be no electrons with $p_T > 20$ GeV, no muons with $p_T > 10$ GeV, and there may be at most 4 central jets with $|\eta| <2.8$ and $p_T > 30$ GeV in the event.  Fig. \ref{mono}, left, shows the signal cross-section at the 13 TeV LHC after the IM1 cuts have been applied as a function of the mass of the down-like quark partner $h$, for the cases of $h$ coupling individually to each of $q=d,s,b$ as well as the universal coupling case, taking $\lambda = 1$ in all cases. The coupling to the first generation dominates the production cross section in the universal case, and the two are nearly equal for $m_h \gtrsim 3$ TeV.

Since the cross section scales as $\lambda^4$, the 36.1 $\textrm{fb}^{-1}$ ATLAS monojet search can be used to constrain $\lambda$ for the various coupling scenarios. Using the tightest limit on the signal cross section from the IM1-IM10 searches of ref \cite{aaboud2018search}, we place an upper bound on $\lambda$ for each coupling scenario as a function of the $h$ mass, shown in Fig. \ref{mono} right. We note that the searches with higher missing $E_T$ thresholds tend to be more constraining, with the IM7 search, corresponding to $E^{miss}_T > 700$ GeV, the most constraining for $m_h = 1$ TeV, and the IM10 search, with $E^{miss}_T > 1$ TeV, becoming the tightest constraint as $m_h$ increases.  The constraints on $\lambda$ from this search are not very stringent, especially at larger $m_h$ and for the cases of $h$ coupling to the second and third generations. For $m_h = 1$ TeV, we find an upper limit $\lambda < 0.49$ for the universal coupling case, while the upper limits on $\lambda$ for $h$ coupling to $d,s,$ and $b$ are 0.59, 0.76, and 0.81, respectively. These limits weaken considerably as $m_h$ increases, as can be seen in Fig. \ref{mono}.  One may expect increased sensitivity in this search channel at the $\sqrt{s}=14$ TeV HL-LHC, but even a gain of $\sim$23 in sensitivity only translates in to a factor of $\sim$ 2.2 in $\lambda_{max}$, since the cross section scales as $\lambda^4$.

\begin{figure}[htbp]
\centerline{\includegraphics[width=3.5in,angle=0]{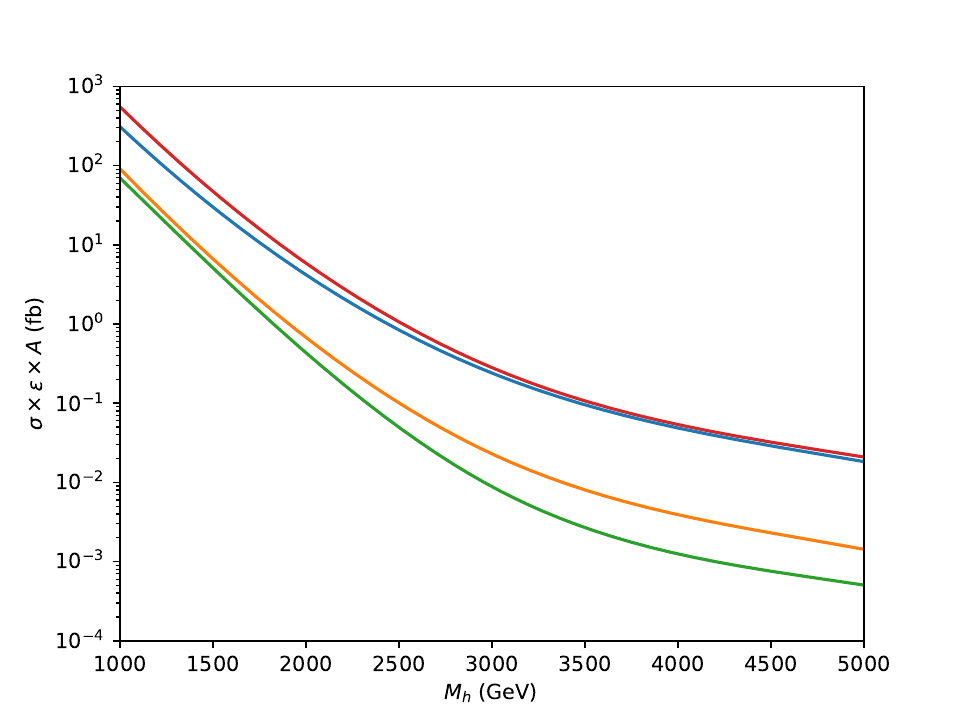}
\hspace*{-0.85cm}
\includegraphics[width=3.5in,angle=0]{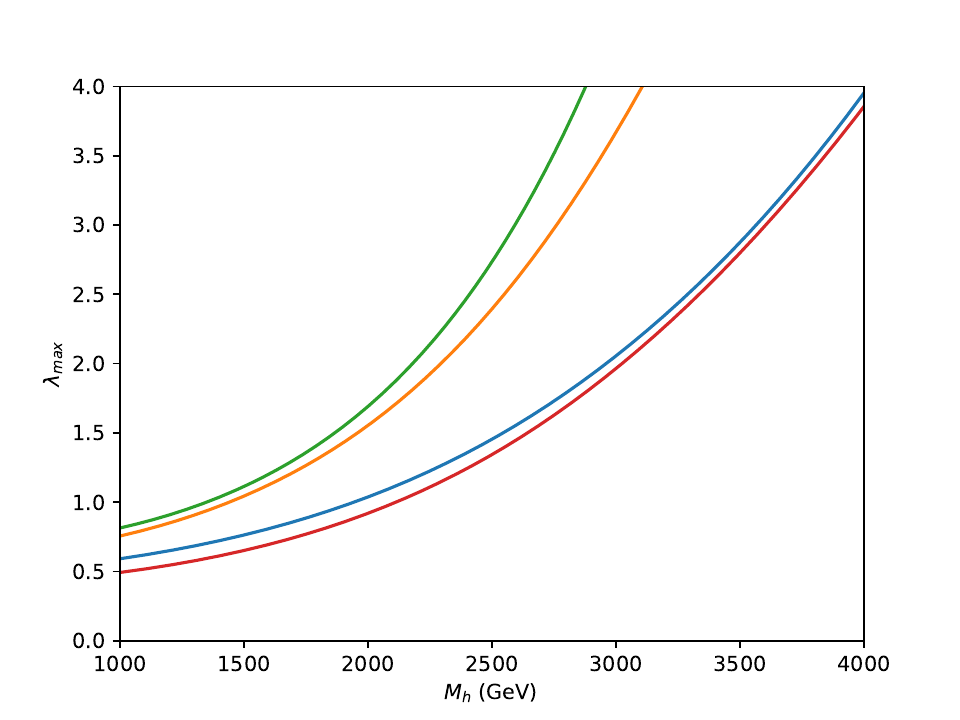}}
\caption{(Left) The signal cross section $\sigma \times \epsilon \times A$ for IM1 for $pp \rightarrow 2A_I, 2S, A_I S + 1-4j$ as a function of $m_{h}$ up to an overall factor of $\lambda^4$ at the $\sqrt s=13$ TeV LHC in the 5FNS. IM1 corresponds to $E_T^{miss} > 250$ GeV in addition to the cuts on jets and leptons described in the text. The red (blue, gold, green) line corresponds to $h$ coupling to all generations (only $d$, only $s$, only $b$). As $m_h$ increases, the case of universal couplings becomes increasingly dominated by the first generation.
(Right) Upper bounds on $\lambda$ from the monojet search of Ref.\cite{aaboud2018search}. As $m_h$ increases, the case of universal couplings is dominated by the first generation and the upper bounds on $\lambda$ converge. }
\label{mono}
\end{figure}

\section{Dark Matter?}

An issue that we have not yet addressed is the identity of the DM state in the present setup{\footnote {For simplicity in the discussion that follows, we will consider only the simpler 
situation of a single set of exotic fermions that mix/couple to only one of the SM generations but the discussion can be directly generalized.}}.  By definition, we require that this DM 
state be a stable, SM singlet which predominantly couples to us via the dark photon and, hence, must necessarily have $Q_D\neq 0$. Among the set of exotic fermions introduced in 
our discussion so far, we see that the $Q_D=-1$ state $S_1$, which sits in an $SU(2)_I \times U(1)_{Y_I}$ isodoublet and is vector-like with respect to this gauge group, would naturally 
fulfill these basic desired requirements. Considered in isolation, $S_1$, together with $S_2$, will have a common Dirac mass, $M$, but radiative corrections will split these two states 
with the $S_1$ state ending up being the {\it heavier} of the two and thus disqualifying it as a stable DM possibility. Since $Q_D(S_1)=0$, it too is not a DM candidate since it does not couple 
to $A_I$. Furthermore, given the set of Higgs fields that we have already introduced above, $H_i$, most of the neutral fermion states $\nu, N, S_{1,2}$ will generally obtain Dirac masses 
and mix amongst themselves such that their gauge interactions will allow for their eventual decays down to SM particles and thus {\it none} of them can be the required stable DM. 
Interestingly, at the renormalizable level so far discussed, one linear combination of these states will necessarily remain massless (due to a lack of a partner with the opposite chirality) 
and may only 
obtain its mass via the introduction of Majorana mass terms or, more generally,  via the introduction of higher dimensional operators; these possibilities will warrant further study elsewhere.     
The only other SM neutral fields that we have seen above are the chiral doublet and triplet fields, $D,T$, for which we would need to produce Majorana mass terms and that were  
introduced simply to cancel the gauge/gravity anomalies. Such states are not likely to be present in a top-down approach or in the next step up the ladder to a more UV-complete theory 
where such cancellations are generally more subtle. 

The implications of these considerations in the bottom-up approach are that we need to introduce a further new neutral, SM singlet state to play the role of DM without upsetting either the 
anomaly cancellation conditions or the requirements associated with the finiteness of the $\epsilon$ parameter. One tempting choice is to add to the fermion 
spectrum a new vector-like, SM singlet, $S_3$,  with a mass $M_3$ which (we assume to be somewhat fine-tuned) which is  of order $ \sim O$(1 GeV), but which is also an $SU(2)_I$ 
singlet and carries a dark charge $Q_D(S_3)=-Q_D(S_1)=1=Y_{I'}$ since such a state also has $T_{3I}=0$. Given the set of Higgs fields $H_i$ above, $S_3$ will not mix with any of the 
other neutral fields that we already have introduced (at the renormalizable level) and hence it has will have no obvious decay paths allowing it to be stable. Such states can arise 
naturally, or at least in a less {\it ad hoc} fashion, at the next step towards a more UV-complete model\cite{tomorrow}.

The state $S_3$, by virtue of having a non-zero $Q_I$ charge, will couple to the $Z_I$ in addition to $A_I$. Since the $Z_I$ is much more massive than the DM, with $m_{Z_I} \simeq 5$ TeV, it will be a subleading contribution to DM-electron scattering. At high temperatures in the early universe, the $Z_I$ interaction may help bring the DM into thermal equilibrium with the SM, $Z_I$ mediated reactions will freeze-out long before $A_I$ mediated reactions, so the latter will control the relic density. However, it is well known that scenarios with light Dirac fermion states with mass less than a few GeV are constrained by measurements of the CMB power spectrum \cite{aghanim2018planck}. These constraints arise from DM annihilating via $s$-wave processes into SM electromagnetic final states (\eg, $e^+e^-$) so that light $S_3$ states annihilating via dark photons are excluded \cite{liu2016contributions, madhavacheril2014current}, as the annihilation cross section is guaranteed to be $s$-wave, either proceeding through $s$-channel $A_I$ exchange for $m_{S_3} < m_{A_I}$ or through the $t$-channel process $S_3 S_3 \rightarrow A_I A_I$ for $m_{S_3} > m_{A_I}$. 

One alternative to these vector-like fermionic DM states is that a (purely) dark Higgs plays the role of dark matter. In the simplest scenario a complex scalar $\phi$ with mass $m_\phi \sim O$(1 GeV) $< m_{A_I}$, charged only under $U(1)_I$ so that its interaction with the SM is mediated only by $A_I$ and $Z_I$, will have $p$-wave annihilations into SM electromagnetic final states, thus avoiding the stringent CMB constraints. We note that $m_\phi < m_{A_I}$ is required for the annihilation cross section to be $p$-wave suppressed, as the process $\phi \phi \rightarrow A_I A_I$ is $s$-wave, and would dominate for $m_\phi > m_{A_I}$. 

While several physical scalar states remain after the symmetry breaking, as described at the end of Section \ref{sec:gauge_sec}, none fit the requirements of a stable, light, complex scalar thermal relic with interactions primarily mediated by $A_I$. The three physical charged states obviously cannot be dark matter, and in the neutral sector we do not have the proper degrees of freedom to form a light $U(1)_D$ charged scalar which only interacts via the dark gauge bosons $A_I$ and $Z_I$. Three of the four electrically neutral states which carry $U(1)_D$ charge in the gauge eigenbasis, $h_{1,4}^0$ and $a_{1,4}^0$, are likely eaten by the $A_I$ and $W_I^{(\dagger)}$, so that the remaining physical neutral states only have a single real CP-even $U(1)_D$ charged component. Indeed, since it is likely that $W_I^{(\dagger)}$ eats a goldstone which is predominantly $h_4$, the remaining state would likely have an appreciable interaction with the SM $Z$ as well. In any event, this state cannot be combined with the presumably $U(1)_D$ neutral CP-odd field which remains after the symmetry breaking to form a complex scalar with $Q_D=1$, so we see that the scalar DM described above must lie outside of the scalar field content proposed in Section \ref{sec:bas_mod} to provide appropriate fermion masses.

One may well ask how the conventional DM physics associated with the canonical KM scenario, where DM-SM interactions are mediated by the exchange of a dark photon with purely 
vectorial couplings alone, is altered by the additional structure introduced above. In particular, we have in mind the calculation of the thermal relic cross section and the DM-electron 
scattering cross section, $\sigma_e$, as relevant for DM direct detection in this mass range $\sim 1$ GeV. From the above discussions we know that this simple interaction picture 
is modified not only by the new interactions introduced by the additional gauge bosons and their mixing with the corresponding SM fields but also by the mass mixing of the portal fields 
and the corresponding SM partners. While some of these effects will influence all of the SM fermions, other possible effects will, of course, be influenced by which SM generation mixes 
with the new exotic fields. It is interesting to note that the coupling of a single set of exotic fields to a specific 
SM generation would produce non-universal couplings which might be explored in flavor experiments, but a study of such effects is beyond the scope of this work. The case considered in 
Section \ref{sec:ferm_mix} dealt with a single exotic generation, but in principle there may be several exotic generations as was briefly discussed above. In this case there would be a 
rather more complex mass mixing structure, but the overall effect would remain roughly of the same order of magnitude. Certainly, if the single set of exotics couple only to the second 
or third generation, the influence of all the new portal matter fields on $\sigma_e$, will be completely absent at tree-level. Fortunately, more generally,  since both DM annihilation and 
electron scattering are dominated by the DM-DP $\sim$ GeV mass scales, or below, they are generally protected from much of this new physics, even when the exotics mix with the first 
SM generation,  which is generally seen to decouple as is certainly the case with the new gauge interactions. For example, the influence of $Z_I$ exchange is clearly inconsequential since 
its mass is likely (at least) several TeV, making a relative contribution to the amplitudes of both processes which is at most $\sim 10^{-3}$. The contributions of the $W,W_I$ gauge bosons 
are corresponding either mass mixing, mass ratio or loop suppressed or some combination of these.  However, due to the $Z-A_I$ mixing discussed above, and the canonical kinetic 
mixing, all SM fields will experience a coupling to the DP given by the combination $e\epsilon Q -\sigma \frac{g}{c_w}(T_{3L}-s_w^2Q)$ where 
$\sigma = (g g_I s_I) (v_1/m_Z)^2 / (2 c_w) \sim \epsilon$ as previously defined; note that $\sigma$ have either sign relative to $\epsilon$.  This $Z-A_I$ mixing term induced a 
parity-violating interaction between all the SM fermions and the DP.  On top of this general effect, the mass mixing of, \eg, the first generation SM fermions with the exotics can produce 
some additional non-trivial effects.  As was discussed above and in {\bf I}, $e-E$ mixing induces a new contribution to the $A_I$ coupling which is purely left-handed and whose 
(relative) magnitude is controlled by the parameter ratio $y\simeq -e_I(v_4/v_3)^2/(e\epsilon)$, which may also have either sign.  Thus for electrons, explicitly, one finds the DP coupling to 
be
\begin{equation}
e\epsilon ~\bar e \gamma_\mu (v_l-a_l\gamma_5)eA_I^\mu\,,
\end{equation}
where $v_l = -1 - \frac{y}{2} -\frac{\sigma g}{2 c_w \epsilon e} (-\frac{1}{2}+2s_w^2)$ and $a_l = -\frac{y}{2} + \frac{\sigma g}{4 c_w \epsilon e}$, respectively. 
The first term in the vector coupling is the canonical one from KM, the second term arises from the $A_I-Z$ mass mixing induced by the light vev of the bidoublet $H_2$, and the final 
term is induced by the $e-E$ mass mixing due to the light vev of the $SU(2)_I$ doublet $H_3$. (This last term will, of course, be absent, \ie, $y\to 0$, if the exotics mix with the second 
or third generation SM fermions.)  DM scattering off of electrons will be then be modified by an overall factor of $v_l^2+a_l^2$ relative to the canonical KM result. For DM pair annihilating into 
$e^+e^-$, since we are far above threshold there is also an identical rescaling factor. However, if the exotics mix with the second SM generation, near but above the 
$\mu^+\mu^-$ threshold, the impact is a bit more complex with a rescaling of the canonical result for this final state by a factor of $v_l^2+2\beta^2a_l^2/(3-\beta^2)$, where to lowest 
order in the velocity expansion $\beta^2=1-(m_\mu/m_{DM})^2$.  Once above the hadronic threshold, the existence of both $v_l,a_l\neq 0$ will re-weight the usual annihilation cross 
sections in a complex manner as various specific particle thresholds are crossed but these effects will remain $O(1)$.

\section{Discussion and Conclusions} 

The dark photon as a mediator between the dark sector and the SM makes a compelling scenario for extending the WIMP idea to smaller DM masses. However, the generation of 
the necessary kinetic mixing at 1-loop requires the existence of new states, portal matter, which are charged under both the SM and the dark $U(1)_D$ gauge group. In order to satisfy 
anomaly freedom, constraints from precision Higgs/electroweak data as well as those on particle lifetimes from nucleosynthesis, these portal matter states must be vector-like (with respect to the SM gauge symmetries) copies of (at least some of) the SM fermions. Furthermore, if the strength of the kinetic mixing is a finite quantity, as might be expected in a UV-framework, then the various couplings of the portal 
matter states must be in some way correlated with one another as might be expected within a non-abelian group structure. 

In this work we take a bottom-up approach to building a theory of the portal matter necessary for 1-loop kinetic mixing, beginning with fermionic matter inspired by $E_6$, but we augment the $SU(5) \times SU(2)_I$ subgroup of $E_6$ by an additional $U(1)_I$, which kinetically mixes with the SM hypercharge. This allows us to avoid a scenario where the SM fermions are charged under the eventual $U(1)_D$ group, as they would be if we identified $U(1)_D$ as a subgroup of $SU(2)_I$. The $SU(2)_I \times U(1)_I$ breaks to $U(1)_D$ at a scale of several TeV, leading to the generation of masses for the gauge bosons $Z_I$ and $W_I^{(\dagger)}$ as well as the exotic fermions at the TeV scale. This TeV scale symmetry breaking can be probed at the LHC, primarily through the production of $Z_I$, $W_I^{(\dagger)}$, or the down-like quark partner $h$, and the absence of such signals to date allows us to place some constraints on the strength of the dark coupling constants and $s_I^2$, the analog of the weak mixing angle associated with the $SU(2)_I \times U(1)_I \rightarrow U(1)_D$ breaking pattern. The exotic and SM fermions are both charged under $U(1)_I$, so that both contribute to the kinetic mixing parameter $\epsilon$. The relatively degenerate masses of the exotic fermions at the TeV scale and the SM fermions at the GeV scale allow the contributions to $\epsilon$ to be suppressed, so that the leading contributions go like log$(\frac{m_E}{m_h})$ and log$(\frac{m_e}{m_d})$. This cancellation generically gives $\epsilon \sim 10^{-(3-4)}$, an interesting portion of parameter space for dark matter.

The gauge charges of the exotic fermions impose requirements on the Higgs sector of the model, both to generate their TeV scale masses and to ensure mass mixings with SM fermions that allow them to decay promptly. Three scales emerge from the pattern of vevs required to give the fermionic content of the theory appropriate masses while producing a GeV scale dark photon: the few -10 TeV scale vev $v_3$, responsible for the $SU(2)_I \times U(1)_I \rightarrow U(1)_D$ breaking and exotic fermion masses, the weak scale vevs $v,v_2$ responsible for the EWSB of the SM and the SM fermion masses, and the GeV scale vevs $v_1,v_4$ which are responsible for breaking $U(1)_D$ as well as generating mass mixings which allow exotic fermion decay. Interestingly, since the light vevs lie in the $SU(2)_I$ doublet and the $SU(2)_L\times SU(2)_I$ bidoublet, this symmetry breaking patter creates a non-standard phenomenology of the dark photon $A_I$, which now acquires a new coupling to the SM through mass mixing with the $Z$ and a new SM-exotic coupling through mass mixing with the hermitian combination $W_I + W_I^{\dagger}$. These mixing effects produce parity-violating couplings to the $A_I$ in addition to the traditional vector $\epsilon e Q$ coupling. The additional $Z$-like coupling modifies cross sections in direct detection experiments, and the SM-exotic coupling can produce a rich collider phenomenology, for instance allowing pair production of $h \bar{h}$ through $t$-channel $A_I$ exchange, which is explored in Section 4. The mass mixing between exotic fermions and their SM counterparts can also induce new diagonal couplings to $A_I$, though these are subdominant and enter at order $\epsilon^2$. For general SM-VLF mixings, FCNCs can yield some important constraints on the details of the model structure.

The extensive exotic field content of this model, and its original $E_6$ inspiration, beg the question of a top-down unified theory approach, which is the topic of a future work. The construction of such a model clearly requires a unifying group larger than $E_6$, and the vector-like dark fermionic states $S_{1,2,3}$ hint at a unification of $SU(2)_I \times U(1)_{I}\rightarrow SU(3)_I$ in a GUT context. Furthermore, the GUT symmetry breaking pattern may yield a more predictive phenomenology by fixing the relative coupling strengths and mixing angles which were treated as free parameters above. This further work should also seek to provide a more compelling dark matter candidate, as there are quite tight constraints for DM coupling to a kinetically mixed dark photon in the GeV mass range.

In summary, we have analyzed the implications of fermionic portal matter inspired by the exotic content of $E_6/SU(2)_I$-based models, which gives rise to a kinetic mixing between a massive dark photon $A_I$ and the SM photon. Requiring that the exotic fermions have TeV scale masses and decay into SM particles implies a Higgs and gauge sector which has many non-standard couplings outlined in Section 3, and a rich phenomenology which may be explored in low-energy experiments and at colliders as detailed in Section 4. The question of thermal dark matter and the implications of the non-standard couplings of the dark photon for direct detection experiments are summarized in Section 5, and a further exploration of a unified theory approach is reserved for future work.

\section*{Acknowledgements}
TGR would like to thank J.L. Hewett for discussions related to this work.  This work was supported by the Department of Energy, Contract DE-AC02-76SF00515.




\begin{thebibliography}{99}

\bibitem{Kawasaki:2013ae} 
  M.~Kawasaki and K.~Nakayama,
  Ann.\ Rev.\ Nucl.\ Part.\ Sci.\  {\bf 63}, 69 (2013)
  [arXiv:1301.1123 [hep-ph]].

\bibitem{Graham:2015ouw} 
  P.~W.~Graham, I.~G.~Irastorza, S.~K.~Lamoreaux, A.~Lindner and K.~A.~van Bibber,
  Ann.\ Rev.\ Nucl.\ Part.\ Sci.\  {\bf 65}, 485 (2015)
  [arXiv:1602.00039 [hep-ex]].

\bibitem{Arcadi:2017kky} 
  For a recent review of WIMPs, see G.~Arcadi, M.~Dutra, P.~Ghosh, M.~Lindner, Y.~Mambrini, M.~Pierre, S.~Profumo and F.~S.~Queiroz,
  arXiv:1703.07364 [hep-ph].

\bibitem{Alexander:2016aln} 
  J.~Alexander {\it et al.},
  arXiv:1608.08632 [hep-ph].

\bibitem{Battaglieri:2017aum} 
  M.~Battaglieri {\it et al.},
  arXiv:1707.04591 [hep-ph].

\bibitem{Aghanim:2018eyx} 
  N.~Aghanim {\it et al.} [Planck Collaboration],
  arXiv:1807.06209 [astro-ph.CO].

\bibitem{vectorportal} 
 There has been a huge amount of work on this subject; see, for example, 
  D.~Feldman, B.~Kors and P.~Nath,
  Phys.\ Rev.\ D {\bf 75}, 023503 (2007)
  [hep-ph/0610133];
  D.~Feldman, Z.~Liu and P.~Nath,
  Phys.\ Rev.\ D {\bf 75}, 115001 (2007)
  [hep-ph/0702123 [HEP-PH]].;
  M.~Pospelov, A.~Ritz and M.~B.~Voloshin,
  Phys.\ Lett.\ B {\bf 662}, 53 (2008)
  [arXiv:0711.4866 [hep-ph]];
  M.~Pospelov,
  Phys.\ Rev.\ D {\bf 80}, 095002 (2009)
  [arXiv:0811.1030 [hep-ph]]; 
  H.~Davoudiasl, H.~S.~Lee and W.~J.~Marciano,
  Phys.\ Rev.\ Lett.\  {\bf 109}, 031802 (2012)
  [arXiv:1205.2709 [hep-ph]] and 
  Phys.\ Rev.\ D {\bf 85}, 115019 (2012)
  doi:10.1103/PhysRevD.85.115019
  [arXiv:1203.2947 [hep-ph]];
  R.~Essig {\it et al.},
  arXiv:1311.0029 [hep-ph];
  E.~Izaguirre, G.~Krnjaic, P.~Schuster and N.~Toro,
  Phys.\ Rev.\ Lett.\  {\bf 115}, no. 25, 251301 (2015)
  [arXiv:1505.00011 [hep-ph]];
  M.~Khlopov,
  Int.\ J.\ Mod.\ Phys.\ A {\bf 28}, 1330042 (2013)
  [arXiv:1311.2468 [astro-ph.CO]];
 For a general overview and introduction to this framework, see  
  D.~Curtin, R.~Essig, S.~Gori and J.~Shelton,
  JHEP {\bf 1502}, 157 (2015)
  [arXiv:1412.0018 [hep-ph]].
 
\bibitem{KM}
  B.~Holdom,
  Phys.\ Lett.\  {\bf 166B}, 196 (1986) and
  Phys.\ Lett.\ B {\bf 178}, 65 (1986); 
  K.~R.~Dienes, C.~F.~Kolda and J.~March-Russell,
  Nucl.\ Phys.\ B {\bf 492}, 104 (1997)
  [hep-ph/9610479];
  F.~Del Aguila,
  Acta Phys.\ Polon.\ B {\bf 25}, 1317 (1994)
  [hep-ph/9404323];
  K.~S.~Babu, C.~F.~Kolda and J.~March-Russell,
  Phys.\ Rev.\ D {\bf 54}, 4635 (1996)
  [hep-ph/9603212];
  T.~G.~Rizzo,
  Phys.\ Rev.\ D {\bf 59}, 015020 (1998)
  [hep-ph/9806397].


\bibitem{Rizzo:2018vlb} 
  T.~G.~Rizzo,
  Phys.\ Rev.\ D {\bf 99}, no. 11, 115024 (2019)
  doi:10.1103/PhysRevD.99.115024
  [arXiv:1810.07531 [hep-ph]].
    
 
  \bibitem{GUT}
 See, for example, the last two papers in Ref.~\cite{KM}.
 
 \bibitem{vlf}
 For a recent overview of vector-like quarks and original references, see 
  C.~Y.~Chen, S.~Dawson and E.~Furlan,
  Phys.\ Rev.\ D {\bf 96}, no. 1, 015006 (2017)
  [arXiv:1703.06134 [hep-ph]];
  for a corresponding recent overview of vector-like leptons and original references, see
  Z.~Poh and S.~Raby,
  Phys.\ Rev.\ D {\bf 96}, no. 1, 015032 (2017)
  [arXiv:1705.07007 [hep-ph]].
For a general recent review of VLF, see 
  V.~Peralta,
  arXiv:1712.06193 [hep-ph].

\bibitem{searches}
  M.~Aaboud {\it et al.} [ATLAS Collaboration];
  arXiv:1808.02343 [hep-ex];
  A.~M.~Sirunyan {\it et al.} [CMS Collaboration],
  arXiv:1805.04758 [hep-ex];
  CMS Collaboration [CMS Collaboration],
  CMS-PAS-EXO-18-005.

\bibitem{Tanabashi:2018oca} 
  M.~Tanabashi {\it et al.} [Particle Data Group],
  Phys.\ Rev.\ D {\bf 98}, no. 3, 030001 (2018).
  
 \bibitem{however}
  See, however, some earlier general discussions in
  B.~Patt and F.~Wilczek,
  hep-ph/0605188; 
  H.~Davoudiasl, R.~Kitano, T.~Li and H.~Murayama,
  Phys.\ Lett.\ B {\bf 609}, 117 (2005)
  [hep-ph/0405097];
  D.~McKeen, M.~Pospelov and A.~Ritz,
  Phys.\ Rev.\ D {\bf 86}, 113004 (2012)
  [arXiv:1208.4597 [hep-ph]];
  B.~Batell, M.~Pospelov, A.~Ritz and Y.~Shang,
  Phys.\ Rev.\ D {\bf 81}, 075004 (2010)
  [arXiv:0910.1567 [hep-ph]];
 J.~H.~Kim,  S.~D.~Lane, H.~S.~Lee, I.~M.~Lewis, and M.~Sullivan
 [arXiv:1904.05893].

\bibitem{Hewett:1988xc} 
For a review, see 
  J.~L.~Hewett and T.~G.~Rizzo,
  Phys.\ Rept.\  {\bf 183}, 193 (1989).

\bibitem{gherghetta2019price}
  T.~Gherghetta, J.~Kersten, K.~Olive, and M.~Pospelov,
  {\it The Price of Tiny Kinetic Mixing} [arXiv:1909.00696]


\bibitem{London:1986dk} 
  D.~London and J.~L.~Rosner,
  Phys.\ Rev.\ D {\bf 34}, 1530 (1986).
  doi:10.1103/PhysRevD.34.1530
    
 \bibitem{LRM}
For a classic review and original references, see R.N. Mohapatra, {\it Unification and Supersymmetry}, 
(Springer, New York,1986). 

\bibitem{tomorrow}
T.~D.~ Rueter and T.~G. ~Rizzo, work in progress.

\bibitem{Branco:2011iw} 
  G.~C.~Branco, P.~M.~Ferreira, L.~Lavoura, M.~N.~Rebelo, M.~Sher and J.~P.~Silva,
  Phys.\ Rept.\  {\bf 516}, 1 (2012)
  doi:10.1016/j.physrep.2012.02.002
  [arXiv:1106.0034 [hep-ph]].

 \bibitem{GBET}
  M.~S.~Chanowitz and M.~K.~Gaillard,
  Nucl.\ Phys.\ B {\bf 261}, 379 (1985);
  B.~W.~Lee, C.~Quigg and H.~B.~Thacker,
  Phys.\ Rev.\ D {\bf 16}, 1519 (1977);
  J.~M.~Cornwall, D.~N.~Levin and G.~Tiktopoulos,
  Phys.\ Rev.\ D {\bf 10}, 1145 (1974)
  Erratum: [Phys.\ Rev.\ D {\bf 11}, 972 (1975)];
  G.~J.~Gounaris, R.~Kogerler and H.~Neufeld,
  Phys.\ Rev.\ D {\bf 34}, 3257 (1986).
 

\bibitem{ishiwata2015new}
  K.~Ishiwata, Z.~Ligeti, and M.~B.~Wise,
  JHEP {\bf 2015}, no. 10, 27 (2015).
  [arXiv:1506.03484]

\bibitem{bobeth2017patterns}
  C.~Bobeth, A.~J.~Buras, A.~Celis, and M.~Jung,
  JHEP {\bf 2017}, no. 4, 79 (2017).
  [arXiv:1609.04783]  

\bibitem{bobeth2017yukawa}
  C.~Bobeth, A.~J.~Buras, A.~Celis, and M.~Jung,
  JHEP {\bf 2017}, no. 7, 124 (2017).
  [arXiv:1703.04753]

\bibitem{leutwyler1984determination}
  H.~Leutwyler and M.~Roos,
  Z. Phys. C {\bf 25}, no. 1, 91 (1984).

\bibitem{ball2005new}
  P.~Ball and R.~Zwicky,
  Phys. Rev. D {\bf 71}, no. 1, 014015 (2005).
  arXiv:0406232 [hep-ph].

\bibitem{tanabashi2018review}
  M.~Tanabashi {\it et al.} [Particle Data Group]
  Phys. Rev. D {\bf 98}, no. 3, 030001 (2018).

\bibitem{babar2013search}
  BABAR Collaboration,
  [arXiv:1303.7465]


\bibitem{Bolton:2019bou} 
  P.~D.~Bolton, F.~F.~Deppisch, C.~Hati, S.~Patra and U.~Sarkar,
  arXiv:1902.05802 [hep-ph].
 
\bibitem{Aad:2019fac} 
  G.~Aad {\it et al.} [ATLAS Collaboration],
  arXiv:1903.06248 [hep-ex].

\bibitem{ATLAStau36}
  M.~Aaboud, {\it et al.} [ATLAS collaboration],
  JHEP {\bf 2018}, no. 1, 55 (2018).
  [arXiv:1709.07242]

\bibitem{ATLASNote}
ATLAS Collaboration, ``Prospects for searches for heavy Z' and W' bosons in fermionic final states with the ATLAS experiment at the HL-LHC'',  ATL-PHYS-PUB-2018-044.

\bibitem{ATLASNote2}
ATLAS Collaboration,  ``Prospects for the search for additional Higgs bosons in the ditau final state with the ATLAS detector at HL-LHC'',  ATL-PHYS-PUB-2018-050

\bibitem{hpair}
We employ numerical estimates based on 
  M.~Czakon and A.~Mitov,
  Comput.\ Phys.\ Commun.\  {\bf 185}, 2930 (2014)
  [arXiv:1112.5675 [hep-ph]] and also 
  M.~Aliev, H.~Lacker, U.~Langenfeld, S.~Moch, P.~Uwer and M.~Wiedermann,
  Comput.\ Phys.\ Commun.\  {\bf 182}, 1034 (2011)
  doi:10.1016/j.cpc.2010.12.040
  [arXiv:1007.1327 [hep-ph]].

\bibitem{LHCTop}
We use as a check of our numerical calculations the results of the LHC Top Physics Working Group, \url{https://twiki.cern.ch/twiki/bin/view/LHCPhysics/LHCTopWG}.


\bibitem{ljet}
  N.~Arkani-Hamed and N.~Weiner,
  JHEP {\bf 0812}, 104 (2008)
  [arXiv:0810.0714 [hep-ph]];
  M.~Baumgart, C.~Cheung, J.~T.~Ruderman, L.~T.~Wang and I.~Yavin,
  JHEP {\bf 0904}, 014 (2009)
  [arXiv:0901.0283 [hep-ph]];
  A.~Falkowski, J.~T.~Ruderman, T.~Volansky and J.~Zupan,
  Phys.\ Rev.\ Lett.\  {\bf 105}, 241801 (2010)
  [arXiv:1007.3496 [hep-ph]] and 
  JHEP {\bf 1005}, 077 (2010)
  [arXiv:1002.2952 [hep-ph]];
  C.~Cheung, J.~T.~Ruderman, L.~T.~Wang and I.~Yavin,
  JHEP {\bf 1004}, 116 (2010)
  [arXiv:0909.0290 [hep-ph]];
  G.~Barello, S.~Chang, C.~A.~Newby and B.~Ostdiek,
  Phys.\ Rev.\ D {\bf 95}, no. 5, 055007 (2017)
  [arXiv:1612.00026 [hep-ph]].
 
 \bibitem{mores} 
 For some related LHC searches, see
  M.~Aaboud {\it et al.} [ATLAS Collaboration],
  arXiv:1811.07370 [hep-ex];
  A.~M.~Sirunyan {\it et al.} [CMS Collaboration],
  [arXiv:1810.10069 [hep-ex]];
  A.~M.~Sirunyan {\it et al.} [CMS Collaboration],
  arXiv:1811.07991 [hep-ex].
  
 \bibitem{ljet2}
  G.~Aad {\it et al.} [ATLAS Collaboration],
  JHEP {\bf 1411}, 088 (2014)
  [arXiv:1409.0746 [hep-ex]];
  G.~Aad {\it et al.} [ATLAS Collaboration],
  JHEP {\bf 1602}, 062 (2016)
  [arXiv:1511.05542 [hep-ex]];
  M.~Del Gaudio [ATLAS Collaboration],
  PoS EPS {\bf -HEP2017}, 690 (2018);
  ATLAS note ATLAS-CONF-2016-042;
  V.~Khachatryan {\it et al.} [CMS Collaboration],
  Phys.\ Lett.\ B {\bf 752}, 146 (2016)
  [arXiv:1506.00424 [hep-ex]];
  CMS Collaboration [CMS Collaboration],
  CMS-PAS-HIG-18-003.

\bibitem{Alpigiani:2018fgd} 
  C.~Alpigiani {\it et al.},
  ``A Letter of Intent for MATHUSLA: a dedicated displaced vertex detector above ATLAS or CMS.,''
  CERN-LHCC-2018-025, LHCC-I-031.
  
\bibitem{Ariga:2018zuc} 
  A.~Ariga {\it et al.} [FASER Collaboration],
  arXiv:1811.10243 [physics.ins-det].

\bibitem{Bhattiprolu:2019vdu} 
  P.~N.~Bhattiprolu and S.~P.~Martin,
  arXiv:1905.00498 [hep-ph].

\bibitem{Willenbrock:1985tj} 
  S.~S.~D.~Willenbrock and D.~A.~Dicus,
  Phys.\ Lett.\  {\bf 156B}, 429 (1985).
  doi:10.1016/0370-2693(85)91638-7

\bibitem{Gunion:1987xi} 
Cross sections can be easily obtainable with small modifications to the results as presented in 
  J.~F.~Gunion, J.~L.~Hewett, E.~Ma, T.~G.~Rizzo, V.~D.~Barger, N.~Deshpande and K.~Whisnant,
  Int.\ J.\ Mod.\ Phys.\ A {\bf 2}, 1199 (1987).
  doi:10.1142/S0217751X87000582

\bibitem{Coutinho:1991pd} 
  Y.~do Amaral Coutinho, J.~A.~Martins Simoes and M.~C.~Pommot Maia,
  Phys.\ Rev.\ D {\bf 45}, 771 (1992).
  doi:10.1103/PhysRevD.45.771

\bibitem{alloul2014feynrules}
  A.~Alloul, N.~D.~Christensen, C.~Degrande, C.~Duhr, and B.~Fuks,
  Computer Physics Communications {\bf 185}, 8 pp. 2250--2300 (2014)
  doi:10.1016/j.cpc.2014.04.012
  [arXiv:1310.1921]

\bibitem{alwall2014automated}
  J.~Alwall, R.~Frederix, S.~Frixione, V.~Hirschi, F.~Maltoni, O.~Mattelaer, H.~S.~Shao, T.~Stelzer, P.~Torrielli, and M.~Zaro,
  JHEP {\bf 07}, 79 (2014) 
  [arXiv:1405.0301 [hep-ph]].

\bibitem{sjostrand2015introduction}
  T.~Sj{\"o}strand, S.~Ask, J.~R.~Christiansen, R.~Corke, N.~Desai, P.~Ilten, S.~Mrenna, S.~Prestel, C.~O.~Rasmussen, and P.~Z.~Skands,
  Computer Physics Communications {\bf 191}, 159--177 (2015)
  doi:10.1016/j.cpc.2015.01.024
  [arXiv:1410.3012]

\bibitem{de2014delphes}
  J.~De Favereau, C.~Delaere, P.~Demin, A.~Giammanco, V.~Lemaitre, A.~Mertens, M.~Selvaggi, and Delphes 3 Collaboration
  JHEP {\bf 2}, 57 (2014)
  doi:10.1007/JHEP02(2014)057
  [arXiv:1307.6346]

\bibitem{aaboud2018search}
  M.~Aaboud {\it et. al} [ATLAS Collaboration]
  JHEP {\bf 1}, 126 (2018)
  doi:10.1007/JHEP01(2018)126
  [arXiv:1711.03301]

\bibitem{aghanim2018planck}
  N,~Aghanim {\it et al.} [Planck Collaboration]
  {\it Planck 2018 results. VI. Cosmological parameters} [arXiv:1807.06209]

\bibitem{madhavacheril2014current}
  M.~S.~Madhavacheril, N.~Sehgal, and T.~R.~Slatyer
  Phys. Rev. D, {\bf 89}, no. 10, 103508 (2014).
  doi:10.1103/PhysRevD.89.103508
  [arXiv:1310.3815]

\bibitem{liu2016contributions}
  H.~Liu, T.~R.~Slatyer, and J.~Zavala
  Phys. Rev. D, {\bf 94}, no. 6, 063507 (2016).
  [arXiv:1604.02457]

\end{thebibliography}
\end{document}